\documentclass[a4paper,fleqn]{cas-dc}

\usepackage[authoryear,longnamesfirst]{natbib}

\usepackage{bm}
\usepackage{algorithm}
\usepackage{algorithmic}
\usepackage{subcaption}
\usepackage{float}
\usepackage{amsmath}
\usepackage{textgreek} 

\graphicspath{{figs/flaredCone/},{figs/}}

\def\tsc#1{\csdef{#1}{\textsc{\lowercase{#1}}\xspace}}
\tsc{WGM}
\tsc{QE}
\tsc{EP}
\tsc{PMS}
\tsc{BEC}
\tsc{DE}

\begin{document}
\let\WriteBookmarks\relax
\def\floatpagepagefraction{1}
\def\textpagefraction{.001}

\shorttitle{Discontinuous Galerkin Methods for Hypersonic Flows}

\shortauthors{Hoskin, Heyningen, Nguyen, Vila-Pérez, Harris, Peraire}

\title [mode = title]{Discontinuous Galerkin Methods for Hypersonic Flows}                      



%

\author[1]{Dominique S. Hoskin}

\ead{dhoskin@mit.edu}

\credit{Data curation, Writing - Original draft preparation}

\author[1]{R. Loek Van Heyningen}
\ead{rloekvh@mit.edu}

\credit{Data curation, Writing - Original draft preparation}

\author[1]{Ngoc Cuong Nguyen}[orcid=0000-0001-9167-5780]

\cormark[1]


\ead{cuongng@mit.edu}

\ead[url]{https://www.mit.edu/~cuongng/}

\credit{Conceptualization of this study, Methodology, Software, Data curation, Writing - Original draft preparation}

\affiliation[1]{organization={Department of Aeronautics and Astronautics, Massachusetts Institute of Technology},
    addressline={77 Massachusetts Avenue}, 
    city={Cambridge},
    postcode={MA 02139}, 
    country={United States}}

\author[1]{Jordi Vila-P\'erez}[orcid=0000-0003-3164-0863]

\ead{jvilap@mit.edu}

\credit{Data curation, Writing - Original draft preparation}


\author[1]{Wesley L. Harris}
\ead{weslhar@mit.edu}

\credit{ Writing - Original draft preparation}

\author[1]{Jaime Peraire}[orcid=0000-0002-8556-685X]
\ead{peraire@mit.edu}
\ead[URL]{https://aeroastro.mit.edu/people/jaime-peraire/}

\credit{ Methodology, Writing - Original draft preparation}


\cortext[cor1]{Corresponding author}



 
\begin{abstract}
In recent years, high-order discontinuous Galerkin (DG) methods have emerged as an attractive approach for numerical simulations of compressible flows. This paper presents an overview of the recent development of DG methods for compressible flows with particular focus on hypersononic flows. First, we survey state-of-the-art DG methods for computational fluid dynamics. Next, we discuss both matrix-based and matrix-free iterative methods for the solution of discrete systems stemming from the spatial DG discretizations of the compressible Navier-Stokes equations. We then describe various shock capturing methods to deal with strong shock waves in hypersonic flows. We discuss adaptivity techniques to refine high-order meshes, and synthetic boundary conditions to simulate free-stream disturbances in hypersonic boundary layers. We present a few examples to demonstrate the ability of high-order DG methods to provide accurate solutions of hypersonic laminar flows. Furthermore, we present direct numerical simulations of hypersonic transitional flow past a flared cone at Reynolds number $10.8 \times 10^6$, and hypersonic transitional shock wave boundary layer interaction flow over a flat plate at Reynolds number $3.97 \times 10^6$. These simulations run entirely on hundreds of graphics processing units (GPUs) and demonstrate the ability of DG methods to directly resolve hypersonic transitional flows, even at high Reynolds numbers, without relying on transition or turbulence models. We end the paper by offering our perspectives on error estimation,  turbulence modeling, and real gas effects in hypersonic flows.
\end{abstract}





\maketitle

\section{Introduction}

 
Many critical decisions in the design of hypersonic vehicles require the ability to predict accurately surface skin friction and aerodynamic heating which, in turn, depend on complex physical processes such as shock wave/boundary layer interaction, laminar-to-turbulent transition, boundary layer separation,  real gas effects, surface chemistry, and heat transfer. Despite considerable efforts in experimental, theoretical, and numerical studies, many critical physical mechanisms in hypersonic boundary layers are not well understood \citep{Zhong2012,Schneider2015}. A unique feature of a hypersonic boundary layers is the presence of a family of acoustic instability modes, the Mack modes \citep{Mack1984a}, in addition to the Tollmien-Schlichting (T-S) waves and the vorticity modes encountered in lower speed flows \citep{STETSON1992}. The Mack modes consist of high frequency, large amplitude density fluctuations and can dominate the transition process. The different instability modes combined with the nature of the free stream disturbances lead to many different paths to flow transition including natural transition, disturbance-induced transition, crossflow-induced transition, separation-induced transition, shock-induced transition, roughness-induced transition, nose bluntness entropy-layer on transition, bypass transition, and transition-reversal phenomena \citep{Fedorov2011a}. These unique characteristics of hypersonic boundary layers make numerical prediction a very difficult task. 

Numerical tools for  turbulence studies still rely heavily on Reynolds--Averaged Navier–-Stokes (RANS) models. However, RANS models are limited in their ability to model laminar-to-turbulent transition,  shock wave boundary layer interaction, and separated turbulence  due to the complex and multifaceted characteristics of hypersonic flows \citep{Roy2006}. Direct numerical simulation (DNS) resolves the whole range of spatial and temporal scales of the turbulence. Although DNS can accurately predict complex physical phenomena encountered in hypersonic flows, it is computationally prohibitive for most practical applications. Large eddy simulation (LES) is an alternative approach which holds the promise to address the shortcomings of both RANS and DNS. With the advent of modern low-energy consumption highly scalable architectures, DNS and LES become accessible for a wide variety of turbulent flows.

Although low-order numerical methods remain frequent use in computational fluid dynamics, high-order methods are increasingly used for LES and DNS computations of turbulent flows. Indeed, the prediction of transitional turbulent flows relies on resolving wave propagation phenomena and small-scale flow features for which high-order accuracy is absolutely needed. Consider for instance, natural transition to turbulence, in which small perturbations are exponentially amplified in the unstable boundary layer. These small perturbations are ultimately responsible for the nonlinear breakdown and transition to turbulence. The amplitude of these perturbations at the location in which the boundary layer becomes unstable is several orders of magnitude (typically 5 to 8) below the free-stream velocity \cite{Fernandez2017a}. Attempting to capture such small  perturbations inside the boundary layers with low-order methods requires very fine meshes that render them less efficient than higher-order methods.  As a result, high-order methods are often used to predict unsteady turbulent flows. 

High-order DG methods possess a number of unique features that make them well-suited to predicting turbulent flows past complex geometries. First, DG methods are based on a strong mathematical foundation that can be exploited for error estimation and mesh adaptation, as presented by \cite{Yano2012,Ceze,Kast2013,Dahm2014a}. Second, DG methods provide local conservation and a stable discretization of the convective operator, which is important for convection-dominated flows \citep{Cockburn1998,Nguyen2012}. And third, DG methods allow for high-order discretizations on complex geometries and unstructured meshes, which is crucial to simulate turbulent flows past complex three-dimensional geometries \citep{Froehle2014,CuongNguyen2022}. For moderate polynomial degrees (between 2 and 4),  DG methods introduce numerical dissipation in under-resolved computations of convection-dominated flows, which acts as an implicit filter to dissipate the unresolved turbulent features  \citep{Fernandez2019}. In addition, they are well-suited to emerging computing architectures, including graphics processing units (GPUs) and other many-core architectures, due to their high flop-to-communication ratio \citep{CuongNguyen2022,Vila-Perez2022}. Indeed, the use of discontinuous Galerkin (DG) methods for computational fluid dynamics has gained considerable attention from multiple researchers, such as \cite{Uranga2011b,Gassner2013,Beck2014,Renac2015,DeWiart2015a,Murman2016,Fernandez2017a,Frere2017}. 


DG methods have been widely used in conjunction with explicit time-stepping schemes for unsteady calculations \citep{Cockburn1998,cockburn01:_rkdg,StanglmeierNguyenPeraireCockburn16}. However, explicit time integration is often impractical for resolving boundary layers due to the severe time-step size restriction --- an issue that can be overcome by using implicit time integration \citep{PerssonPeraire08}. When they are paired with implicit time-marching schemes, most DG methods yield large systems of equations due to the duplication of degrees of freedom along the element faces \citep{PerairePersson08,PerssonPeraire08,Nguyen2012}. The hybridizable DG (HDG) methods were introduced in \cite{CockburnGopalakrishnanLazarov09HDG} as part of the effort of devising efficient implicit DG methods and subsequently extended to CFD problems  \citep{Peraire2010,Moro2011a,Nguyen2011h,Nguyen2012,Schutz2013}. Indeed, the HDG methods guarantee that only the degrees of freedom  of the approximation of the scalar variable on the element boundaries are globally coupled and that the approximate gradient attains optimal order of convergence for diffusion-dominated problems \citep{CockburnDongGuzman08,CockburnGuzmanWang09SDG,CockburnGopalakrishnanSayas09}. A novel variant of the HDG method is the embedded DG (EDG) method \citep{Nguyen2015c}, which has the same global degrees of freedom as the static condensation of the continuous finite element method. Hence, the EDG method has the lowest global degrees of freedom compared to any other DG methods. The significant reduction in the number of global unknowns results in savings in terms of computation times and memory storage for the EDG method. However, unlike the HDG method, the EDG method does not possess superconvergence properties. Blending  EDG and HDG methods in a single discretization yields the IEDG method \citep{Fernandez2017a}. The IEDG method allows for the HDG discretization to be used on any part of the domain to enhance solution accuracy in that region, while using the EDG discretization elsewhere.
 
Iterative methods have been developed to solve discrete systems stemming from DG discretizations. In \cite{Gopalakrishnan2003}, a Gauss-Seidel block smoother (GS) was shown to have optimal behavior for linear convection-diffusion systems combined with the multigrid method. In \cite{Nastase2006}, linear and nonlinear multigrid methods combined with block Jacobi and block GS smoothers were demonstrated for subsonic inviscid flows on fairly isotropic meshes. In \cite{Fidkowski2005}, a multigrid smoother based on the solution of block tridiagonal systems was shown to significantly outperform  block Jacobi smoother for DG discretizations of the Navier-Stokes equations. In \cite{PerssonPeraire08}, block-Jacobi and Gauss-Seidel smoothers were shown to lack robustness for low Mach numbers, stretched grids or high Reynolds number turbulent flows. The block incomplete LU factorization with coarse grid correction was shown to outperform the above-mentioned preconditioners \citep{PerssonPeraire08}. In \cite{Fernandez2017a}, a restricted additive Schwarz preconditioner combined with block ILU0 was developed for Newton-GMRES solution of nonlinear systems arising from the HDG discretization of the Navier-Stokes equations and applied to implicit LES of transitional flows at moderate Reynolds. 
Finally, in recent years, \cite{CuongNguyen2022} introduced a reduced basis (RB) preconditioning technique suitable for matrix-free iterative methods.

The formation, propagation, and interaction of shock waves represent one of the most challenging problems in hypersonic flows. Difficulties in simulating shock flows  are that (1) at the very moment a shock is formed it poses a source of instability in the shock region, which then leads to numerical instabilities if no treatment of shock waves is introduced; (2) it is hard to predict when and where new shocks arise, and track them as they propagate through the physical domain and interact with each other and with boundary layers and vortices; and (3) numerical treatment of shock waves should not cause deterioration in resolution and reduction of accuracy in domains where the solution is smooth. In addition to shock waves, a number of other sharp features such as contact discontinuities, high thermal gradients, and thin shear layers may also appear in  turbulent hypersonic flows.  For high-order methods, insufficient resolution or an inadequate treatment to capture these shocks often results in  oscillations, which can grow rapidly and contribute to numerical instabilities. These challenges have been a driving force behind the development of shock capturing methods designed to detect and stabilize shocks. 

A number of shock detection methods rely on the non-smoothness of the numerical solution to detect shocks as well as other sharp features \cite{Cook2004,Cook2005,Fiorina2007,Kawai2008,Kawai2010,Klockner2011,MR2056921,Mani2009,Olson2013,persson06:_shock_capturing,Persson2013}. Among them, the sensor by \cite{MR2056921}, devised in the context of DG methods, takes advantage of the theoretical convergence rate of DG schemes for smooth solutions in order to detect discontinuities. The shock sensor by \cite{persson06:_shock_capturing,Persson2013} is based on the decay rate of the coefficients of the DG polynomial approximation. Other methods that rely on high-order derivatives of the solution include \cite{Cook2004,Cook2005,Fiorina2007,Kawai2008,Kawai2010,Klockner2011,Mani2009,Olson2013,Premasuthan2010b_tmpfix,Premasuthan2014}, and apply to numerical schemes for which such derivatives can be computed, such as spectral methods and finite difference methods. The most simple shock-detection method is to take advantage of the strong compression that a fluid undergoes across a shock wave and use the divergence of the velocity field as a shock sensor \cite{Fernandez2018,Moro2016,Nguyen2011a}.  Shock stabilization methods lie within one of the following two categories: limiters and artificial viscosity. Limiters, in the form of flux limiters \cite{Burbeau2001,Cockburn1989,krivodonova2007}, slope limiters \cite{Cockburn1998a,MR2056921,Lv2015,Sonntag2017}, and WENO-type schemes \cite{Luo2007,Qiu2005,Zhu2008,Zhu2013} pose implementation difficulties for implicit time integration schemes and high-order methods on complex geometries. As for artificial viscosity methods, Laplacian-based \cite{Barter2010,Hartmann2013,Lv2016,Moro2016,Nguyen2011a,persson06:_shock_capturing,Persson2013} and physics-based \cite{Abbassi2014,Chaudhuri2017,Cook2004,Cook2005,Fernandez2018,Fiorina2007,Kawai2008,Kawai2010,Mani2009,Olson2013,persson06:_shock_capturing} approaches have been proposed. An assessment of artificial viscosity methods for LES is presented in \cite{Johnsen2010}. A more recent shock capturing scheme \cite{Zahr2020} aims to align mesh elements along shocks and employ DG methods to handle shocks. 

The quality of meshes has a significant impact on the accuracy of numerical solutions. This is particularly so for hypersonic flows because numerical solutions of hypersonic flows are much more sensitive to mesh resolutions, element types, and element orders than those of transonic and supersonic flows \cite{Barter2010}. For DG methods, it is important to generate high-order meshes that have sufficient resolutions to resolve shock waves and boundary layers. Insufficient mesh resolution can result in inaccurate predictions or even lead to numerical instability due to the oscillatory behavior of the numerical solution. Mesh adaptation encompasses a wide array of techniques, each has its advantages and complexities. The $r$-adaptivity methods \cite{Zahr2018, Nguyen2023c, ameur2021r} move the nodes of the mesh elements without changing the mesh topology. The advantage of $r$-adaptivity is that the numerical accuracy can be enhanced for a fixed number of elements. In contrast to $r$-adaptivity, $h$- and $p$-adaptivity methods do not move the nodes of the mesh, but instead locally refine/coarse individual elements or vary polynomial degrees.  DG methods are compatible with automatic mesh refinement (AMR) techniques \cite{panourgias2016discontinuous} and anisotropically adapted meshes \cite{hecht1998bamg}, the latter of which can be especially useful for adapting to shocks \cite{Barter2008, Moro2016}. The advantage of $h$- and $p$-adaptivity  is that each element can be refined to obtain arbitrary resolution in any region of the mesh regardless of the shape and size of the original mesh. While combinations of the individual adaptation techniques such as $h/p$ \cite{BeyOden96, brazell20133d}, 
$h/r$ \cite{edwards1993h, antonietti2009hr, dobrev2022hr}, and $r/p$ \cite{bhatia20132} adaptation can improve the quality of meshes even further, their implementation is complicated due to several reasons. Designing algorithms that dynamically adjust both mesh sizes and polynomial degrees is complex. Managing data structures to accommodate variable mesh sizes and polynomial degrees while ensuring compatibility with iterative solvers can be challenging. Transferring solutions between different meshes with variable polynomial degrees demands sophisticated interpolation schemes to maintain accuracy.
These methods require some form of indicator to drive mesh refinement. 
Feature-based indicators use features of the solution such as flow field gradients \cite{panourgias2016discontinuous} or Hessians \cite{Moro2016}, while
goal-oriented methods use error estimates of quantities of interest to drive adaptation \cite{yano2011importance, rangarajan2020adjoint}

Numerical results are presented to illustrate DG methods for a wide range of hypersonic flow phenomena. We present numerical solutions of hypersonic flow past a circular cylinder at $M_\infty = 17.6$ and $Re=376,000$, and type IV shock-shock interaction over a cylinder in free stream Mach number 8.03 and $Re=388,000$, which are computed using the HDG method with polynomial degree 4 on $r$-adaptive meshes. These examples serve to demonstrate the robustness and accuracy of the high-order HDG discretization equipped with shock capturing and mesh adaptivity. Furthermore, we present direct numerical simulations of hypersonic transitional flow past a flared cone at  $Re= 10.8 \times 10^6$, and hypersonic transitional shockwave boundary layer interaction flow over a flat plate at $Re= 3.97 \times 10^6$. These simulations run entirely on hundreds of GPUs and demonstrate the ability of DG methods to directly resolve hypersonic transitional flows, even at high Reynolds numbers, without relying on transition or turbulence models. 

The remainder of the paper is organized as follows. In Section 2, we survey state-of-the-art DG methods in computational fluid dynamics. In Section 3, we discuss iterative methods and preconditioners for implicit DG methods. We describe shock capturing algorithms in Section 4 and mesh adaptation techniques in Section 5. In Section 6, we discuss boundary-layer instabilities, receptivity mechanisms, and boundary conditions for introducing disturbances into boundary layers. In Section 7, we present numerical results to demonstrate DG methods for hypersonic flows. The paper is concluded with our perspectives on error estimation, turbulence modeling, and real gas effects in hypersonic flows. 

\section{Discontinuous Galerkin Methods}

\subsection{Governing equations}

We consider the conservation laws of $m$  state variables, defined on a physical domain $\Omega \in \mathbb{R}^d$ and subject to appropriate  boundary conditions, as follows
\begin{equation}
\label{eq1}
\frac{\partial \bm u}{\partial t} +  \nabla \cdot \bm F(\bm u, \nabla \bm u) = 0  \quad \mbox{in }\Omega \times (0, T],    
\end{equation}
where $\bm u(\bm x, t) \in \mathbb{R}^m$ is the solution of the system of conservation laws at $(\bm x, t) \in \Omega \times [0, T]$ and the  physical fluxes $\bm F = (\bm f_1(\bm u, \nabla \bm u), \ldots, \bm f_d(\bm u, \nabla \bm u)) \in \mathbb{R}^{m \times d}$ include $d$ vector-valued functions of the solution. The initial condition is $\bm u(\bm x, 0) = \bm u_0(\bm x)$, where $\bm u_0$ is the initial state. This paper focuses on compressible flows in the hypersonic regime. 

For the compressible Euler equations, the state vector and physical fluxes are given by
\begin{equation}
    \bm u = \begin{pmatrix}
        \rho \\
        \rho v_i \\
        \rho E \\
    \end{pmatrix}, \qquad  
    \bm F(\bm u) = \begin{pmatrix}
        \rho v_j \\ 
        \rho v_i v_j + \delta_{ij} p \\ 
        \rho v_j H
    \end{pmatrix}
    \label{eq:euler}
\end{equation}
with density $\rho$, velocity $\bm v$, total energy $E$, total specific enthalpy $H = E + p/\rho$ and pressure $p$ given by the ideal gas law $p = (\gamma - 1)  \rho ( E-\frac{1}{2}v_i \ v_i)$. Let $\Gamma_{\rm wall} \subset \partial \Omega$ be the wall boundary. The boundary condition at the wall boundary $\Gamma_{\rm wall}$ is $\bm v \cdot \bm n = 0$, where $\bm v$ is the velocity field and $\bm n$ is the unit normal vector outward the boundary. Supersonic inflow and outflow conditions are imposed on the inflow and outflow boundaries, respectively. For supersonic inflow, a free-stream boundary condition is imposed at the inflow boundary using the free-stream state $\bm u_\infty$. The free-stream Mach number $M_\infty$ enters through the non-dimensional free-stream pressure $p_{\infty} = 1 / (\gamma M^2_{\infty})$, where $\gamma = c_p/c_v$ denotes the specific heat ratio, $c_p$ and $c_v$ are the specific heats at constant pressure and volume. 

For the compressible Navier-Stokes equations, the fluxes are given by
\begin{equation}
\label{flux}
\bm{F}(\bm{u},\nabla \bm u) = \left( \begin{array}{c}
\rho v_j \\
\rho v_i v_j + \delta_{ij} p \\
 \rho v_j H
\end{array}
\right) - \left( \begin{array}{c}
0 \\
\tau_{ij}  \\
v_i \tau_{ij} + f_j
\end{array}
\right) .
\end{equation}
For a Newtonian, calorically perfect gas in thermodynamic equilibrium, the non-dimensional viscous stress tensor and heat flux are given by
$$
\tau_{ij} = \mu_f \bigg[ \Big( \frac{\partial v_i}{\partial x_j}+\frac{\partial v_j}{\partial x_i} \Big) -\frac{2}{3}\frac{\partial v_k}{\partial x_k}\delta_{ij} \bigg] ,  \quad f_j = \kappa_f \ \frac{\partial T}{\partial x_j} ,
$$
respectively. Here, $T$ denotes the temperature, $\mu_f$ the dynamic (shear) viscosity, $\kappa_f = c_p \, \mu_f / Pr$ the thermal conductivity,  $Pr$ the Prandtl number, and $\beta_f = 0$ the bulk viscosity under Stokes' hypothesis. For high Mach number flows, Sutherland's law is used to obtain the dynamic viscosity which is dependent on the temperature. The boundary conditions at the wall impose zero velocity and either isothermal or adiabatic conditions for the temperature. Other boundary conditions are similar to those of the compressible Euler equations. In Section 8, we extend our discussion to turbulence modeling and real-gas effects. The DG formalism described in this section is applicable to these models either without or with little modifications.

\subsection{General weak formulation}

We denote by $\mathcal{T}_h$ a collection of disjoint regular elements $K$ that partition $\Omega$, and set $\partial \mathcal{T}_h := \{ \partial K : K \in \mathcal{T}_h \} $ to be the collection of the boundaries of the elements in $\mathcal{T}_h$. Let $\mathcal{F}_h$ be a collection of faces in $\mathcal{T}_h$.  Let $\mathcal{P}^{k}(D)$ denote the space of complete polynomials of degree $k$ on a domain $D \in \mathbb{R}^n$, let $L^2(D)$ be the space of square-integrable functions on $D$. We  introduce the following discontinuous finite element spaces:
$$\bm{\mathcal{Q}}_{h}^k  = \big\{\bm{q} \in [L^2(\mathcal{T}_h)]^{m \times d} \ : \ \bm q|_K \in \bm W(K),   \ \ \forall K \in \mathcal{T}_h \big\} , $$
$$\bm{\mathcal{V}}_{h}^k  = \big\{\bm{v} \in [L^2(\mathcal{T}_h)]^m \ : \ \bm v|_K \in \bm V(K), \ \ \forall K \in \mathcal{T}_h \big\} , $$
$$\bm{\mathcal{M}}_{h}^k  = \big\{\bm{\mu} \in [L^2(\mathcal{F}_h)]^m \ : \ \bm \mu|_F \in \bm V(F), \ \ \forall F \in \mathcal{F}_h \big\} , $$
where $\bm{W}(K) \equiv [\mathcal{P}^k(K)]^{m \times d}$ and $\bm{V}(K) \equiv [\mathcal{P}^k(K)]^{m}$. Next, we define several inner products associated with these finite element spaces as
\begin{subequations}
\label{innerProducts}
\begin{alignat}{3}
& (\bm{w},\bm{v})_{\mathcal{T}_h} && = \sum_{K \in \mathcal{T}_h} (\bm{w}, \bm{v})_K && = \sum_{K \in \mathcal{T}_h} \int_{K} \bm{w} \cdot \bm{v}  , \nonumber \\
& (\bm{W},\bm{V})_{\mathcal{T}_h} && = \sum_{K \in \mathcal{T}_h} (\bm{W}, \bm{V})_K && = \sum_{K \in \mathcal{T}_h} \int_{K} \bm{W} : \bm{V} , \nonumber \\
& \left\langle \bm{w}, \bm{v} \right\rangle_{\partial \mathcal{T}_h} && = \sum_{K \in \mathcal{T}_h} \left\langle \bm{w},\bm{v} \right\rangle_{\partial K} && = \sum_{K \in \mathcal{T}_h} \int_{\partial K} \bm{w} \cdot \bm{v} , \nonumber
\end{alignat}
\end{subequations}
for $\bm{w}, \bm{v} \in \bm{\mathcal{V}}_{h}^k$, $\bm{W}, \bm{V} \in \bm{\mathcal{Q}}_{h}^k$,  where $\cdot$ and $:$ denote the scalar product and Frobenius inner product, respectively. 

The DG discretization of the governing equations reads as follows: find $\big( \bm{q}_h,\bm{u}_h \big) \in \bm{\mathcal{Q}}_h^k \times \bm{\mathcal{V}}_h^k$ such that
\begin{subequations}
\label{IEDG}
\begin{alignat}{2}
\label{IEDGa}
\big( \bm{q}_h, \bm{r} \big) _{\mathcal{T}_h} + \big( \bm{u}_h, \nabla \cdot \bm{r} \big)  _{\mathcal{T}_h} -  \big< \widehat{\bm{u}}_h, \bm{r} \cdot \bm{n} \big> _{\partial \mathcal{T}_h}  & =  0, \\
\label{IEDGb}
\Big( \frac{\partial \, \bm{u}_h}{\partial t}, \bm{w} \Big)_{\mathcal{T}_h} - \Big(  \bm{F}, \nabla \bm{w} \Big) _{\mathcal{T}_h}  +  \left\langle \widehat{\bm{f}}_h, \bm{w} \right\rangle_{\partial \mathcal{T}_h}  & = 0, 
\intertext{for all $(\bm{r},\bm{w}) \in \bm{\mathcal{Q}}^k_h \times \bm{\mathcal{V}}^k_h$ and all $t \in (0,T]$, as well as}
\label{IEDGd}
\big( \bm{u}_{h}|_{t=0} - \bm{u}_0 , \bm{w} \big) _{\mathcal{T}_h} & =  0, 
\end{alignat}
\end{subequations}
for all $\bm{w} \in \bm{\mathcal{V}}^k_h$. Here  $\widehat{\bm{u}}_h$ is the numerical trace and $\widehat{\bm{f}}_h$ is the numerical flux. For DG methods, both the numerical trace and flux must be continuous across element boundaries. 

\subsection{Numerical trace and flux}

The general form of the numerical trace and flux on the interior faces that satisfies the continuity requirement is given by
\begin{equation}
\label{numflux}
\begin{split}
\widehat{\bm{u}}_h  &=  \{\bm u_h\} + \bm{C}_{21}  [\bm u_h] +  \bm{C}_{22} [ \bm{F}(\bm u_h , \bm{q}_h)] \cdot \bm n, \\
\widehat{\bm{f}}_h & = \{\bm{F}(\bm u_h , \bm{q}_h)  \}  \cdot \bm{n} + \bm{C}_{11}  [\bm u_h] + \bm{C}_{12} [\bm{F}(\bm u_h , \bm{q}_h)]  \cdot \bm{n} ,
\end{split}
\end{equation}
where $\{\bm u_h\} = (\bm u_h + \bm u_h^-)/2$, $[\bm u_h] = \bm u_h - \bm u_h^-$, $\{\bm{F}(\bm u_h , \bm{q}_h)  \} = (\bm{F}(\bm u_h , \bm{q}_h) + \bm{F}(\bm u_h^- , \bm{q}_h^-))/2$, $[\bm{F}(\bm u_h , \bm{q}_h)] = \bm{F}(\bm u_h , \bm{q}_h) - \bm{F}(\bm u_h^- , \bm{q}_h^-)$, and $\bm C_{11}, \bm C_{12}, \bm C_{21}, \bm C_{22}$ are  matrix-valued stabilization functions. Note that $(\bm u_h^-, \bm q_h^-)$ denotes the numerical solution from the neighboring elements that share the same faces as the element $K$. On the boundary faces, the definition of the numerical trace and flux depends on the boundary conditions. For supersonic inflow condition, we can set $\bm u_h^- = \bm u_\infty, \bm q_h^- = 0$. For supersonic outflow condition, we can set $\bm u_h^- = \bm u_h, \bm q_h^- = \bm q_h$. We refer to \cite{Fernandez2017a,Nguyen2012} for the implementation of the iso-thermal and adiabatic boundary conditions.

\subsection{Another form of the numerical flux} 

A number of DG methods define the numerical flux as follows
\begin{equation}
\label{numflux2}
\widehat{\bm{f}}_h  = \{\bm{F}(\bm u_h , \widehat{\bm{q}}_h)  \}  \cdot \bm{n} + \bm{C}_{11}  [\bm u_h] + \bm{C}_{12} [\bm{F}(\bm u_h , \widehat{\bm{q}}_h)]  \cdot \bm{n} ,
\end{equation}
where
\begin{equation}
\label{numflux3}
\widehat{\bm{q}}_h  = \nabla \bm u_h +  l([\bm u_h ] \bm n) .
\end{equation}
Here $l(\cdot)$ is known as the lifting operator. Different definitions of the lifting operator yield different DG methods, such as the interior penalty (IP) method \cite{DouglasDupont76}, the second Bassi Rebay (BR2) method \cite{bassi2005discontinuous}, the compact DG (CDG) method \cite{PerairePersson08}.  These DG methods have compact connectivity in the sense that only the degrees of freedom belonging to neighboring elements are connected in the discretization. DG methods with compact connectivity require less memory storage and operation counts to solve the resulting system of equations than their non-compact counterparts.  Note that the LDG and BR1 methods do not have compact connectivity.

\subsection{Stabilization functions}

%

The stabilization functions play an important role in the stability and accuracy of the resulting DG discretization. Indeed, different choices of the stabilization functions result in different DG methods. The choice $\bm C_{22} = 0$ yields the LDG method \cite{Cockburn1998}, while $\bm C_{11}  = \bm C_{12} = \bm C{21}  = \bm C_{22} = 0$ leads to the first Bassi-Rebay (BR1) method \cite{Bassi97:_dfem}. It is known that the LDG method is stable  for elliptic problems, whereas the BR1 method is only weakly stable \cite{arnold2002unified}. The stability of the LDG method comes from the stabilization function $\bm C_{11}$.

For general convection-diffusion systems, the stabilization function $\bm C_{11}$ should include both the convection-stabilizing term $\bm C_{11}^{\rm conv}$ and the diffusion-stabilizing term $\bm C_{11}^{\rm diff}$, namely $\bm C_{11} = \bm C_{11}^{\rm conv} + \bm C_{11}^{\rm diff}$, so that the resulting DG method can be stable in both pure convection limit and pure diffusion limit.

The convection-stabilizing term is usually computed by using approximate Riemann solvers such as Roe's scheme \cite{Roe1997}, Lax-Friedrich scheme \cite{Nguyen2012} or HLL/HLLEM schemes \cite{Vila-Perez2021}, while the diffusion-stabilizing term is proportional to the diffusion coefficient. See \cite{Fernandez2017a,Nguyen2012,Vila-Perez2021,Terrana2019a} for additional discussion on the stabilization of DG methods.

\subsection{Hybridization of DG methods}

With $\bm C_{22} = 0$, the numerical trace $\widehat{\bm{u}}_h$ depends solely on $\bm u_h$. In this case, we can locally solve (\ref{IEDGa}) for $\bm q_h$ in terms of $\bm u_h$ and substitute it into (\ref{IEDGb}) to obtain a global system in terms of the degrees of freedom of $\bm u_h$ only. Hence, the DG method with $\bm C_{22} = 0$ is referred to as the local DG (LDG) method. For $\bm C_{22} \neq 0$, because $\widehat{\bm{u}}_h$ depends on both $\bm u_h$ and $\bm q_h$, (\ref{IEDGa})-(\ref{IEDGb}) is a globally coupled system of the degrees of freedom of both $\bm u_h$ and $\bm q_h$. As a result, the DG method with $\bm C_{22} \neq 0$ can be much more expensive than the LDG method if both $\bm u_h$ and $\bm q_h$ are solved together. It is known that, for diffusion-dominated problems, the DG method with $\bm C_{22} \neq 0$ yields the optimal convergence rate for $\bm q_h$ and superconvergence properties that can be exploited to obtain a postprocessed solution $\bm u_h^*$ that converges faster than the approximate solution $\bm u_h$ \cite{CockburnGuzmanWang09SDG, Nguyen2009c, Nguyen2009d}. Furthermore, the DG method with $\bm C_{22} \neq 0$ can be efficiently implemented by using the hybridization technique as discussed below.  

The choice $\bm C_{12} = \bm C_{21} = 0$ is known as the hybridizable DG (HDG) method \cite{CockburnGopalakrishnanLazarov09HDG}.  In the HDG method, the numerical trace $\widehat{\bm{u}}_h$ becomes a dependent variable to be solved together with $(\bm{u}_h, \bm q_h)$ by introducing another equation that  weakly imposes the continuity of the numerical flux. This process is known as hybridization \cite{CockburnGopalakrishnanLazarov09HDG} and is widely used to implement mixed, CG, and DG finite element methods. The HDG discretization of the governing equations reads as follows: Find $\big( \bm{q}_h,\bm{u}_h,  \widehat{\bm{u}}_h\big) \in \bm{\mathcal{Q}}_h^k \times \bm{\mathcal{V}}_h^k \times \bm{\mathcal{M}}_h^k$ such that
\begin{subequations}
\label{HDG}
\begin{alignat}{1}
\label{HDGa}
\big( \bm{q}_h, \bm{r} \big) _{\mathcal{T}_h} + \big( \bm{u}_h, \nabla \cdot \bm{r} \big)  _{\mathcal{T}_h} -  \big< \widehat{\bm{u}}_h, \bm{r} \cdot \bm{n} \big> _{\partial \mathcal{T}_h}  & =  0, \\
\label{HDGb}
\Big( \frac{\partial \, \bm{u}_h}{\partial t}, \bm{w} \Big)_{\mathcal{T}_h} - \Big(  \bm{F}, \nabla \bm{w} \Big) _{\mathcal{T}_h}  +  \left\langle \widehat{\bm{f}}_h, \bm{w} \right\rangle_{\partial \mathcal{T}_h}  & = 0, \\
\label{HDGd}
  \left\langle \widehat{\bm{f}}_h, \bm{\mu} \right\rangle_{\partial \mathcal{T}_h\backslash \partial \Omega} + \left\langle \widehat{\bm{b}}_h, \bm{\mu} \right\rangle_{\partial \Omega}  & =  0, 
\end{alignat}
\end{subequations}
for all $\big( \bm{r},\bm{w}, \bm \mu\big) \in \bm{\mathcal{Q}}_h^k \times \bm{\mathcal{V}}_h^k \times \bm{\mathcal{M}}_h^k$. Here $\widehat{\bm{b}}_h$ is the boundary flux whose precise definition depends on the boundary conditions. For instance, we define $\widehat{\bm{b}}_h = \widehat{\bm u}_h - \bm u_\infty$ for inflow condition, and $\widehat{\bm{b}}_h = \widehat{\bm u}_h - \bm u_h$ for outflow condition. We refer to \cite{Nguyen2012} for the definition of $\widehat{\bm{b}}_h$ for wall boundary conditions.

The HDG method is computationally efficient for implicit time integration because it results in a global system of the degrees of freedom of $\widehat{\bm u}_h$. By locally solving (\ref{HDGa})-(\ref{HDGb}) for $(\bm q_h, \bm u_h)$ in terms of $\widehat{\bm u}_h$ and substituting it into (\ref{HDGd}), we obtain the desired global system. Since $\widehat{\bm u}_h$ often has much fewer degrees of freedom than $\bm u_h$, the HDG method can be more efficient than the LDG method. Due to its superconvergnce properties, the HDG method is more accurate than the LDG method for diffusion-dominated problems.

A novel variant of the HDG method is the embedded DG (EDG) method \cite{Nguyen2015c}. The EDG method results from (\ref{HDG}) by replacing $\bm{\mathcal{M}}_{h}^k$ with the following space
$$\widetilde{\bm{\mathcal{M}}}_{h}^k  = \big\{\bm{\mu} \in [C^0(\mathcal{F}_h)]^m \ : \ \bm \mu|_F \in \bm V(F), \ \ \forall F \in \mathcal{F}_h \big\} . $$
The EDG method has the same global degrees of freedom as the static condensation of the continuous finite element method. Hence, it has the lowest global degrees of freedom compared to other DG methods. The significant reduction in the number of global unknowns results in savings in terms of computation times and memory storage for the EDG method. However, unlike the HDG method, the EDG method does not possess superconvergence properties. Combining both the EDG and HDG methods in a single discretization yields the IEDG method \cite{Fernandez2017a}. The IEDG method allows HDG discretization to be used on any part of the domain to enhance solution accuracy in that region while using EDG discretization everywhere else.



\section{Solution Methods}

\subsection{Implicit time integration}

For any DG method, the spatial discretization of the governing equations yields the following index-1 differential-algebraic equation (DAE) system
\begin{subequations}
\label{ch5appendx1:eq8}
\begin{alignat}{2}
\mathbf M \frac{d \mathbf u}{d t} + \mathbf f(\mathbf u, \mathbf w) & = 0 , \qquad t > 0 , \\
\mathbf g(\mathbf u, \mathbf w) & = 0 , \qquad t \geq 0 , 
\end{alignat}
\end{subequations}
with initial condition $\mathbf u(t = 0) = \mathbf u_0$ and $\mathbf M$ being mass matrix. Here $\mathbf f$ and $\mathbf g$ are non-linear vector-valued functions which result from spatial discretization and depend on the DG method used. Note that $\mathbf u$ is the vector of degrees of freedom of $\bm u_h$ for all DG methods. While $\mathbf w$ is the vector of degrees of freedom of $\bm q_h$ for LDG, BR1, BR2, CDG, IP methods, whereas it is the vector of degrees of freedom of both $\bm q_h$ and $\widehat{\bm u}_h$ for HDG, EDG, IEDG methods. In this paper, we focus on implicit time-marching methods to integrate the above DAE system in time \cite{Nguyen2011k,Nguyen2012}. 

\subsubsection{Linear multistep methods}

We denote by $\mathbf u^n$ an approximation for the function $\mathbf u(t)$ at discrete time $t^n = n \, \Delta t$, where $\Delta t$ is the time step and $n$ is an integer.  Linear multistep (LM) methods use information from the previous $s$ steps, $\{\mathbf u^{n+i}\}_{i=0}^{s-1}$, to calculate the solution at the next step $\mathbf u^{n+s}$.  When we apply a general LM method to the differential part (\ref{ch5appendx1:eq8}a) and treat the algebraic part (\ref{ch5appendx1:eq8}b) implicitly, we arrive at the following algebraic system:
\begin{subequations}
\label{ch5appendx1:eq9}
\begin{alignat}{2}
 \sum_{i=0}^s \left( a_i \mathbf M \mathbf u^{n+i} + \Delta t \, b_i \mathbf f(\mathbf u^{n+i},\mathbf w^{n+i}) \right)  & =  0, \\
 \label{ch5appendx1:eq9_}
\mathbf g(\mathbf u^{n+s}, \mathbf w^{n+s}) & = 0 . 
\end{alignat}
\end{subequations}
The coefficient vectors $\mathbf a = (a_0,a_1,\ldots,a_s)$ and $\mathbf b = (b_0,b_1,\ldots,b_s)$ determine the method. If $b_s=0$, the method is called {\em explicit}; otherwise, it is called {\em implicit}. Backward difference formula (BDF) schemes with $s$ steps have $\mathbf b = (0,\ldots,0,b_s)$ with $b_s > 0$.

\subsubsection{Implicit Runge-Kutta methods}

The coefficients of an $s$-stage Runge-Kutta (RK) method, $a_{ij}, \, b_i, \, c_i , \ 1 \le i,j \le s$, are usually arranged in the form of a {\em Butcher tableau}:
\begin{equation}
\begin{array}{c|cccc}
c_1 & a_{11} & a_{12} & \ldots   & a_{1s} \\
c_2 & a_{21} & a_{22} & \ldots & a_{2s} \\
\vdots & &  \ldots & \vdots & \vdots \\
 c_s & a_{s1} & a_{s2} &  \ldots & a_{ss} \\
\hline
& b_1 & b_2 &  \ldots & b_s 
\end{array}
\end{equation}
For the family of implicit RK (IRK) methods, the RK matrix $a_{ij}$ must be invertible. Let $d_{ij}$ denote the inverse of $a_{ij}$, and let $\mathbf u^{n,i}$ be the approximation of $\mathbf u(t)$ at discrete times $t^{n,i} = (t_n + c_i \Delta t) , \ 1 \le i \le s$. The $s$-stage IRK method for the DAE system (\ref{ch5appendx1:eq8}) can be sketched as follows. First, we solve the following $2s$ coupled systems of equations
\begin{subequations}
\label{ch5appendx1:eq10}
\begin{equation}
\label{ch5appendx1:eq10_1}
\begin{split}
\sum_{j=1}^s d_{ij}\mathbf M \left(\mathbf u^{n,j} - \mathbf u^{n} \right) +  \Delta t  \mathbf f(\mathbf u^{n,i}, \mathbf w^{n,i})  &=  0,  \\
\mathbf g(\mathbf u^{n,i}, \mathbf w^{n,i}) & =  0, 
\end{split}
\end{equation}
for $(\mathbf u^{n,i}, \mathbf w^{n,i})$ and $i = 1,\ldots, s$. Then we compute $\mathbf u^{n+1}$ from
\begin{equation}
\label{ch5appendx1:eq10_2}
\mathbf u^{n+1}  =  \left(1 -  \sum_{j=1}^s e_j \right)\mathbf u^{n} +  \sum_{j=1}^s e_{j} \mathbf u^{n,j} ,
\end{equation}
where $e_j =  \sum_{i=1}^s b_i d_{ij}$. Finally, we solve the following system of equations for $\mathbf w^{n+1}$:
\begin{equation}
\label{ch5appendx1:eq10_3}
\mathbf g(\mathbf u^{n+1}, \mathbf w^{n+1})  =  0 .
\end{equation}
\end{subequations}
If the RK matrix $a_{ij}$ is a lower-triangular matrix, then the method is called {\em diagonally implicit} RK (DIRK) scheme \cite{Alexander77}. In the case of a DIRK method, each stage of the system \eqref{ch5appendx1:eq10} can be viewed as the backward Euler step due to the fact the matrix $d_{ij}$ is lower-triangular.

\subsection{Matrix-based solvers}

Implicit time integration of the DAE system (\ref{ch5appendx1:eq8}) yields a nonlinear system of equations. Newton’s method is usually used to linearize the nonlinear system. At each Newton's iteration, we need to solve a linear system. For DG methods, the resulting linear system can be statically condensed to yield a smaller linear system 
\begin{equation}
\label{LS}
\mathbf J \delta \mathbf x = \mathbf r    
\end{equation}
where $\mathbf J$ is the Jacobian matrix, $\mathbf r$ is the residual vector, and $\delta \mathbf x$ is the Newton increment vector. The size and sparsity pattern of $\mathbf J$ depends on which DG method is used for spatial discretization. The size of $\mathbf J$ is equal to the degrees of freedom of $\bm u_h$ for LDG, CDG, BR, IP methods, and of $\widehat{\bm u}_h$ for HDG, EDG, IEDG methods. The sparsity pattern of $\mathbf J$ is compact, meaning that only degrees of freedom in neighboring elements are connected, for BR2, CDG, IP methods. For the HDG method, only degrees of freedom in neighboring faces are connected. For the EDG method, $\mathbf J$ has the same sparsity pattern as that of statically condensed continuous FE methods.

Krylov subspace methods are widely used to solve the linear system (\ref{LS}). In general, Krylov methods must be preconditioned to perform well. This amounts to applying Krylov methods to the preconditioned linear system 
\begin{equation}
\label{LS2}
\mathbf P^{-1} \mathbf J \delta \mathbf x = \mathbf P^{-1}  \mathbf r    
\end{equation}
where  $P$ is an appropriate preconditioner. Popular choices for $\mathbf P$ include block Jacobi, block Gauss Seidel, and block ILU preconditioners \cite{PerssonPeraire08}. Although it is harder to implement, block ILU is significantly more effective than block Jacobi and Gauss Seidel. The performance of an ILU factorization depends on the ordering of the unknowns, assuming the elimination is done from top to bottom. The minimum discarded fill algorithm is often used to reorder the unknowns \cite{PerssonPeraire08}. 

Multigrid methods are often used to solve the linear system (\ref{LS}) by introducing coarser level discretizations \cite{Gopalakrishnan2003,Fidkowski2005,Nastase2006,Shahbazi2009}. This coarser discretization can be obtained either by using a coarser mesh ($h$-multigrid) or, for hig-horder methods, by reducing the polynomial degree ($p$-multigrid) \cite{Nastase2006}. For DG methods, it is natural to consider coarser scales obtained by reducing the polynomial degree \cite{PerssonPeraire08}. The residual is restricted to the coarse-scale discretization (lower polynomial degree), where an approximate error is computed, which is then applied as a correction to the fine-scale discretization. Some iterations of a smoother are applied to the residual before and after the correction to reduce the high-frequency errors. Block Jacobi,  Gauss Seidel, and ILU0 are commonly used as smoothers for multigrid methods. Multigrid methods can be used either as an independent iterative solver or a preconditioner in Krylov subspace methods.

For large-scale problems, it is necessary to solve the linear systems on many processors. Restricted additive Schwarz (RAS) method \cite{Cai1999} with $\beta$-level overlap is widely used as a parallel preconditioner. This approach is based on the distribution of the unknowns among parallel workers. The RAS preconditioner for $N$ subdomains is defined as
\begin{equation}
 \mathbf{P}_{RAS}^{-1} := \sum_{i=1}^{N} \mathbf{R}_{i}^{0} \ \mathbf{J}_{i}^{-1} \ \mathbf{R}_{i}^{\beta} ,
\end{equation}
where $\mathbf{J}_{i} = \mathbf{R}_{i}^{\beta} \ \mathbf{J} \ \mathbf{R}_{i}^{\beta}$ is the Jacobian matrix on the $i$th subdomain, and $\mathbf{R}_{i}^{\beta}$ is the restriction operator onto the subspace associated to the nodes in the $\beta$-level overlap of the $i$th subdomain.  One-level overlap provides the best balance between the communication cost and the computational cost. The subdomain problems $\mathbf{J}_{i}^{-1} \ \mathbf{R}_{i}^{\beta}$ can be solved using Krylov or multigrid methods as discussed earlier. In practice, we replace the exact subdomain problems with $\mathbf{P}_{i}^{-1} \ \mathbf{R}_{i}^{\beta}$ to reduce the computational cost, where $\mathbf{P}_{i}$ denotes the preconditioner of $\mathbf{J}_{i}$ \cite{Fernandez2017a}.

\subsection{Matrix-free solvers}

Iterative methods require calculation of matrix-vector products, which can be expensive if we have to form the Jacobian matrix and perform matrix-vector multiplication. Instead, the product of the Jacobian matrix $\mathbf J(\mathbf x)$ with any vector $\mathbf y$ can be approximately computed by the Taylor expansion as follows
\begin{equation}
\label{eq7v}
 \mathbf J(\mathbf x) \, \mathbf y \approx \frac{\mathbf r (\mathbf x + \epsilon \mathbf y) -  \mathbf r (\mathbf x) }{\epsilon} \ , 
\end{equation}
for small enough $\epsilon$, where $\mathbf x$ is the state vector at which the Jacobian matrix is evaluated. With this approach, matrix-vector products can be computed approximately by  evaluating the residual vector without the need to compute and store the Jacobian matrix. However, the choice of the parameter $\epsilon$ can affect the performance of iterative methods. Automatic differentiation (AD) approach can be used to compute matrix-vector products exactly \cite{Vila-Perez2022}.

The main challenge in matrix-free iterative methods is to construct an effective preconditioner without forming and storing the Jacobian matrix. Block Jacobi is a popular choice for matrix-free methods since it only requires the block diagonal of the Jacobian matrix, whereas block Gauss Seidel and ILU require the entire Jacobian matrix. The recent work \cite{CuongNguyen2022} introduces a matrix-free preconditioner based on the reduced basis (RB) approximation of the linear system \cite{Grepl2007,Nguyen2008d}, where the reduced basis subspace comprises the solutions at previous time steps. This RB preconditioning technique demonstrates high effectiveness for unsteady simulations and can be easily implemented on graphics processors  \cite{Nguyen2020gpu,CuongNguyen2022,Nguyen2023a}.

\section{Shock Capturing Methods}

The existence of strong shock waves in hypersonic flows is one of the major challenges for DG methods. Many shock capturing methods have been developed to handle shock waves effectively while maintaining accuracy in smooth solution regions. These methods generally fall into three categories: limiting, fitting, and artificial viscosity. 


\subsection{Limiting methods}
Limiting methods involve modifying the numerical solution to prevent un-physical oscillations near shocks without overly damping the solution in smooth regions. Slope limiters control the steepness of gradients of the numerical solution within elements, while flux limiters regulate the numerical fluxes across element interfaces. Flux and slope limiters, as are often used in finite volume methods, have been extensively developed for DG methods. 
The classical RKDG method of Cockburn and Shu uses a slope limiter that reverts to a linear expansion around shocks and applies a slope limiter that preserves the TVD property \cite{Cockburn1998a}.
This strategy was extended to higher order in \cite{Burbeau2001, krivodonova2007} using the moment limiter approach of \cite{BiswasDevineFlaherty94}.  
The application of this approach to unstructured grids is involved, though it has been demonstrated recently in \cite{giuliani2019moment, giuliani2020moment}.

Oscillations can also be dealt with using WENO-based reconstruction procedures. 
In WENO-based limiters, oscillatory solutions are replaced with a WENO reconstruction. 
They were introduced in \cite{Qiu2005}, with recent work going towards simplifying the method for structured and unstructured grids \cite{zhong2013simple, Zhu2013} and developing limiters with increased accuracy \cite{li2020p}. 
Hermite WENO reconstructions, which use values of the solution and its gradient can also be used \cite{qiu2004hermite, qiu2005hermite, Luo2007}. 
These reconstructions have a smaller stencil, making them more complementary to the compact stencil of DG methods. 

Limiters are particularly well suited for explicit time stepping, making them somewhat less often used for hypersonic flow than for other applications. One exception is the application of the RKDG minmod limiter to hypersonic flow in thermochemical nonequilibrium over a double cone shown in \cite{papoutsakis2014discontinuous}.
The hMLP limiter introduced in \cite{park2014higher} and improved in \cite{you2018high} for unstructured grids was demonstrated on hypersonic with 11-species air in equilibrium for blunt body flows in \cite{You2022artificial}.

\subsection{Shock fitting methods}
An advantage of discontinuous Galerkin methods is that they can capture discontinuities in a solution exactly given an appropriate Riemann solver and a mesh that is exactly aligned with the shock. 
This has been taken advantage of in low-order simulations of hypersonic flow for years, where techniques such as grid tailoring iteratively deform the mesh to align with a bow shock \cite{saunders2007approach}. 
This approach still finds use in state-of-the-art hypersonics codes including NASA's DPLR, LAURA, and FUN3D codes, US3D \cite{candler2015development} and the Sandia Parallel Aerodynamics Reentry Code (SPARC) \cite{howard2017towards}.

High-order elements are more readily able to capture curved and intersecting shocks. 
This has led to a recent surge of activity in developing shock tracking DG methods. 
The high-order implicit shock tracking (HOIST) method forms a PDE-constrained optimization statement that iteratively aligns the mesh and solves the PDE, with regularization terms to promote good mesh quality \cite{zahr2018optimization, Zahr2020, shi2022high}. 
It has been used in 2D and 3D, for supersonic and hypersonic problems, including problems with chemical reactions \cite{zahr2021high}, and for steady and unsteady problems using the method of lines \cite{shi2022implicit_lines} or space-time discretizations \cite{naudet2023space}. 
Recent work has focused on improving the robustness of the method for high-speed flows \cite{huang2022robust, huang2023high} and improving high-order mesh operations in higher dimensions \cite{shi2023local}. 

Developed around the same time, the moving discontinuous Galerkin method with interface conservation enforcement (MDG-ICE) methods enforce the interface or Rankine-Hugoniot conditions across element faces and treat the geometry as a solution variable to be solved along with the flow \cite{corrigan2019moving}. 
The method has been modified to target improved boundary layer meshes with a discontinuous Petrov-Galerkin formulation in \cite{kercher2021moving}.
A reformulation proposed by Luo decreases the number of interface conservation equations needed by treating the geometry with a continuous variational formula \cite{luo2021moving}. 
These methods have been demonstrated for viscous  hypersonic flows \cite{kercher2021moving, ching2023moving}, unsteady flows using a space-time approach \cite{corrigan2019moving_unsteady}, and flows in thermochemical nonequilibrium \cite{luo2023moving}.

For both HOIST and MDG-ICE methods, an optimization solver is used to solve for the degrees of freedom of the numerical solution and the computational mesh simultaneously.  Although the mesh movement problem is challenging, especially in higher dimensions, the potential of these methods lies in their ability to retain high-order accuracy in the presence of shocks. 
This means that significantly fewer elements can be used around shock waves, which will hopefully make the cost of mesh motion more manageable. 

\subsection{Artificial viscosity methods}

Shock capturing using artificial viscosity may date back as early as 1950~\cite{VonNeumann1950}. The main idea is to add an artificial viscous term into the governing equations to stabilize shock waves without affecting the solution away from the shock region. When the amount of viscosity is properly added in a neighborhood of shocks, the solution can converge uniformly except in the region around shocks, where it is smoothed and spread out over some length scale. On the one hand, excessive addition of artificial viscosity may negatively affect  the computed solution not only in the shock region but also in other parts of the domain where the solution is smooth. On the other hand, inadequate addition of  artificial viscosity can lead to oscillatory solutions and even numerical instability. Artificial viscosity has been widely used DG methods~\cite{HartmannHoustonCompressible02, persson06:_shock_capturing,Barter2010,Ching2019,Bai2022a,Vila-Perez2021}. Both Laplacian-based \cite{persson06:_shock_capturing,Hartmann2013,Lv2016,Nguyen2011a,Moro2016,Persson2013} and physics-based \cite{Cook2005,Cook2007,Fiorina2007,Kawai2008,Bhagatwala2009,Mani2009,Kawai2010,Premasuthan2010b_tmpfix,Olson2013,Abbassi2014} artificial viscosity methods have been used for shock capturing.

\subsubsection{Physics-based artificial viscosity}

In order to deal with shock waves and discontinuities, we add artificial viscosities to the physical ones as follows:
\begin{equation}
\label{eq4w}
\beta= \beta_f + \bar{\beta}^\ast,\qquad \mu = \mu_f + \bar{\mu}^\ast, \qquad \kappa = \kappa_f + \bar{\kappa}^\ast,
\end{equation}
where $\bar{\beta}^*$, $\bar{\mu}^*$, and $\bar{\kappa}^*$ are the artificial bulk viscosity, artificial shear viscosity, and artificial thermal conductivity, respectively. In the governing equations, the physical viscosities of the fluid are replaced with those in (\ref{eq4w}). The artificial viscosities are defined below.
\begin{equation}\label{artificial}
\begin{split}
\beta^\ast  & = \hat{s} \frac{k_\beta h}{k}\sqrt{|\bm{v}|^2+a^{\ast 2}}, \\
\mu^\ast  & =  k_\mu \beta^\ast, \\
\kappa^* & =  k_{\kappa} c_p \, \mu^\ast / Pr.    
\end{split}
\end{equation}
Here $h$ is a mesh size field, $k$ is the polynomial degree, $k_{\beta,\mu, \kappa}$ are parameters that control the amount of artificial viscosities,  $a^*$ is the critical speed of sound,  and $\hat{s}$ denotes the smoothly bounded value of the shock sensor  
\begin{equation}
\label{e:sbeta}
\hat{s} = \ell_{\min}(\ell_{\max}(s-s_0) - s_{\max}) + s_{\rm max}  ,
\end{equation}
where
\begin{equation}
\begin{split}
s & = -\frac{h}{k}\frac{\nabla\cdot\bm{v}}{a^\ast}  \frac{\left(\nabla\cdot\bm{v}\right)^2}{\left(\nabla\cdot\bm{v}\right)^2+|\nabla\times\bm{v}|^2+10^{-8}}, \\
\ell_{\max}(s) & =  \frac{s}{\pi} \arctan (100 s) + \frac{s }{2} - \frac{1}{\pi} \arctan(100) + \frac{1}{2} , \\
\ell_{\min}(s) & = s  - \ell_{\max}(s) , 
\end{split}
\end{equation}
Here the first parameter $s_{\rm 0}$ represents the starting point of the limiting function $\ell$ where it begins to increase with $s$, while the second parameter $s_{\rm max} > 0$ is the upper bound of the non-negative variable $s$. The  parameters are chosen as $s_{\max} = 2$ and $s_{0} = 0.01$ according to Fernandez et al. \cite{Fernandez2018}. 

Since the  artificial viscosity (AV) fields $\left(\beta^\ast,\mu^\ast, \kappa^*\right)$ are discontinuous, we use a node-averaging operator  to make them $\mathcal{C}^0$ continuous by averaging all the  multiple values along the element boundaries to obtain continuous fields. The proposed reconstruction is particular to the DG discretization. Let $\bm x_n, 1 \le n \le  N_k N_e,$ be DG nodes of $\mathcal{T}_h$, where $N_k$ is the number of nodes per element and $N_e$ is the number of elements. For every node $\bm x_n$, $\bar{\beta}^*(\bm x_n) = \frac{1}{J_n}\sum_{j=1}^{J_n} \beta^\ast(\bm x_n)|_{K_j}$, where $K_j, 1 \le j \le J_n,$ are all the elements in which $\bm x_n$ is located. If a mesh node $\bm x_n$ is located inside an element then $J_n = 1$, and it is located on a face then $J_n = 2$. If it is located on an edge or at an element vertex, then $J_n$ is equal to the number of elements connected to that edge or that vertex, respectively. In essence, $\bar{\beta}^*$ is a polynomial of degree $k$ on every element and continuous across element boundaries. 

This physics-based AV method has been successfully applied to unsteady simulations \cite{Fernandez2018a,Fernandez2018,Ciuca2020,Nguyen2020gpu,CuongNguyen2022,Nguyen2023a,VanHeyningen2023}. In this method, the AV field is computed from the flow state at the previous time step or at the previous DIRK stage.  While this simplifies implementation, especially for implicit time integration methods, it might not be as well-suited for steady-state problems. The reliance on previous time steps introduces a dependency on the temporal evolution of the solution. This dependence can impact convergence and stability, posing challenges in achieving steady-state solutions.

\subsubsection{Adaptive viscosity regularization}

In recent work \cite{Nguyen2023c}, an adaptive viscosity regularization method is introduced for steady-state compressible flows 
\begin{equation}
\label{eqcl}
 \nabla \cdot \bm F(\bm u, \nabla \bm u) = 0  \quad \mbox{in }\Omega ,
\end{equation}
with appropriate boundary conditions. The method involves solving the following minimization problem
\begin{subequations}
\label{eq5}
\begin{alignat}{2}
\min_{\lambda_1 \in \mathbb{R}^+, \lambda_2 \ge 1, \bm u, \eta} & \quad \lambda_1 \lambda_2  \\
\mbox{s.t.} & \quad \mathcal{L}(\bm u, \eta, \bm \lambda) = 0 \\
 & \quad \bm u \in \mathcal{C} .
\end{alignat}
\end{subequations}
Here $\mathcal{C}$ represents a set of constraints on the solution $\bm u$, and $\mathcal{L}$ represents the spatial discretization of the following regularized problem 
\begin{subequations}
\label{eq3}
\begin{alignat}{1}
 \nabla \cdot \bm F(\bm u, \nabla \bm u) - \lambda_1 \nabla \cdot \bm G(\bm u,   \nabla \bm u,  \eta)  = 0  \quad \mbox{in }\Omega, \\
\eta - \lambda_2^2 \nabla  \cdot \left( 
\ell^2 \nabla  \eta  \right)- s(\bm u, \nabla \bm u) = 0 \quad \mbox{in }\Omega ,
\end{alignat}
\end{subequations}
where $\eta(\bm x)$ is the solution of the Helmholtz  equation (\ref{eq3}b) with homogeneous  Neumann boundary conditions 
\begin{equation}
\eta = 0 \quad \mbox{on } \Gamma_{\rm wall}, \qquad 
\ell^2 \nabla  \eta \cdot \bm n = 0 \quad \mbox{on }\partial \Omega \backslash \Gamma_{\rm wall} \ .
\end{equation}
Here $\lambda_1$ is the first regularization parameter that controls the amplitude of artificial viscosity, and $\lambda_2$ is the second regularization parameter that controls the thickness of artificial viscosity. Furthermore, $\ell$ is an appropriate length scale which is chosen as the smallest mesh size $h_{\rm min}$. For convenience of notation, we denote $\bm \lambda = (\lambda_1, \lambda_2)$.

A homotopy continuation method is used to solve (\ref{eq5}). The key idea is to solve the regularized system (\ref{eq3}) with a large value of $\bm \lambda$ first and then gradually decrease $\bm \lambda$ until any of the constraints on the solution is violated. At this point, we take the value of $\bm \lambda$ from the previous iteration, where the solution still satisfies all the constraints. This procedure is summarized in the following algorithm:

\begin{itemize}
 \item Given an initial value $\bm \lambda_0 = (\lambda_{0,1}, \lambda_{0,2})$ and $\eta_0$ such that $\|\eta_0\|_{\infty} = 1$, solve (\ref{eq3}a) with $\lambda_1 = \lambda_{0,1}, \eta = \eta_0$ to obtain the initial solution $\bm u_0$. 
 \item   Set $\lambda_{n,1} = \zeta^{n-1} \lambda_{{n-1},1}$ and $\lambda_{n,2} = 1 + \zeta^{n-1}(\lambda_{{n-1},2} -1)$ for some constant $\zeta \in (0,1)$; solve the Helmholtz equation (\ref{eq3}b) with $\lambda_2 = \lambda_{n,2}$ and the source term from $\bm u_{n-1}$ to obtain $\eta_{n}$; and solve (\ref{eq3}a) with $\lambda_1 = \lambda_{n,1}, \eta = \eta_{n}$ to obtain the solution $\bm u_{n}$ for $n = 1, 2, \ldots$ until $\bm u_{n}$ violates any of the constraints.
 \item Finally, we accept $\bm u_{n-1}$ as the numerical solution of the compressible Euler/Navier-Stokes equations.
\end{itemize}
The initial AV field $\eta_0$ is set to 1 in most of the physical domain $\Omega$ except near the wall boundary, where it smoothly vanishes to zero on the wall. The initial value $\bm \lambda_0$ is chosen large enough to make the initial solution $\bm u_0$ very smooth. 

The artificial flux operator $\bm G$ provides a viscosity regularization  to smooth out discontinuities in the shock region. There are several different options to define $\bm G$. In \cite{Nguyen2023c}, we use the Laplacian flux of the form
\begin{equation}
\bm G(\bm u,   \nabla \bm u,  \eta) = \mu(\eta) \nabla \bm u  ,  
\end{equation} 
where 
\begin{multline}
\mu(\eta) = (\bar{\eta}-\bar{\eta}_{\rm T})\left(\frac{\arctan(100(\bar{\eta} -\bar{\eta}_{\rm T}))}{\pi} + \frac{1}{2} \right)  \\ - \frac{\arctan(100)}{\pi} + \frac{1}{2}  
\end{multline} 
Here $\bar{\eta} = \eta/\|\eta\|_\infty$ is the normalized function with $\|\eta\|_{\infty} = \max_{\bm x \in \Omega}  |\eta(\bm x)|$ being the $L_\infty$ norm. Note that $\bar{\eta}_{\rm T}$ is the artificial viscosity threshold that makes $\mu(\eta)$ vanish to zero when $\bar{\eta} \le \bar{\eta}_{\rm T}$. In other words, the artificial viscosity is added only to the shock region where $\bar{\eta}$ exceeds $\bar{\eta}_{\rm T}$. Therefore, the threshold $\bar{\eta}_{\rm T}$ will help remove excessive artificial viscosity. Since $\|\bar{\eta}\|_\infty = 1$, $\bar{\eta}_{\rm T} = 0.2$ is a sensible choice. Note that the artificial viscosity field is equal to $\lambda_1 \mu(\bm x)$, where  $\mu(\bm x)$ is bounded by $\mu(\bm x) \in [0, 1 - \bar{\eta}_{\rm T}]$ for any $\bm x \in \Omega$. We can also consider a more general form $\bm G = \mu(\eta) \nabla \bm u^*$ \cite{Barter2010,Nguyen2011a}, where $\bm u^*$ is a modified state vector.  Another option is the physics-based artificial viscosities described earlier.  

The source term $s$ in (\ref{eq3}b) is required to determine $\eta$ and is defined as follows
\begin{equation}
\label{avsource}
s(\bm u, \nabla \bm u) =  {g}(S(\bm u, \nabla \bm u)) 
\end{equation}
where  $g(S)$ is a smooth approximation of the following step function 
\begin{equation}
\label{eq8g}
\tilde{g}(S) = \left\{
\begin{array}{cl}
   0  & \mbox{if } S < 0, \\
   S  & \mbox{if } 0 \le S \le s_{\rm max}, \\
   s_{\rm max} & \mbox{if } S > s_{\rm max} . 
\end{array}
\right.
\end{equation}
The quantity $S(\bm u, \nabla \bm u)$ is a measure of the shock strength which is given by
\begin{equation}
S(\bm u, \nabla \bm u) = -\nabla \cdot \bm v \ ,
\end{equation}
where $\bm v$ is the non-dimensional velocity field that is determined from the state vector $\bm u$.  The use of the velocity divergence as shock strength for defining an artificial viscosity field follows from \cite{Nguyen2011a,Moro2016,Fernandez2018}. The parameter $s_{\max}$ is used to put an upper bound on the source term when the divergence of the velocity becomes too negatively large. Herein we choose $s_{\max} = 0.5 \|S\|_\infty$, where $\|S\|_{\infty} = \max_{\bm x \in \Omega}  |S(\bm x)|$ is the $L_\infty$ norm. Since $S$ depends on the solution, its norm may not be known prior. In practice, we employ a homotopy continuation scheme to  iteratively solve the problem (\ref{eq3}).  Hence, $s_{\max}$ is computed by using the numerical solution at the previous iteration of the homotopy continuation.

The constraint set $\mathcal{C}$ is introduced to ensure the quality of the  solution. First and foremost, both pressure and density must be positive. In order to impose a smoothness constraint on the  solution, we express an approximate scalar variable $\xi$ of degree $k$ within each element in terms of an orthogonal basis and its truncated expansion of degree $k-1$ as
\begin{equation}
\xi = \sum_{i=1}^{N(k)} \xi_i \psi_i, \qquad \xi^* = \sum_{i=1}^{N(k-1)} \xi_i \psi_i    
\end{equation}
where $N(k)$ is the total number of terms in the $k$-degree expansion and $\psi_i$ are the basis functions \cite{persson06:_shock_capturing}. Here $\xi$ is chosen to be either density, pressure, or local Mach number.  We introduce the following quantity.
\begin{equation}
\label{eq10}
\sigma(\bm \lambda) = \max_{K \in \mathcal{T}_h^{\rm shock}} \sigma_K(\bm \lambda), \quad  \sigma_K(\bm \lambda) \equiv   \frac{\int_K  |\xi/\xi^* - 1| d \bm x}{\int_K d \bm x} ,
\end{equation}
where $\mathcal{T}_h^{\rm shock}$ is the set of elements defining the shock region
\begin{equation}
\mathcal{T}_h^{\rm shock} = \{K \in \mathcal{T}_h \ : \ \int_K  \bar{\eta}  d \bm x  \ge \bar{\eta}_{\rm T} |K|  \} 
\end{equation}
The constraint set $\mathcal{C}$ in (\ref{eq5}) consists of the following contraints    
\begin{equation}
 \rho(\bm x) > 0, \quad  p(\bm x) > 0,   \quad   \sigma(\bm \lambda)  \le C_0 \, \sigma(\bm \lambda_0) ,
\end{equation}
with $C_0 = 5$. The first two constraints enforce the positivity of density and pressure, while the last constraint guarantees the smoothness of the numerical solution. The smoothness constraint imposes a degree of regularity on the numerical solution and plays a vital role in yielding sharp and smooth solutions. 

\section{Mesh Adaptation Techniques}

The quality of meshes has a significant impact on the accuracy of numerical simulations. For DG methods, it is important to generate high-order meshes which are capable of resolving highly localized features such as  shock waves and boundary layers. This is particularly so for hypersonic flows because numerical solutions of hypersonic flows tend to be more sensitive to mesh quality than those of subsonic flows. Various mesh adaptation techniques such as $h$-adaptivity, $r$-adaptivity, $p$-adaptivity, and their combinations such as $h/p$-adaptivity, $h/r$-adaptivity can be used to enhance mesh quality. The application of these techniques is critical in achieving reliable and accurate numerical simulations of hypersonic flows.

\subsection{$h/p$-adaptivity}
The $h$-adaptivity refines the mesh by selectively adjusting the mesh size or resolution in specific regions based on solution behavior or error indicators. 
Of particular interest for steady hypersonic flows is the use of goal-oriented mesh adaptation by way of dual-weighted residual (DWR) error estimates and solution adjoints. See Section \ref{sec:err} for further discussion on error estimation. 

Goal-oriented mesh adaptation has also been applied to 2D inviscid hypersonic flow over an airfoil in \cite{may2021hybridized}. 
In \cite{Barter2008}, anisotropic mesh adaptation using the method of \cite{Fidkowski2007} is coupled with the PDE-based AV detailed in \cite{Barter2010}.
The combination of PDE-based AV, anisotropic mesh adaptation, and a BR2 discretization is used to simulate a Holden compression ramp at Mach 11.68
and the Gnoffo Mach 17.605 cylinder. 
More recently, the Mesh Optimization via Error Sampling and Synthesis (MOESS) algorithm \cite{Yano2012} for goal-oriented mesh adaptation has been applied to steady hypersonic flow using the Variational Multiscale with Discontinuous Subscales (VMSD) method, a stabilized CG method that shares similarities with EDG and is identical to EDG in many cases. 
It has been used to study uncertainty in high-speed RANS modeling over a compressions corner at Mach 9.22 \cite{waligura2022investigation} and a flat plate at Mach 11.1 \cite{onyeador2022comparison}. 
This method was also used to analyze inviscid flow in thermochemical nonequilibrium, comparing the use of more standard QoIs, like drag and total enthalpy, to one specific to nonequilibrium flow, a local measure of thermochemical nonequilibrium \cite{sabo2022investigation}. 
The work of \cite{coderoutput} analyzes the difference between shock-aligned meshes and meshes attained with MOESS for hypersonic flows over a cylinder. 
The authors point out that, when compared to a shock-alignment approach, a goal-oriented method using drag as an output will more heavily refine along the stagnation streamline and the sonic line, while placing less emphasis on the shock towards the outflow boundaries.

Similar patterns are pointed out in the study of \cite{Bai2022a}, using a variant of MOESS \cite{fidkowski2016local} and DG and HDG discretizations for high-speed flows over a cylinder. 
They note that adapting to outputs based on the total enthalpy provides reasonable refinement along the shock, while targeting the drag on the cylinder can sometimes lead to spurious solutions owing to a lack of resolution near the outflow boundaries. 

These works have shown the potential of goal-oriented mesh adaptation with DG for high-speed flows, although some works mention a lack of robustness in the adaptation procedures when applied to high-speed flows \cite{Barter2008, may2021hybridized, Bai2022a}.

The $p$-adaptivity refines the mesh by adjusting polynomial orders within elements to enhance accuracy in critical regions while maintaining lower orders elsewhere.  It often relies on error indicators derived from the solution, gradients, or other criteria to identify areas where higher or lower polynomial orders are required. This technique is commonly used in DG methods \cite{Sonntag2017, huerta2012simple} due to its ability to fine-tune accuracy with a relatively straightforward implementation compared to other mesh refinement techniques. 
This simplicity has allowed it to be extended to 3D hypersonic flows in \cite{brazell20133d}. 
DG methods with $h/p$ adaptivity were used to simulate inviscid hypersonic flow over a cylinder in \cite{wangMavriplis2009adjoint} and viscous flow in \cite{burgess2012computing}, while an $r/p$ adaptivity was used for hypersonic flow in thermochemical nonequilibrium in \cite{bhatia20132}, all in 2D.
Recent works have demonstrated the ability to perform $h/p$ adaptation for 3D for transonic and supersonic flows \cite{panourgias2016discontinuous, hpMunz}.


\subsection{$r$-adaptivity}
In $r$-adaptivity, mesh points are neither created nor destroyed, data structures do not need to be modified in-place, and complicated load-balancing is not necessary \cite{Aparicio-Estrems2023}. There has been considerable interest in $r$-adaptive mesh generation by the optimal transport theory via solving the Monge–Amp\`ere equation \cite{Delzanno2008,Budd2009a,Budd2015,Browne2014,Chacon2011,Weller2016,McRae2018,Sulman2011,Sulman2021} due to its ability to avoid mesh entanglement and sharp changes in mesh resolution.
 Mesh adaptation is based on the principle of equidistribution, which equidistributes a target density function $\varrho'$ on $\Omega$ \cite{Delzanno2008,Chacon2011}. The equidistribution principle leads to a constant source density $\theta = \int_{\Omega'} \varrho'(\bm x') d \bm x' / \int_{\Omega} d \bm x$. The r-adaptive mesh $\bm \phi$ is found by solving the  Monge–Amp\`ere equation:
\begin{equation}
\label{maem}
\begin{split}
\varrho'(\bm \phi)  \det (\nabla \bm \phi) & = \theta, \quad \mbox{in } \Omega, \\
\bm{\phi} - \nabla u & =  0, \quad  \mbox{in } \Omega, \\
 c(\bm \phi) & = 0, \quad  \mbox{on } \partial \Omega ,
 \end{split}
\end{equation}
with the constraint $\int_{\Omega} u(\bm x) d \bm x = 0$. Here $c(\cdot)$ is a function for which the root of the equation $c(\bm x) = 0$ defines $\partial \Omega$. 

In the context of mesh adaptation, $\varrho'(\bm x')$ is the mesh density function and $\mathcal{T}_h$ is the initial mesh. The optimal map $\bm \phi(\bm x) = \nabla u(\bm x)$ drives the coordinates of the initial mesh to concentrate around a region where the mesh density function is high. Therefore, we need to make $\varrho'(\bm x')$ large in the shock region and small in the smooth region. It is also necessary for $\varrho'(\bm x')$ to be sufficiently smooth, so that the numerical approximation of the Monge–Amp\`ere equation (\ref{maem}) is convergent. To this end, we compute $\varrho'(\bm x')$ as solution of the Helmholtz equation 
\begin{equation}
\varrho'(\bm x') - \nabla \cdot \left( \ell^2 \nabla \varrho'(\bm x') \right) = b(\bm x') \quad \mbox{in }\Omega, \label{r-density}
\end{equation}
with homogeneous Neumann boundary condition. Here $b$ is a resolution indicator function based on the physical density gradient 
\begin{equation}
\label{mdf2}
b(\bm x') = \sqrt{1 + \beta g(|\nabla \rho(\bm x')|) } 
\end{equation}
where $g(\cdot)$ is given by (\ref{eq8g}) \cite{nguyen2023optimal}. Other indicator functions are possible, such as those based on some combination of physics-based sensors that can distinguish between shocks, large temperature gradients, and other sharp features. The mesh density function is the  solution of the Helmholtz equation (\ref{r-density}) whose source term depends on the flow state $\bm u$. In practice, we compute the approximate solution of the flow state by using the adaptive viscosity regularization method to solve the problem (\ref{eq5}) on the initial mesh $\mathcal{T}_h$ or on the previous adaptive mesh during the mesh adaptation procedure described below.

We start mesh adaptation with an initial mesh $\mathcal{T}_h$ and compute the initial solution $\bm u_h$. Next, we compute a mesh density function based on $\bm u_h$ and solve the Monge-Amp\`ere equation to obtain an adaptive mesh $\mathcal{T}_h^*$ \cite{nguyen2023hybridizable}. Finally, we interpolate $\bm u_h$ onto $\mathcal{T}_h^*$ and use it as an initial guess to solve for the final solution $\bm u_h^*$ on the adaptive mesh. The mesh adaptation procedure is described in Algorithm 1. The adaptation procedure can be repeated by using the adaptive mesh as an initial mesh in the next iteration until $\|\bm u_h^* - \bm u_h\|_{\Omega}$ is less than a specified tolerance. It should be pointed out that we do not perform the homotopy continuation at every mesh adaptation iteration, but only at the final iteration to reduce the computational cost.

\begin{algorithm}
\begin{algorithmic}[1]
\REQUIRE{The initial mesh $\mathcal{T}_h$.}
\ENSURE{The r-adaptive mesh $\mathcal{T}_h^{*}$  and the numerical solution $\bm u_h^*$ on $\mathcal{T}_h^{*}$.}
\STATE{Solve (\ref{eq5}) for $\bm u_h$ on $\mathcal{T}_h$ using the adaptive viscosity regularization method.}
\STATE{Compute the mesh density $\varrho'_h(\bm x')$ based on $\bm u_h$ by solving  {r-density}.}
\STATE{Solve the Monge-Amp\`ere equation \eqref{maem} on $\mathcal{T}_h$ using the fixed-point HDG method.}
\STATE{Average $\bm q_h$ at duplicate nodes to obtain the adaptive mesh $\mathcal{T}_h^{*}$.}
\STATE{Interpolate $\bm u_h$ onto $\mathcal{T}_h^{*}$ and use it as the initial guess.}
\STATE{Solve (\ref{eq5}) for $\bm u_h^*$ on $\mathcal{T}_h^{*}$ using the adaptive viscosity regularization method.}
\end{algorithmic}
\caption{Mesh adaptation procedure.}
\end{algorithm}

\section{Synthetic Disturbance Generation}

\subsection{Hypersonic boundary layer instabilities}


Significant advancements in predicting transition in low-speed flows can be attributed to the relative simplicity of the disturbance spectrum in such flows. Specifically, transition phenomena observed in low-speed flows are predominantly initiated through the excitation and subsequent amplification of Tollmien-Schlichting (T-S) waves. The analysis and understanding of the growth of T-S waves are essential in predicting the critical point at which a laminar flow will transition into turbulence.  For low-speed flows, other modes of the discrete spectrum remain stable, and their eigenvalues are distinct from those of the T-S waves. Consequently, the interaction between T-S waves and these other modes is often considered negligible, leading to a single-mode analysis approach \cite{Fedorov2011a}. Such a single-mode approach is widely used in linear stability theory (LST) analysis to predict the transition onset \cite{Malik1989}. Central to the LST analysis is the concept of the $N$ factor defined as the logarithmic growth of the amplitude of a disturbance measured from its initial value. The transition from laminar to turbulent flow in a boundary layer is correlated with a critical $N$ factor value. The critical $N$ factor is generally determined through empirical methods and is known to vary according to specific flow conditions and geometric configurations.

%

Compared to low-speed flows, transition mechanisms of hypersonic boundary layers are more complex and  less understood. In contrast with low-speed transition, hypersonic flow transition reveals the following distinctive features:

\begin{itemize}
\item \textbf{Coexistence of many instability modes.} Linear Stability Theory (LST) analysis, as per \cite{Mack1984a}, shows that there exist second and higher modes alongside the first mode (Tollmien-Schlichting waves). These modes are part of the family of trapped acoustic waves. The supersonic mean flow relative to the disturbance phase velocity creates a unique environment where the boundary layer acts as an acoustic waveguide. In this setting, acoustic rays are reflected off the wall and circulate near the sonic line. The Mack second, third, and higher modes are essentially waveguide normal modes and represent inviscid instabilities of an acoustic nature.
\item \textbf{Dominance of modes based on mach Number.} It is also known from LST that while the first mode is dominant at lower supersonic Mach numbers, the second mode becomes predominant at Mach numbers above 4 \cite{Mack1984a}. This shift occurs because the growth rate of the second mode tends to surpass that of the first mode at high Mach numbers.
\item \textbf{Synchronization and branching in the discrete spectrum.} \cite{Gushchin1990} show that second-mode instability occurs in regions where two modes of the discrete spectrum synchronize, meaning their frequencies and wave numbers merge. This synchronization leads to branching in the discrete spectrum, with the Mack second-mode instability resulting from the branching of slow and fast modes near their synchronization point.
\item \textbf{Inadequacy of LST analysis.} Experiments by \cite{Kendall1975,Stetson1991} on stability and transition in high-speed boundary layers reveal that low-frequency disturbances grow, contradicting predictions by linear stability calculations that these should be stable.
\item \textbf{Nonlinear interactions.} When the amplitudes of various instability modes grow to sufficiently high levels, nonlinear interactions among these modes become a dominant factor. Their interactions lead to  breakdown to turbulence and a highly non-unique transition process in hypersonic boundary layers, in which slight changes in disturbance environments or geometry can significantly impact the transition.
\end{itemize}

Given these complexities, LST often falls short in adequately analyzing instabilities in hypersonic boundary-layer flows. Therefore, more sophisticated theories like the Parabolized Stability Equations (PSE) and Nonlinear Parabolized Stability Equations (NPSE) are required for a more precise analysis. These advanced theories incorporate more physics of hypersonic transitions than LST and offer a more accurate description of the transition processes \cite{Herbert1997}.

\begin{figure}[htbp]
	\centering
	\includegraphics[width=\columnwidth]{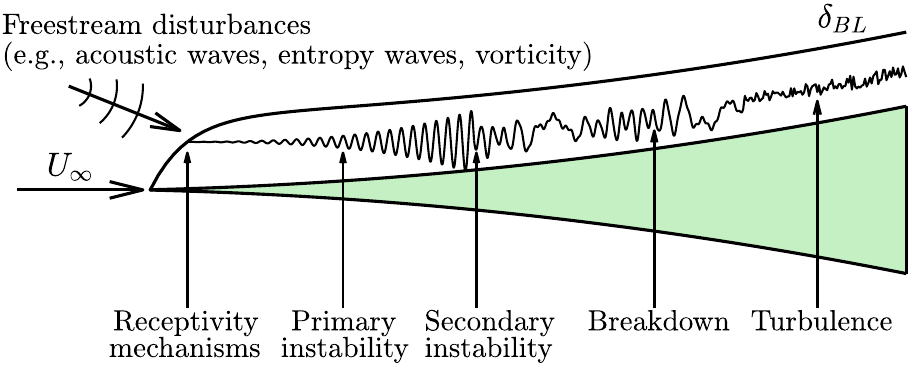}
	\caption{Sketch of the instability mechanisms producing the hypersonic laminar-turbulent transition for the flared cone.}
	\label{fig:cone_problemStructure}
\end{figure}

The path of transition associated with hypersonic boundary-layer instabilities can be broadly divided into three stages (see Figure \ref{fig:cone_problemStructure}): (i) receptivity, (ii) linear eigenmode growth or transient growth, and (iii) nonlinear breakdown to turbulence. Receptivity is the first stage in which external disturbances are sensed by the boundary layer and converted into the initial instability waves (such as first-mode or second-mode instability waves) in the boundary layer.  During the second stage, the boundary-layer instability waves undergo linear eigenmode growth which can be derived according to linear stability
theory. Besides the Mack modes, the Gortler mode may undergo considerable growth in certain cases (e.g. along with a concave surface). As the instability waves grow to certain amplitudes, secondary instability or other nonlinear interactions begin to take effect and eventually lead to the nonlinear breakdown to turbulence. Bypass transition may occur when the receptivity to large-amplitude free-stream disturbances leads to  transition without the second stage.

\subsection{Receptivity mechanisms}

In the context of hypersonic boundary layers, receptivity refers to the process by which external disturbances (e.g., free-stream disturbances due to acoustic and vortical perturbations, surface roughness, or wall disturbances) enter and interact with the boundary layer waves, potentially leading to the amplification of instabilities like Tollmien-Schlichting waves and Mack  modes. The receptivity process converts external forcing into boundary-layer waves and provides their initial conditions of amplitudes, frequencies, and phases. Resonant interactions between external forcing waves and boundary layer waves are the main receptivity mechanisms in hypersonic boundary layers \cite{Zhong2012}. The efficiency of the receptivity mechanisms depends on the amplitudes, frequencies, and phases of the external forcing waves and the characteristics of the hypersonic boundary layers \cite{Choudhari1990, Choudhari1996}. In hypersonic flows, boundary-layer shock wave interactions can greatly amplify the instability waves and potentially lead to an earlier transition to turbulence.

Understanding receptivity mechanisms is vital for designing the shape and surface characteristics of hypersonic vehicles to manage or delay the transition to turbulence. Modeling hypersonic flows and their receptivity mechanisms is computationally demanding due to the need to capture a wide range of scales and complex physics. Direct Numerical Simulation (DNS) or Large Eddy Simulation (LES) are often used to predict and analyze these receptivity mechanisms and their impact on transition \cite{Zhong2012}. In DNS and LES simulations, forcing waves are often introduced at or near the inlet to induce instability waves. Examples of forcing waves include periodic wall disturbances, free-stream disturbances, surface roughness, or boundary-layer waves obtained by LST or PSE. Free-stream disturbances are important disturbances that hypersonic vehicles experience in real flight conditions. The receptivity and instability of hypersonic boundary layers to wall disturbances, including blowing suction, is also important. Blowing suction is not only the most sensitive wall disturbance for hypersonic boundary layers but also widely used to control the boundary-layer transition. In what follows, we discuss means of introducing free-stream disturbances into boundary layers.



\subsection{Generating free-stream disturbances}


Free-stream disturbances include acoustic waves, turbulence, and entropy waves. Physically, acoustic waves are related to pressure disturbances, turbulence is related to vorticity disturbances, and entropy waves are related to temperature or density disturbances. Acoustics waves propagate at the speed of sound with respect to the flow, whereas turbulence and entropy waves travel with the flow. Free-stream disturbances are generally imposed at the inlet.

To model free-stream wind tunnel noise, \cite{Hader2019} introduced random pressure perturbations to the inflow boundary of their computational domain of the form
\begin{equation}
\label{eq:rhop1}
p'(x,y,z,t) = A(2r-1).
\end{equation}
Where $A$ is the amplitude and $r(x,y,z,t)$ is a field of random numbers generated by FORTRAN's pseudo-random number generator. This random forcing was applied at the symmetry line in the azimuthal direction, up to the last five grid points before the top boundary in the wall-normal direction, and from the sixth to tenth grid points in the streamwise direction. No random pressure perturbations were introduced in the first five grid points in the streamwise direction and the last five grid points in the wall-normal direction so as to not interfere with the Dirichlet boundary condition used for the laminar base flow. Random pressure perturbations were only added at the symmetry line in the azimuthal direction to ensure a broad range of azimuthal wavenumbers.

\cite{Sandham2014} added periodic perturbations to the density field at the inflow boundary of the computational domain of the form
\begin{multline}
\label{eq:rhop}
\rho'(y,z,t) = AW(y)\sum_{j=1}^{16}\cos{\left(\frac{2\pi j z}{\ell_z} + \phi_j\right)} \times \\ \sum_{k=1}^{20}\sin{\Big( 0.04\pi kt + \psi_k\Big)}.
\end{multline}
Where $A$ is the amplitude, $\ell_z$ is the spanwise extent of the computational domain, and $\phi_j$ and $\psi_k$ are fixed phases set to random numbers between 0 and $2\pi$. A windowing function $W(y)$ was used to force the magnitude of the perturbations to go to zero at the wall as well as the top boundary. This had the desired effect of adding disturbances inside the boundary layer at the inflow.

To generate free-stream disturbances, we add periodic perturbations to the velocity field at the inlet. We employ the synthetic random Fourier method proposed by \cite{Kraichnan1970} and further developed by \cite{Bechara1994}, and \cite{Bailly1999}. The synthetic turbulence velocity field is given by
\begin{equation}
\label{eq:vp}
\bm v'(\bm x,t) = 2\sum_{n=1}^{N}\hat{u}_n \bm \sigma_n\cos{\left(\bm k_n \cdot (\bm x - \bm v_ct) + \psi_n + \omega_nt\right)}.
\end{equation}
Here $\hat{u}_n$, $\bm \sigma_n$, $\psi_n$, and $\omega_n$ are the amplitude, direction vector, phase angle, and angular frequency of the $n$th Fourier mode, respectively. The convective velocity vector $\bm v_c$ is taken to be the same as the free-stream velocity vector $\bm u_\infty$, and the wave vector $\bm k_n$ is chosen randomly on a sphere with radius equal to the wave number $k_n$ to ensure isotropy of the generated velocity field. The amplitude $\hat{u}_n$ of each mode is computed so that the turbulence energy spectrum $E(k_n)$ correspond to the energy spectrum for isotropic turbulence
\begin{equation}
\hat{u}_n = \sqrt{ E(k_n) \Delta k_n}    
\end{equation}
where $\Delta k_n = k_{n} - k_{n-1}$ is a small interval in the energy spectrum. We use a logarithmic distribution for the wave numbers $k_1, \ldots, k_N,$ because it results in a better resolution of the spectrum for low wave numbers corresponding to the most energy containing eddies. The last wave number $k_N$ is set to $k_\eta$, while the first one $k_1$ is set to $k_1 = k_e/5$. Here $k_\eta$ and $k_e$ correspond to the Kolmogorov wave number and the wave number associated with the most energetic turbulent eddies, respectively. Then $k_n = r k_{n-1}$, where the constant $r$ is determined such that $k_N = k_\eta$ for a specific value of $N$. We refer to \cite{Bailly1999} for the calculation of $\bm \sigma_n, \bm k_n, \psi_n$, and $\omega_n$.

The turbulence energy spectrum $E(k_n)$ is simulated by a von Kármán-Pao spectrum
\begin{equation}
\label{eq:Ekn}
E(k) = \alpha \frac{I^2u_\infty^2}{k_e} \frac{(k/k_e)^4}{[1 + (k/k_e)^2 ]^{17/6}} e^{-2(k/k_\eta)^2},
\end{equation}
where $I$ is the free-stream turbulence intensity, $u_\infty$ is the magnitude of the free-stream velocity,  $k_\eta = \varepsilon^{1/4}\nu^{-3/4}$ is the Kolmogorov wave number corresponding to the smallest turbulent structures, $\nu$ is the molecular viscosity, and $\varepsilon$ is the dissipation rate. The dissipation rate is related to the turbulence length scale $\Lambda$ and the turbulent kinetic energy $\bar{k} = 1.5 (u_\infty I)^2$ as
$$\varepsilon = c_\mu^{3/4} \bar{k}^{3/2} \Lambda^{-1} = (1.5)^{3/2} c_\mu^{3/4} (u_\infty I)^3 \Lambda^{-1}$$  
where $c_\mu = 0.09$ is a typical value. The constant $\alpha$ can be determined by requiring that the integral of the energy spectrum $E(k)$ over all wave numbers should be equal to the total turbulent kinetic energy
$$\bar{k} = \int_{0}^\infty E(k) d k , $$  
which yields $\alpha = 1.453$. Assuming that the turbulence length scale $\Lambda$ is the same as the integral length scale for isotropic turbulence, we get the following relation
$$\Lambda = \frac{\pi}{2 (u_\infty I)^2} \int_{0}^\infty \frac{E(k)}{k} d k . $$  
This relation yields the most energetic wave number $k_e = 9 \pi \alpha / (55 \Lambda)$.  The turbulence length scale is set to $\Lambda = 10 h_{\min}$, where $h_{\min}$ is the minimum grid length at the wall. The free-stream intensity $I$ controls the amplitude of the free-stream disturbance. We will study the effects of this parameter on boundary layer instabilities and the onset of laminar-to-turbulent transition in hypersonic flows.

To limit these disturbances to the boundary layer region, each component of the synthetic turbulence velocity field is multiplied by a windowing function
\begin{equation}
\label{eq:W}
W(y) = e^{-4\left(y/(2y_{ble})\right)^6},
\end{equation}
where $y_{ble}$ is the approximate y-location at the inflow of the computational domain of the boundary layer edge. The velocity components in the streamwise, wall-normal, and spanwise directions for the Dirichlet inflow boundary condition are thus
\begin{subequations}
\begin{align}
u_{in}(\bm x,t)  & = \Bar{u}(\bm x) + W(y)u'(\bm x,t) \label{eq:u}\\
v_{in}(\bm x,t)  & = \Bar{v}(\bm x) + W(y)v'(\bm x,t) \label{eq:v}\\
w_{in}(\bm x,t)  & = \Bar{w}(\bm x) + CW(y)w'(\bm x,t) \label{eq:w},
\end{align}
\end{subequations}
where $\Bar{u}$, $\Bar{v}$, and $\Bar{w}$ are the mean velocity components.  The constant $C\ge1$ in equation (\ref{eq:w}) is used to increase the amplitude of the spanwise synthetic turbulence velocity component relative to the streamwise and wall-normal components. 


\section{Results and Discussion}

\subsection{Laminar hypersonic flows past a cylinder}

The first test case involves the viscous hypersonic flow past a unit circular cylinder at $M_\infty = 17.6$ and $Re=376,000$. The free-stream temperature is $T_\infty = 200$ K. The cylinder surface is isothermal with wall temperature $T_{\rm wall} = 500$ K. Supersonic inflow and outflow boundary conditions are imposed at the inlet and outlet, respectively. This test case serves to demonstrate the high-order HDG method equipped with shock capturing and $r$-adaptivity for very strong bow shocks and extremely thin boundary layers. This problem was studied by Gnoffo and White \cite{Gnoffo2004} comparing the structured code LAURA and the unstructured code FUN3D. The simple geometry and strong shock make it a common benchmark case for assessing the performance of numerical methods and solution algorithms in hypersonic flow predictions \cite{Barter2010,Ching2019,Gnoffo2004,Kitamura2013,Nguyen2020gpu}. 

\begin{figure}[htbp]
\includegraphics[width=0.485\textwidth]{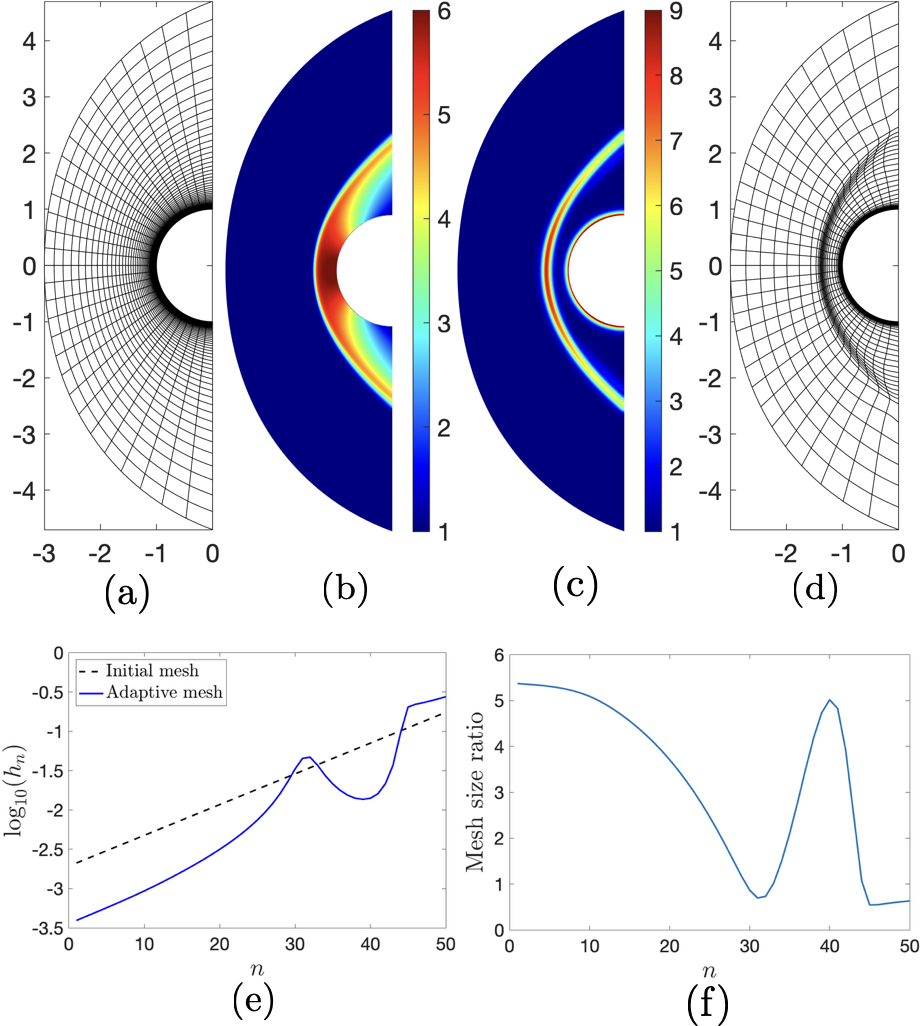}
\caption{(a) Initial mesh $\mathcal{T}_h$, (b)  computed density $\rho_h/\rho_\infty$, (c) target density function $\varrho'$, (d)  adaptive mesh $\mathcal{T}_h^*$, (e) logarithm distribution of the mesh size $h_n$ along the line $y = 0$, and (f) the mesh size ratio $h_n^{\rm initial}/h_n^{\rm adaptive}$ for the hypersonic laminar flow past a circular cylinder. These meshes consist of 1500 P4 quadrilaterals.}
\label{cylfig1}
\end{figure}

The HDG method with polynomial degree 4 is used to solve this problem. Figure \ref{cylfig1} shows the initial and adaptive meshes as well as the mesh density function used to obtain the adaptive mesh.  The mesh density function is computed from the mesh indicator (\ref{mdf2}) with using the numerical solutions on the initial mesh. The optimal transport moves the elements of the initial mesh toward the shock and the boundary layer regions because the mesh density function is high in those regions. As a result, the optimal transport can adapt meshes to capture  shocks and resolve boundary layers. To see this feature more clearly, in Figure \ref{cylfig1}(e), we plot the logarithm of the element size along the horizontal line $y=0$ on both the initial mesh and the adaptive mesh. That is, $\log_{10}(h_n)$, where $h_n$ denotes the element size of the $n$th element starting from the cylinder wall along $y=0$. We see from Figure \ref{cylfig1}(f) that the adaptive mesh has smaller element sizes than the initial mesh near the wall and in the shock region. As a result, the adaptive mesh should be able to resolve the boundary layer and shock better than the initial mesh. 

\begin{figure}[htbp]
\includegraphics[width=0.485\textwidth]{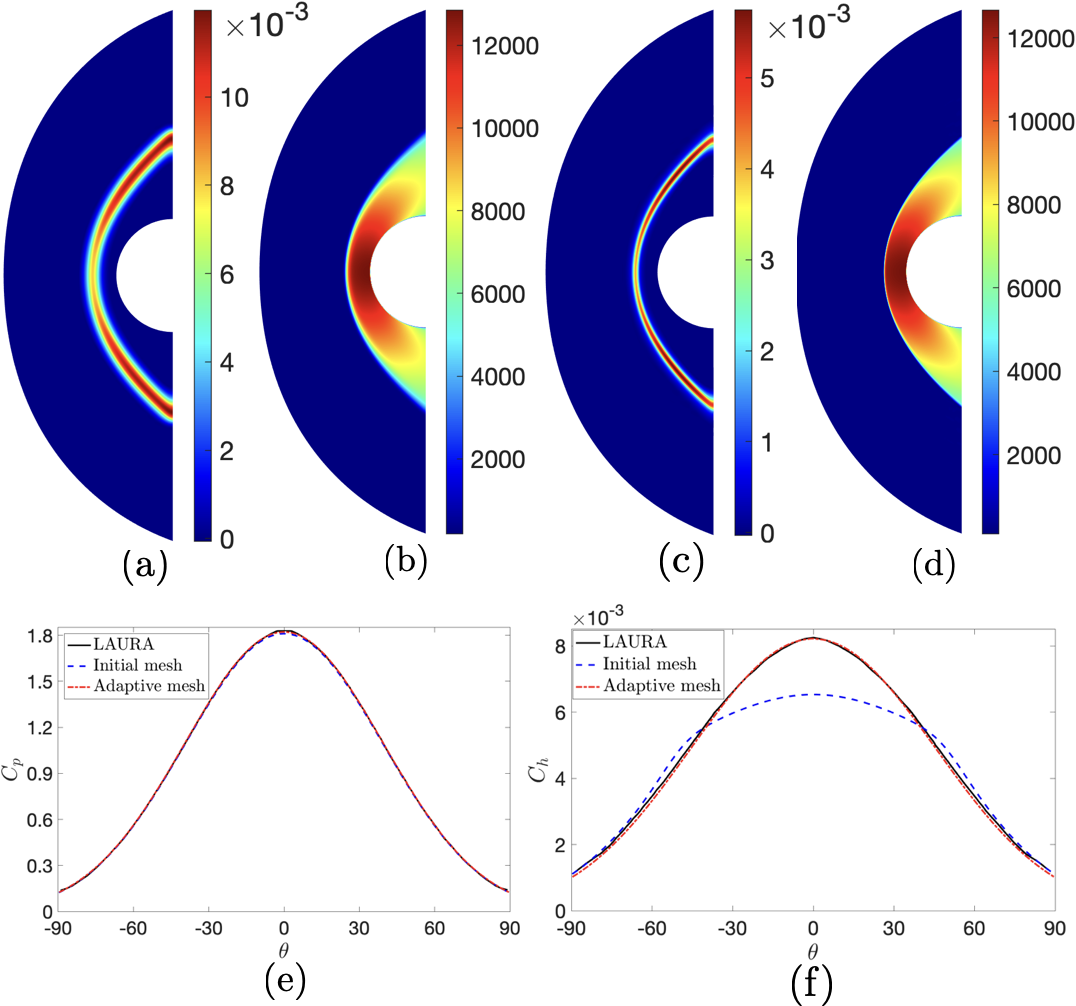}
\caption{(a)-(b) AV field and computed temperature on the initial mesh, (c)-(d) AV field and computed temperature on the adaptive mesh, and (e)-(f) pressure coefficient and Stanton number along the cylinder surface for the hypersonic laminar flow past a circular cylinder.}
\label{cylfig2}
\end{figure}

We present the numerical solution calculated on the initial mesh and on the adaptive mesh in Figure \ref{cylfig2}. We observe that temperature rises rapidly behind the bow shock, creating a very high pressure and temperature environment surrounding the cylinder. We notice that the numerical solution on the adaptive mesh is sharper and more accurate than that on the initial mesh in the shock region and boundary layer. This is because the adaptive mesh has more grid points to resolve those features than the initial mesh. By redistributing the elements of the initial mesh to resolve the bow shock and boundary layer, the optimal transport  can considerably improve the prediction of heating rate as shown in Figure \ref{cylfig2}. We see that while the pressure coefficient on the initial mesh is very similar to that on the adaptive mesh, the heat transfer coefficient on the initial mesh is lower than that on the adaptive mesh. The heat transfer coefficient on the adaptive mesh agrees very well with the prediction by Gnoffo and White \cite{Gnoffo2004}.

\subsection{Hypersonic type IV shock-shock interaction}

Type IV shock-shock interaction results in a very complex flow field with high pressure and heat flux peak in a localized region. It occurs when the incident shock impinges on a bow shock and results in the formation of a supersonic impinging jet, a series of shock waves, expansion waves, and shear layers in a local area of interaction. The supersonic impinging jet, which is bounded by two shear layers separating the jet from the upper and lower subsonic regions, impinges on the body surface, and is terminated by a jet bow shock just ahead of the surface. This impinging jet bow shock wave creates a small stagnation region of high pressure and heating rates. Meanwhile, shear layers are formed to separate the supersonic jet from the lower and upper subsonic regions. 

Type IV hypersonic flows were experimentally studied by \cite{WIETING1989}. Over the years, many numerical methods have been used in the study of type IV shock-shock interaction \cite{Hsu1996,Nguyen2020gpu,Thareja1989,Yamamoto1998,Xu2005,Zhong1994a}. In the present work, we consider an inviscid type IV interaction with free-stream Mach number $M_\infty = 8.03$. Based on the experimental measurement and the numerical calculations, Thareja et al. \cite{Thareja1989} summarized that the position of incident impinging shock on the cylinder can be approximated by the curve $y = 0.3271x + 0.4147$ for the experiment (Run 21) \cite{WIETING1989}. Boundary conditions are the same as those for the previous test case, where the freemstream state $\bm u_\infty$ is represented by a hyperbolic tangent function to account for the incident impinging shock.

\begin{figure}[htbp]
\includegraphics[width=0.485\textwidth]{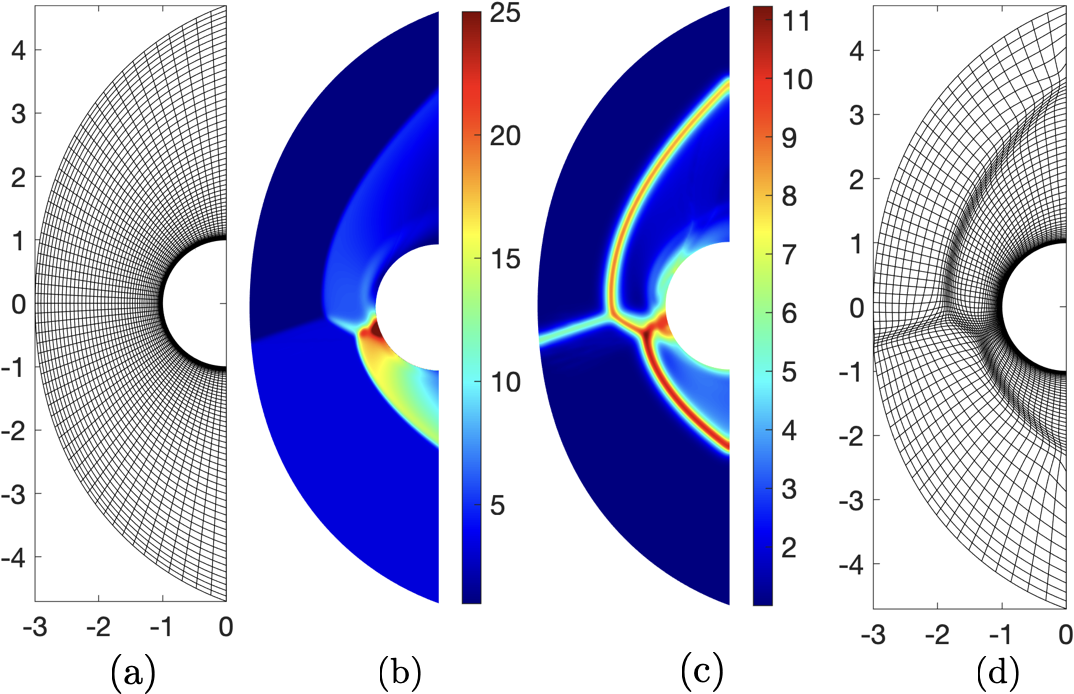}
\caption{(a) Initial mesh $\mathcal{T}_h$, (b)  computed density $\rho_h/\rho_\infty$, (c) target density function $\varrho'$, and (d)  adaptive mesh $\mathcal{T}_h^*$ for the hypersonic type IV shock-shock interaction over a circular cylinder. These meshes consist of 3600 P4 quadrilaterals.}
\label{typeiva}
\end{figure}

The HDG method with polynomial degree 4 is used to solve this problem. Figure \ref{typeiva} shows the initial and adaptive meshes as well as the mesh density function used to obtain the adaptive mesh. The mesh density function is computed from the mesh indicator (\ref{mdf2})  using the numerical solution on the initial mesh. The optimal transport moves the elements toward the shock region and aligns them along the shock curves. Furthermore, it also distributes elements around supersonic impinging jet, jet bow shock, expansion waves, and shear layers according to the mesh density function. As a result, the optimal transport can adapt meshes to capture complicated flow features without increasing the number of elements and modifying data structure.

\begin{figure}[htbp]
\includegraphics[width=0.485\textwidth]{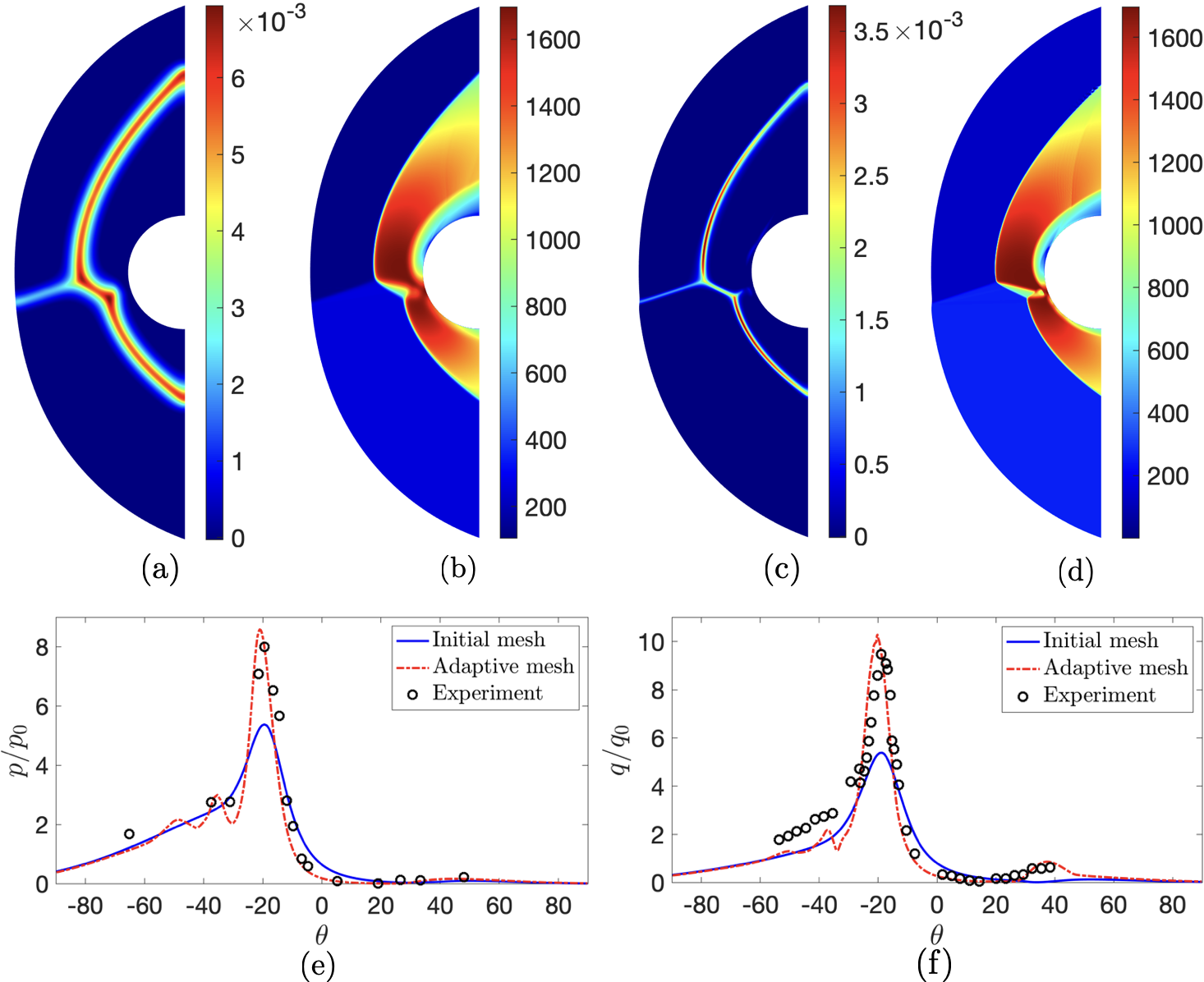}
\caption{(a)-(b) AV field and temperature computed on the initial mesh, (c)-(d) AV field and temperature computed on the adaptive mesh, and (e)-(f) pressure coefficient and Stanton number along the cylinder surface for the hypersonic type IV shock-shock interaction over a circular cylinder.}
\label{typeivb}
\end{figure}

We present the numerical solution computed on the initial mesh and the adaptive mesh in Figure \ref{typeivb}. We notice that the numerical solution on the second adaptive mesh reveals supersonic impinging jet, jet bow shock, expansion waves, and shear layers of the flow, whereas the solution on the initial mesh does not possess some of these features. This is because the initial mesh does not have enough grid points to resolve those features even though it has the same number of elements as the second adaptive mesh. By redistributing the elements of the initial mesh to resolve shocks, impinging jet, jet bow shock, expansion waves, and shear layers, the optimal transport  considerably improves the numerical solution. This test case shows the ability of optimal transport for dealing with complex shock flows.  

Finally, we present in Figure \ref{typeivb}(e)-(f) the profiles of the computed pressure and heat flux along the cylindrical surface, where the symbols $\circ$ are the experimental data from \cite{WIETING1989}. We see that the  profiles on the adaptive mesh have larger peak than those on the initial mesh. This is because the adaptive mesh has a lot more elements in the supersonic jet region than the initial mesh and the first adaptive mesh. As a result, the computed pressure and heat flux on the adaptive mesh agree with the experimental measurements better than those on the initial mesh.

\subsection{Transitional hypersonic flow past a flared cone}
\begin{figure}[htbp]
	\centering
		\includegraphics[width=\columnwidth]{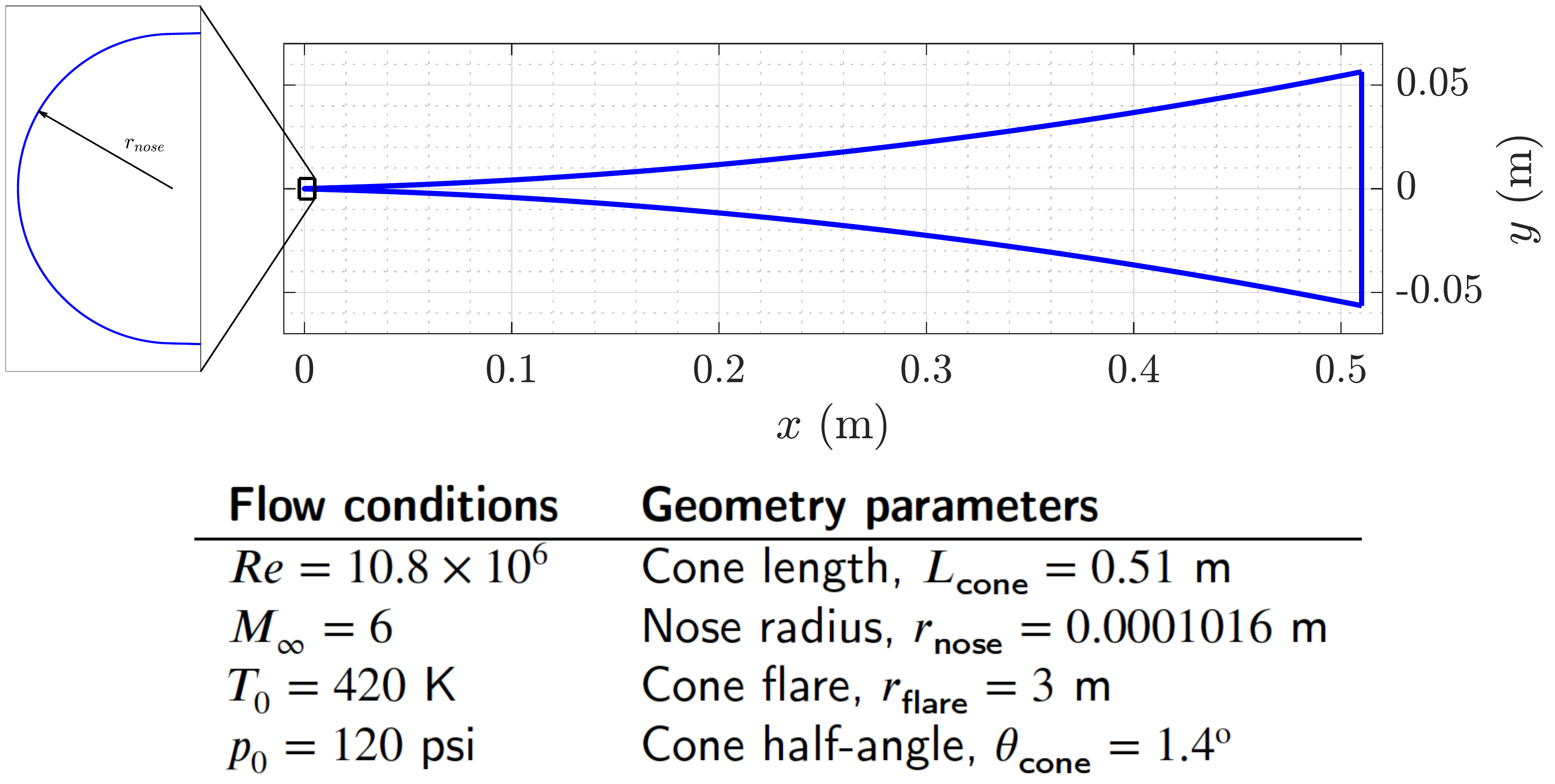}
	\caption{Geometry and flow conditions for the flared cone study.}
	\label{fig:cone_geometry}
\end{figure}

Hypersonic laminar-turbulent transition is one of the major unresolved problems in fluid dynamics, which combines the inherent complexities of transitional flows with the associated increased heat transfer effects of hypersonic flows.
This section presents the study of the hypersonic boundary-layer transition for a flared cone, whose geometry and flow conditions are described in Figure~\ref{fig:cone_geometry}.

This example, which has been experimentally investigated using the Boeing/AFOSR Mach 6 Quiet Tunnel (BAM6QT) at Purdue University \citep{Wheaton2009,Chynoweth2019} and numerically reproduced by means of DNS studies \citep{Chynoweth2019,Hader2019}, describes a hypersonic flow in a low-disturbance environment, a common setting in flight conditions. In this case, boundary-layer transition generally occurs through the linear and nonlinear growth of instabilities, such as the second-mode instability.
This kind of instability occurs after the growth and breakdown of a set of primary streaks, which act as acoustic waves trapped between the surface and the sonic line of the flow, inside the boundary layer. These primary streaks are directly attributed to the steady streamwise vortices, which are generated by fundamental resonance. As they convect in the downstream convection, linear and nonlinear growth occurs until the pressure fluctuations cause the breakdown of such vortices, which are pushed back toward the wall and secondary streaks appear with an increased streak count in the azimuthal direction.
A schematic of such instabilities is displayed in Figure~\ref{fig:cone_problemStructure}.

\begin{figure}[htbp]
	\centering
	\includegraphics[width=\columnwidth]{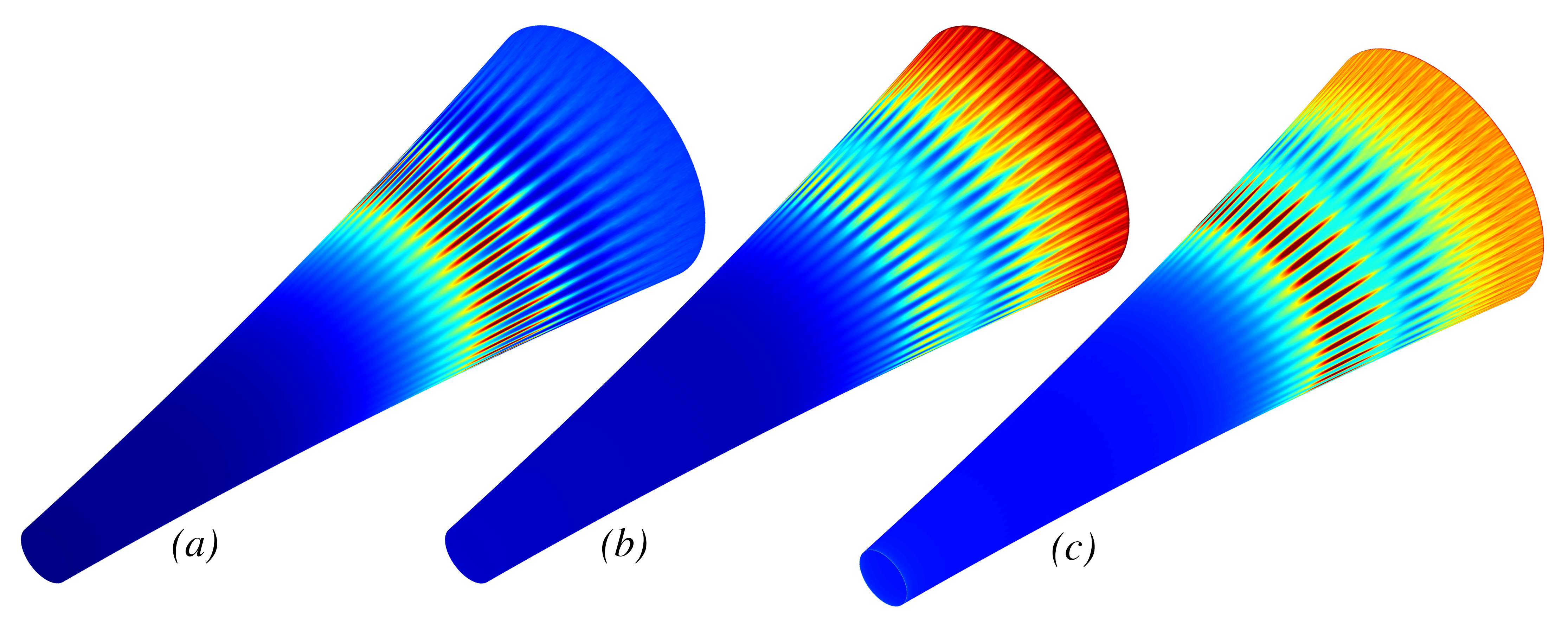}
	\caption{Time-averaged numerical solution of the hypersonic flow past a flared cone: (a) pressure, (b) skin friction coefficient and (c) heat transfer coefficient, computed with a turbulence intensity of $I = 0.125\%$.}
	\label{fig:cone_solution}
\end{figure}

To study the hypersonic boundary-layer transition of the flared cone, we carry out numerical implicit LES (ILES) studies of the Purdue flared cone, employing our high-performance GPU-based DG code, Exasim \citep{Nguyen2020gpu,Vila-Perez2022,Nguyen2023a}.
The synthetic free-stream turbulence approach is employed to analyze the effects of the free-stream disturbances on the transitional flow and thus better characterize the experimental conditions.
Three different turbulent free-stream intensities, namely $I = 0.125 \%$, $0.25 \%$, and $0.5 \%$, are considered.
A computational grid consisting of more than 6 millions of quadratic hexahedral elements is considered. A DIRK(3,3) scheme is used for the temporal discretization, with non-dimensional step size $\Delta t = 5 \times 10^{-5}$, performing 35,000 time steps.
The numerical simulations were performed using 48 GPU nodes at the OLCF's Summit supercomputer (a total of 288 V100 GPUs), taking 15 hours of run time.
A sketch of the numerical solution is depicted in Figure~\ref{fig:cone_solution}, which shows the time-averaged pressure, skin friction coefficient and heat transfer coefficient on the surface of the flared cone for a free-stream disturbance intensity of $I = 0.125\%$.


\begin{figure}[htbp]
	\centering
	\includegraphics[width=0.9\columnwidth]{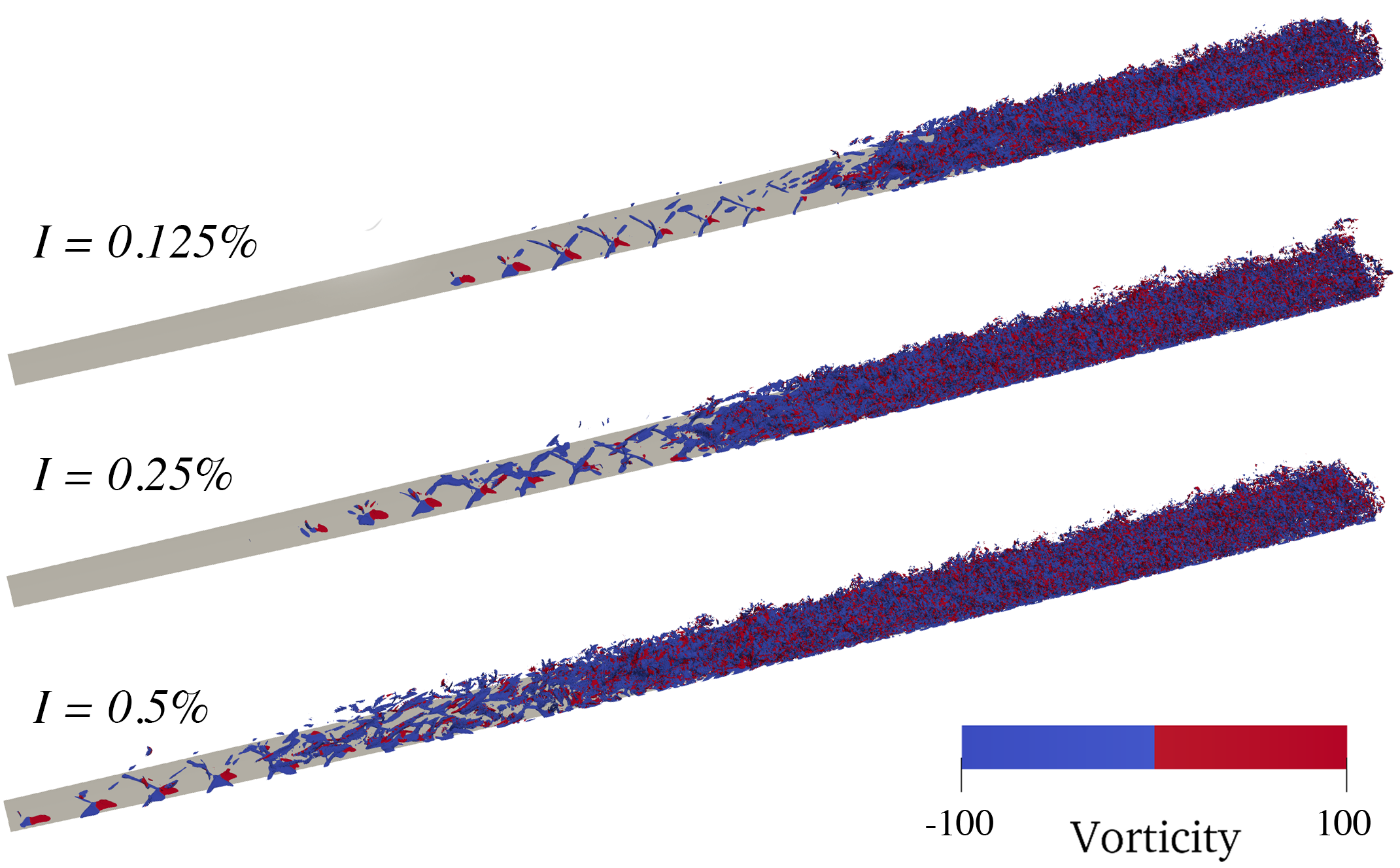}
	\caption{Instantaneous visualization of the $Q$-criterion isocontours in a portion of the flared cone domain, colored by the streamwise vorticity, displaying the structures of the fundamental vortical breakdown of the flow for different turbulent intensities.}
	\label{fig:cone_qcriterion}
\end{figure}

In order to study the laminar-turbulent transition, the instantaneous $Q$-isosurfaces ($Q = 80,000$), colored by the value of the streamwise vorticity, are shown in Figure~\ref{fig:cone_qcriterion} for the three considered turbulence intensities.
The image displays the flow structures, in particular the vortical breakdown of the flow, and reveals the effect of the free-stream disturbances on turbulent transition. We observe an earlier onset of transition for higher turbulence intensities, together with a denser configuration of vortical structures.  Clearly, the amplitude of free-stream disturbances has a very significant impact on the boundary layer instabilities and the subsequent onset of transition to turbulence.

\begin{figure}[htbp]
	\centering
	\includegraphics[width=0.9\columnwidth]{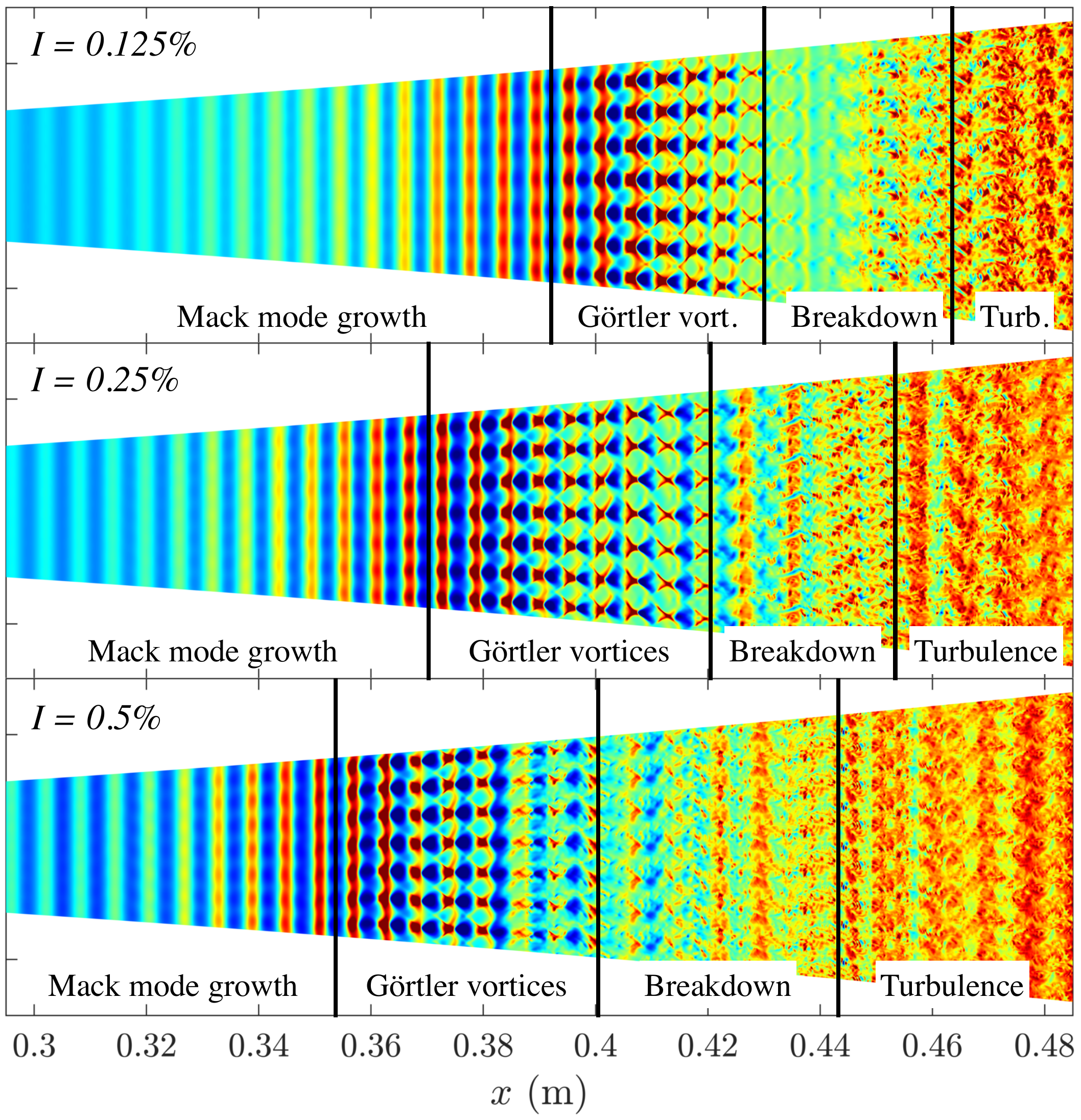}
	\caption{Instantaneous pressure distributions on the flared cone surface, showing the growth of the second-mode waves, breakdown of Görtler vortices, and transition to turbulence for different disturbance intensities.}
	\label{fig:cone_instPressure}
\end{figure}

In this same spirit, Figure~\ref{fig:cone_instPressure} shows a top view of the computed instantaneous pressure  on the flared cone surface for the three different turbulence intensities.
The plot displays the succession of instabilities and mechanisms that intervene in the laminar-turbulent transition.
On the first section, we observe the presence and growth of Mack modes \citep{Mack1984a}, a primary instability of the flow caused by free-stream disturbances which enter the boundary layer and excite the second-mode instability, a two-dimensional instability of acoustic nature that grows between the boundary surface and the sonic line.
As the second-mode instability packet grows downstream, formation and growth of Görtler vortices occurs.
Due to the second-mode instability and the formation of the Görtler vortices, the pressure fluctuations increase rapidly and reach a sufficient magnitude to cause the vortices to break down.
At this point, transition to turbulence begins and the flow becomes fully turbulent.

\begin{figure}[htbp]
	\centering
	\includegraphics[width=0.9\columnwidth]{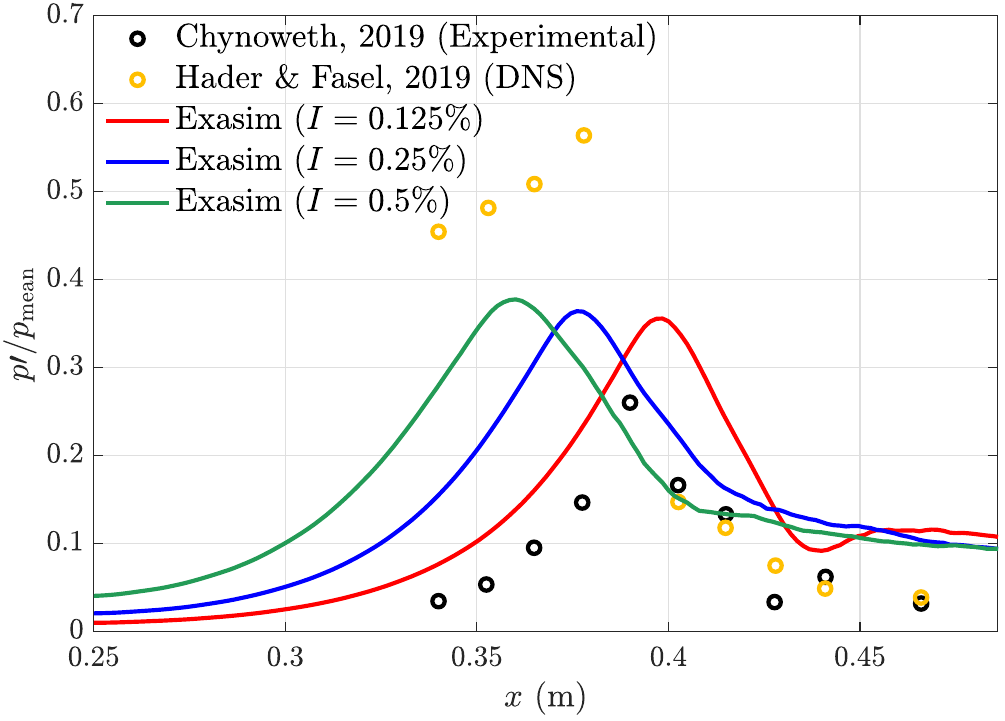}
	\caption{Pressure fluctuations computed numerically with the three considered free-stream turbulence intensities, compared to experimental results by \cite{Chynoweth2019}, and numerical DNS results by \cite{Hader2019}.}
	\label{fig:cone_pComparison}
\end{figure}

Similarly to the previous analysis based on the $Q$-criterion isosurfaces, the pressure fluctuations shown in Figure~\ref{fig:cone_instPressure} shift leftward, closer to the nose tip, as the turbulence intensity increases.
This effect can be better observed in Figure~\ref{fig:cone_pComparison}, which displays the normalized pressure fluctuations of the mean azimuthal flow for the three considered turbulence intensities. The analysis also includes the experimental data from \cite{Chynoweth2019} and the numerical DNS results from \cite{Hader2019}.
The image reveals that an increase on the free-stream disturbance intensity accelerates the onset of transition, which takes place closer to the nose tip of the flared cone.
In particular, the location of the onset of transition predicted by the DG numerical results with $I = 0.125\%$ agrees well with the experimental data, even though the magnitude of the pressure fluctuations is certainly higher.
In turn, the numerical DNS results by \cite{Hader2019} show much stronger pressure fluctuations when compared to both experimental and the numerical DG results.

\begin{figure}[htbp]
	\centering
	\begin{subfigure}{\columnwidth}
		\centering
		\includegraphics[width=\columnwidth]{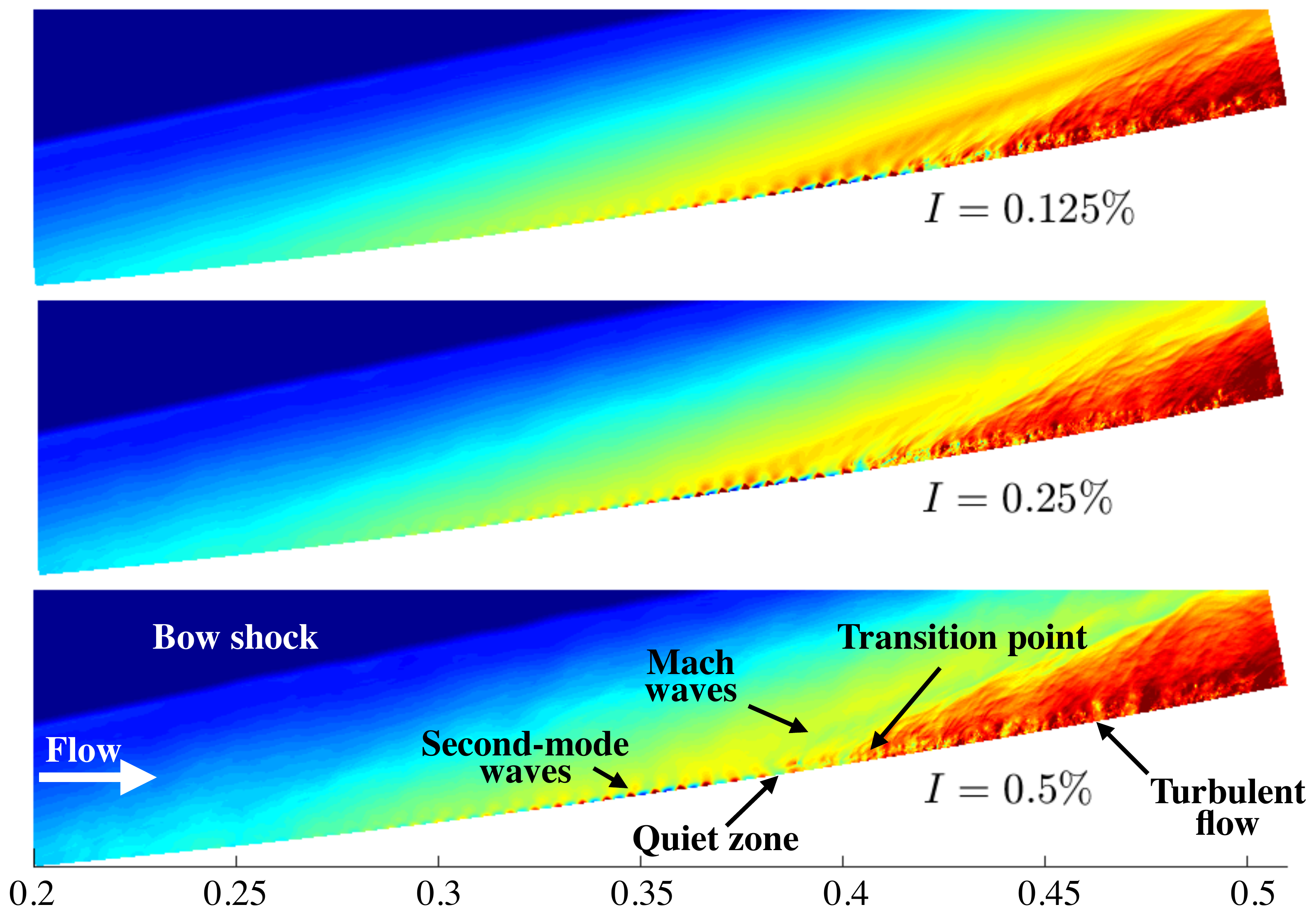}
		\caption{Instantaneous pressure obtained numerically for three different turbulence intensities.}
	\end{subfigure}
	\hfill
	\begin{subfigure}{\columnwidth}
		\centering
		\includegraphics[width=0.9\columnwidth]{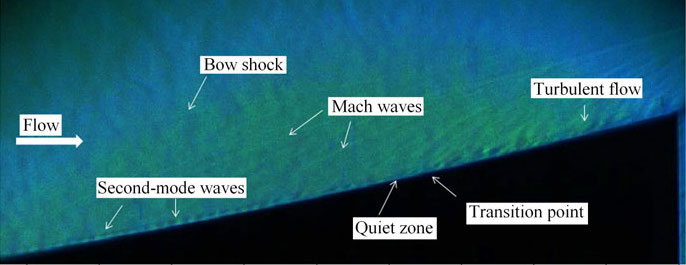}
		\caption{Schlieren image of the flow obtained through experiments. Image taken from \cite{Zhang2013}.}
	\end{subfigure}
	\caption{Comparison of the flow characteristics between numerical ILES results (top) and experiments (bottom).}
	\label{fig:cone_characteristics}
\end{figure}

\begin{figure}[htbp]
	\centering
	\includegraphics[width=0.9\columnwidth]{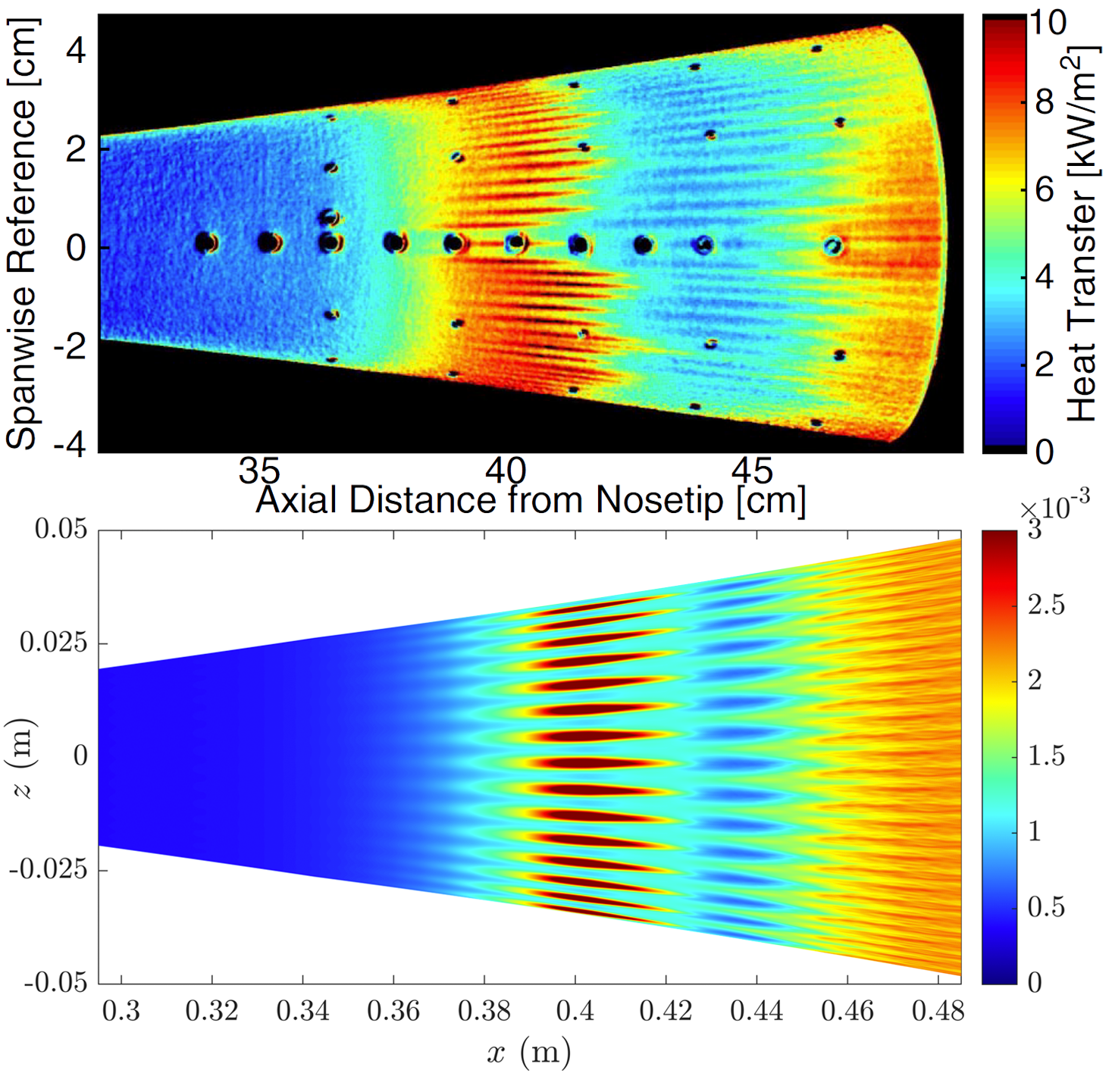}
	\caption{Comparison of the time-averaged heat transfer rate on the flared cone surface obtained experimentally by \cite{Chynoweth2019} (top) with the numerical ILES results with a disturbance intensity of $I = 0.125\%$ (bottom).}
	\label{fig:cone_Ch_experimental}
\end{figure}

\begin{figure}[htbp]
	\centering
	\includegraphics[width=0.9\columnwidth]{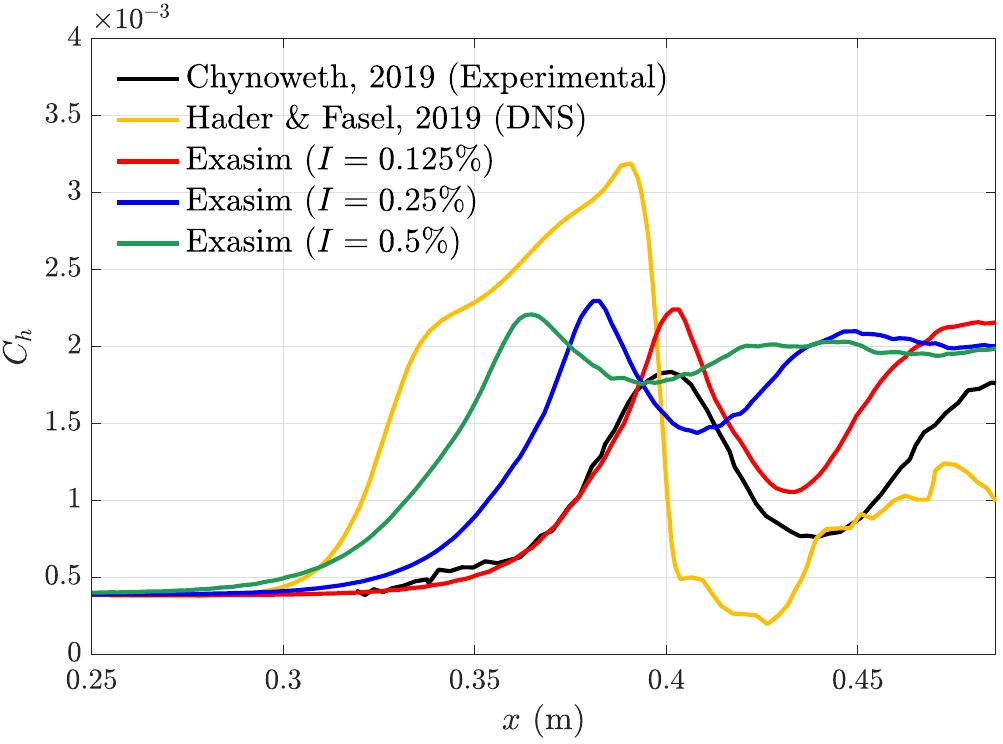}
	\caption{Mean heat transfer coefficient computed numerically with the three considered free-stream turbulence intensities, compared to experimental results by \cite{Chynoweth2019}, and numerical DNS results by \cite{Hader2019}}
	\label{fig:cone_Ch_comparison}
\end{figure}

For a deeper understanding of the different flow features involved in this case, Figure~\ref{fig:cone_characteristics} shows a streamwise view of the computed instantaneous pressure in comparison with an experimental schlieren visualization of the flow obtained by \cite{Zhang2013}.
The images display the structure of the flow and highlight the different instability mechanisms that produce the laminar-turbulence transition within the hypersonic boundary layer, as detailed previously.
In addition, the images allow for a qualitative comparison, revealing a similar flow pattern between experimental and numerical results. An additional qualitative comparison between experimental results and numerical simulations is the one shown in Figure~\ref{fig:cone_Ch_experimental}, which displays the heat transfer coefficient on the flared cone surface for the two cases. The heat transfer coefficient obtained numerically with the ILES approach for a disturbance intensity of $I = 0.125\%$ agrees well, qualitatively, with the experimental heat transfer obtained by \cite{Chynoweth2019}. In particular, a quantitative comparison between the ILES simulation results for the three considered turbulent intensities, the experimental heat transfer obtained by \cite{Chynoweth2019} and the numerical DNS results by \cite{Hader2019} is presented in Figure~\ref{fig:cone_Ch_comparison}.
The image displays the mean heat transfer coefficient, azimuthally averaged along the cone surface, for the different experimental and numerical cases.

Similarly to previous analyses, we observe that higher turbulence intensities accelerate the onset of transition, which moves closer to the leading edge.
In particular, the heat transfer coefficient obtained obtained numerically with the ILES approach for a disturbance intensity of $I = 0.125\%$ agrees very well with the experimental heat transfer obtained by \cite{Chynoweth2019}, especially for $x < 0.4$ m. For $x > 0.4$ m, the ILES results overpredict the experimental data.
Interestingly, the heat transfer rises rapidly during the growth of the second-mode instability and reaches its peak when the Görtler vortices begin to form.
Then it decreases rapidly when the Görtler vortices grow as these vortices carry heat away from the cone surface, and reaches its minimum when the Görtler vortices break up and transition occurs at this location.
The heat transfer rises rapidly again when the flow becomes fully developed turbulent.
Conversely, the numerical DNS results by \cite{Hader2019} show bigger discrepancies with respect to both experimental and the numerical DG results, displaying a higher heat transfer, especially $x<0.4$ m.

\begin{figure}[htbp]
	\centering
	\begin{subfigure}{0.48\columnwidth}
		\centering
		\includegraphics[width=\columnwidth]{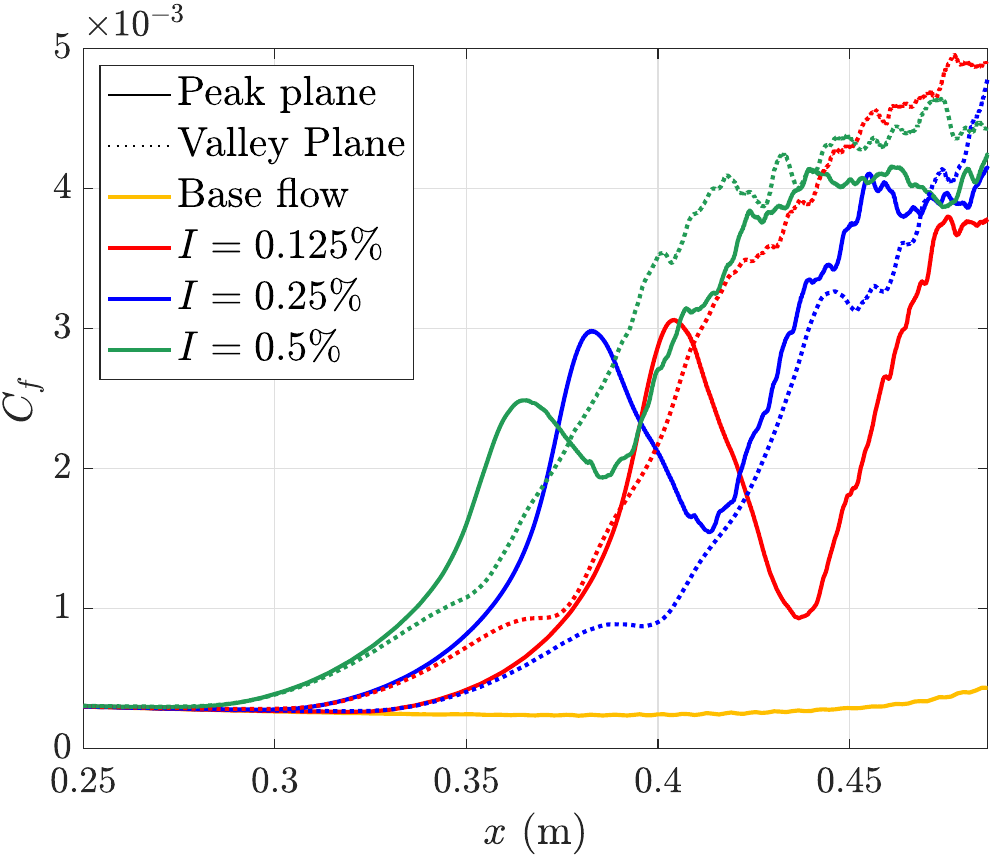}
		\caption{Skin friction coefficient.}
	\end{subfigure}
	\begin{subfigure}{0.48\columnwidth}
		\centering
		\includegraphics[width=\columnwidth]{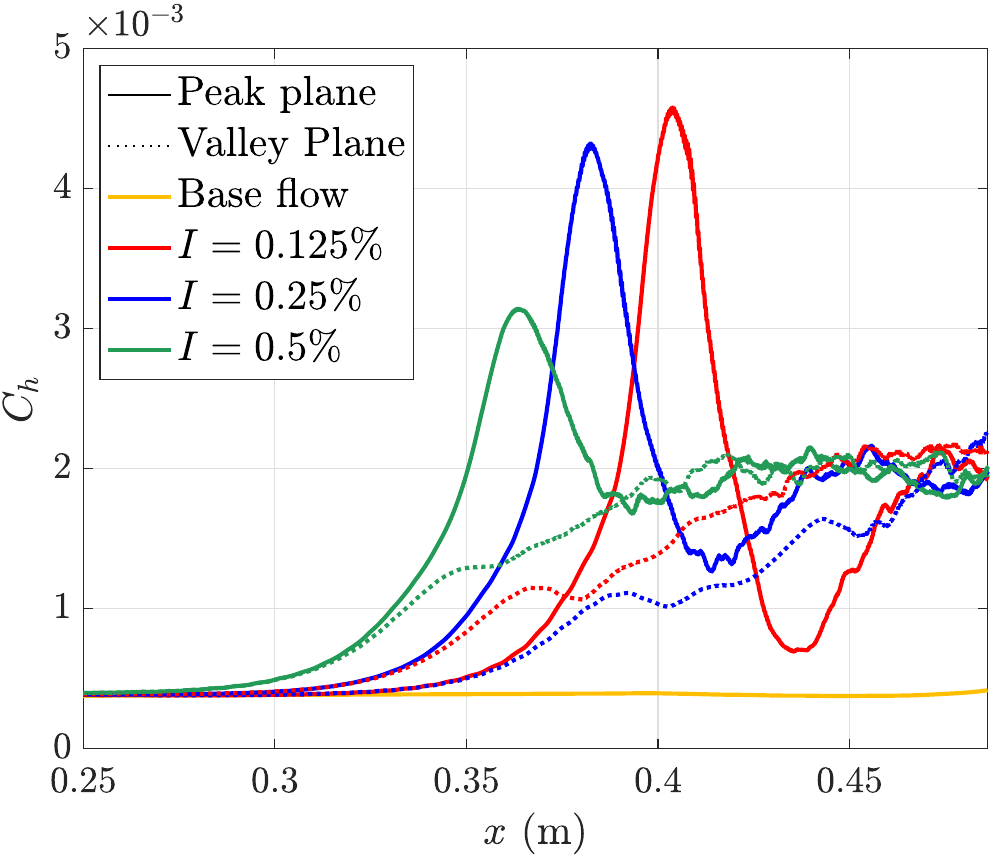}
		\caption{Heat transfer coefficient.}
	\end{subfigure}
	\caption{Surface coefficients computed along the peak and valley planes for the different turbulence intensities. Laminar base flow ($I = 0\%$) is included as a reference.}
	\label{fig:cone_peakvalley}
\end{figure}

A similar analysis can be performed taking into account streamwise cuts both of the skin friction coefficient or the heat transfer coefficient, for instance along the peak and valley planes, as shown in Figure~\ref{fig:cone_peakvalley}.
In these images, we can observe the effect of the free-stream disturbances on the transition to turbulence, especially when compared to the laminar base flow, with $I=0\%$, which serves as a reference.

\begin{figure}[htbp]
	\centering
	\includegraphics[width=0.9\columnwidth]{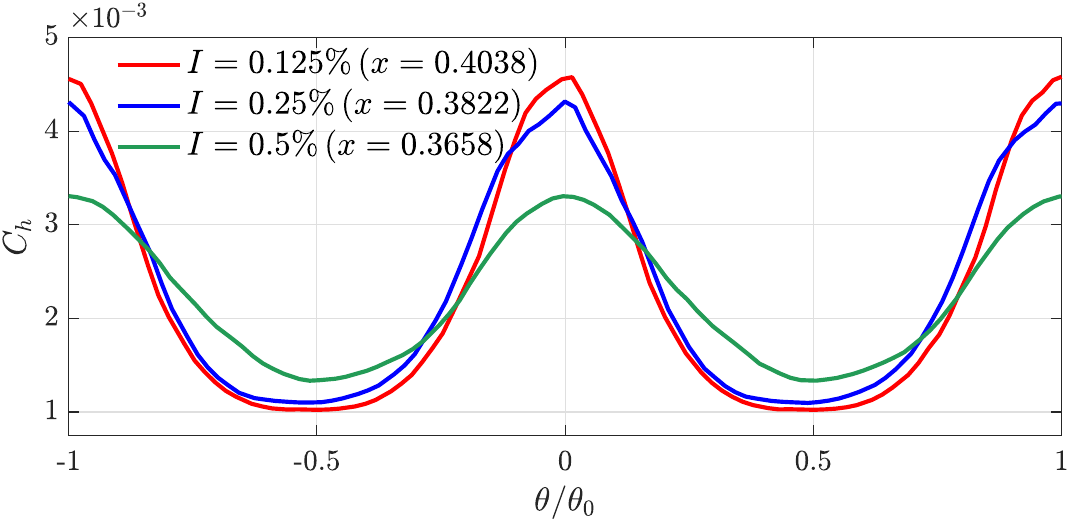}
	\caption{Time-averaged heat transfer coefficient on the azimuthal plane along its maximum value, computed numerically with the three considered free-stream turbulence intensities.}
	\label{fig:cone_Ch_azimuthal}
\end{figure}

Finally, Figure~\ref{fig:cone_Ch_azimuthal} displays azimuthal slices of the time-averaged heat transfer coefficient for the three turbulence intensities.
In particular, for each disturbance intensity, an azimuthal cut along the location where it attains its maximum value is considered.
In particular, smaller turbulence intensities lead to higher amplitudes of the heat transfer coefficient.
In the plot, the longitudinal location of such peak value is indicated for each of the turbulence intensities.

In summary, the ILES simulations show that hypersonic boundary layer transition is sensitive to free-stream disturbances resulting in the development and growth of the second-mode waves and the subsequent breakdown of the Görtler vortices. The good agreement between the computed results of pressure or heat transfer and the experimental data suggests that the natural transition observed in the experiments is initiated by the receptivity of boundary layers to free-stream disturbances. Furthermore, the amplitude of the disturbance has a very significant impact on flow instabilities and the onset of transition. Future studies will investigate how the phases and frequencies of free-stream disturbance impact transition and turbulence in hypersonic flows.

\subsection{Transitional shock wave-boundary layer interaction}

Transitional hypersonic shock wave-boundary layer interactions (SBLI) are one of the most demanding cases for CFD codes. Here, even LES, which resolves all but the smallest turbulent length scales, is insufficient to accurately predict the surface aerodynamic heating downstream of the shock impingement location. At low incidence angles, where free-stream disturbance propagation and amplification dominate the transition process, the course grids of LES are insufficient for numerically supporting the propagation of inlet disturbances along the laminar boundary layer. At higher incidence angles, the boundary layer transitions by the sole action of the shock wave, and the intense spikes in skin friction and aerodynamic heating in the post-shock region were not predicted well by the LES of \cite{Fu2018}. This section presents the study of a Mach 6 flat plate with an impinging shock wave, whose geometry and flow conditions are described in table \ref{tab:geom}.

\begin{table}[H]
\begin{center}
\begin{tabular}{ll}
\textbf{Flow conditions} & \textbf{Geometry parameters}\\
\hline
$Re = 3.9706 \times 10^6/\mbox{m}$ & Plate length, $L_{plate} = 0.595 \mbox{m}$\\
$M_\infty = 6$ & Leading edge radius, $r_{le} = 10 \mu\mbox{m}$\\
$u_\infty = 969.48 \mbox{m/s}$ & SI location, $x = 0.337 \mbox{m}$\\
$T_\infty = 65 \mbox{K}$ & SI wedge angle, $\theta_{sw} = 4^\circ$\\
$T_w = 292.5 \mbox{K}$ & 
\end{tabular}
\end{center}
\caption{Geometry and flow conditions for flat plate SBLI study. Here SI stands for ``shock impingement''.}
\label{tab:geom}
\end{table}

\cite{Sandham2014} previously studied this case both experimentally and numerically through DNS. The experimental results were seen to be sensitive to the leading edge radius. At the computational domain inflow boundary, which is downstream of the flat plate leading edge, a similarity solution was imposed. Density perturbations were added to the inflow boundary layer as shown in Eq. (\ref{eq:rhop}). For the DNS, both the free-stream disturbance amplitude and the shock impingement location were varied. The objectives were to gain a better understanding of transitional SBLI and to explore the effect of boundary layer intermittency. Although the experimental data and the DNS results were similar, there were some quantitative differences. These differences were said to be caused by different disturbance environments in the wind tunnels versus the DNS and the omission of the plate leading edge from the DNS. The highest levels of wall heat transfer were consistently observed for transitional rather than fully turbulent SBLI.

To study the effects of free-stream disturbances on the boundary layer transition of the hypersonic SBLI studied by \cite{Sandham2014}, DNS were performed on a Mach 6 flat plate with an impinging shock wave using our in-house code Exasim \citep{Nguyen2020gpu,Vila-Perez2022,Nguyen2023a}. The 3D simulations were run using the results from 2D laminar simulations as the initial condition. These simulations were run for a total non-dimensional time of $t = 20$ to ensure that the 3D flow was fully developed.
A non-dimensional step size of $\Delta t = 1\times10^{-4}$ was used, thus performing 200,000 time steps. The computational grid consists of more than 10.5 million quadratic hexahedral elements, resulting in a total of almost 300 million grid points. All of the DNS were performed using 92 nodes (552 V100 GPUs) of the Summit supercomputer and took approximately 140 hours of run time.

\begin{figure}[htbp]
\includegraphics[width=0.485\textwidth]{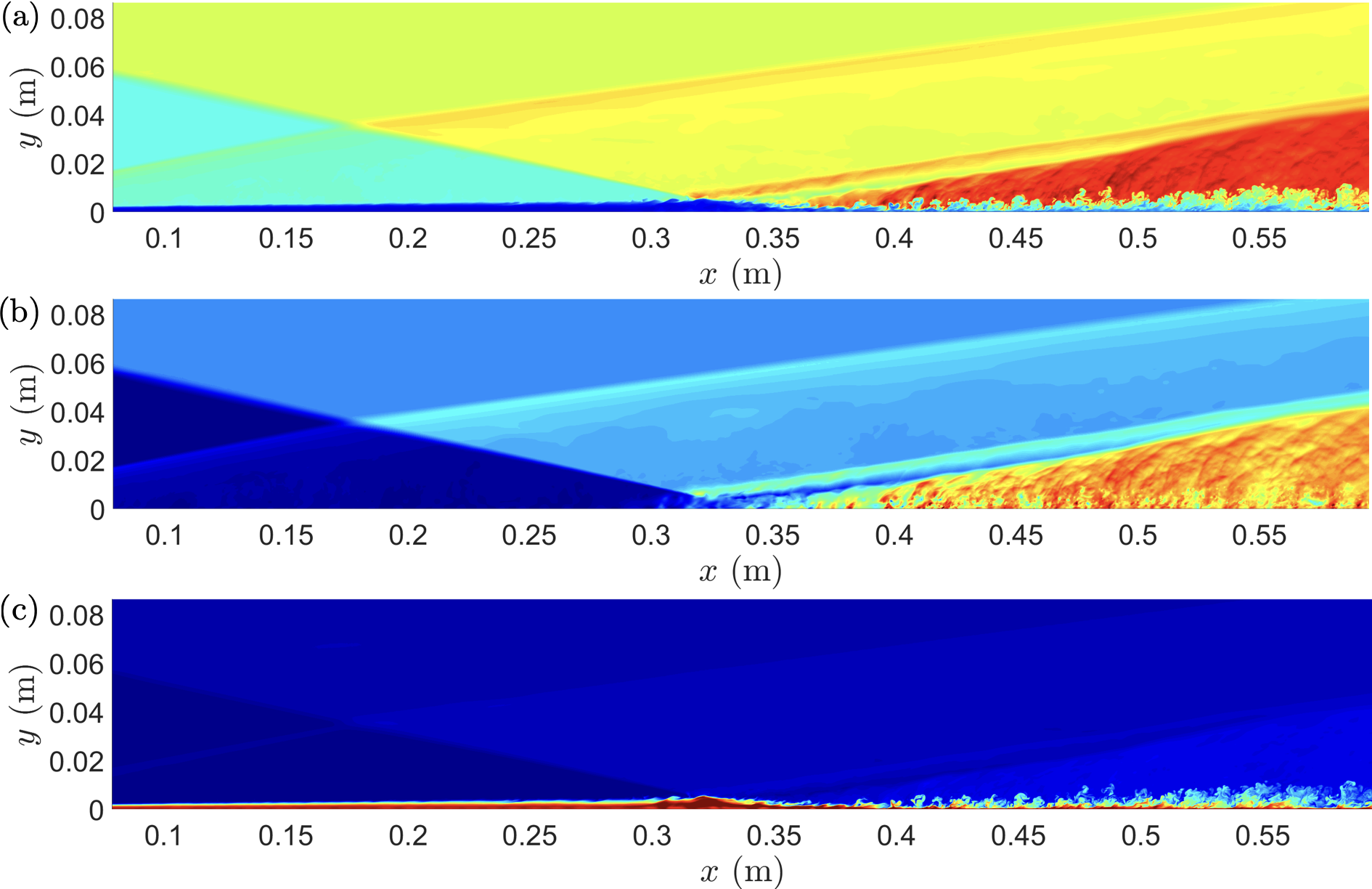}
\caption{(a) instantaneous density $\rho_h/\rho_\infty$, (b) instantaneous pressure $p_h/p_\infty$, (c) instantaneous temperature $T_h/T_\infty$ for the shock wave boundary-layer interaction.}
\label{swbli1}
\end{figure}

\begin{figure}[htbp]
\includegraphics[width=0.485\textwidth]{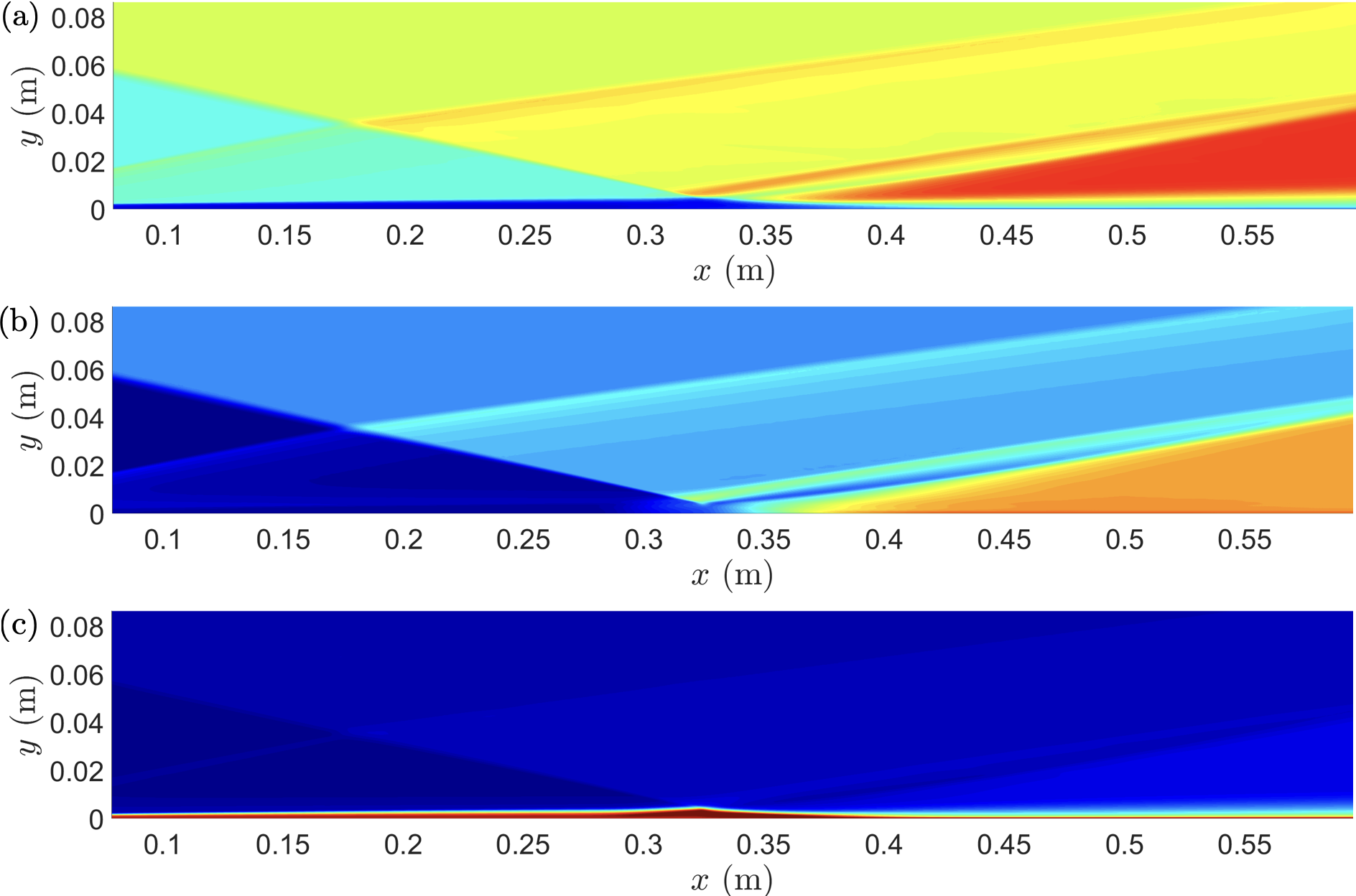}
\caption{(a) time and spanwise average density $\bar{\rho}_h/\rho_\infty$, (b) time and spanwise average pressure $\bar{p}_h/p_\infty$, (c) time and spanwise average temperature $\bar{T}_h/T_\infty$ for the shock wave boundary-layer interaction.}
\label{swbli2}
\end{figure}

Figure \ref{swbli1} shows the instantaneous density, pressure, and temperature fields for a free-stream turbulence intensity of $I = 0.5\%$ and a  spanwise velocity perturbation coefficient of $C = 27.5$. It reveals that the boundary layer has not yet fully transitioned to turbulence before the start of the SBLI. The initial boundary layer disturbances generated at the inflow boundary exhibit a slow amplification rate upstream of the shock impingement location. The impinging shock wave then destabilizes the boundary layer, causing a much more rapid disturbance amplification rate in the boundary layer, ultimately leading to breakdown and transition. Figure \ref{swbli2} shows the time and spanwise average density, pressure, and temperature fields for the same simulation case. The transition to turbulence is observed in the density and temperature plots where the cooler and more dense flow begins to penetrate deeper into the boundary layer just downstream of the reattachment shock. The pressure plot reveals the locations of the leading edge, impinging, separation, and reattachment shock waves involved in this complex flow pattern.

\begin{figure}[htbp]
\includegraphics[width=0.485\textwidth]{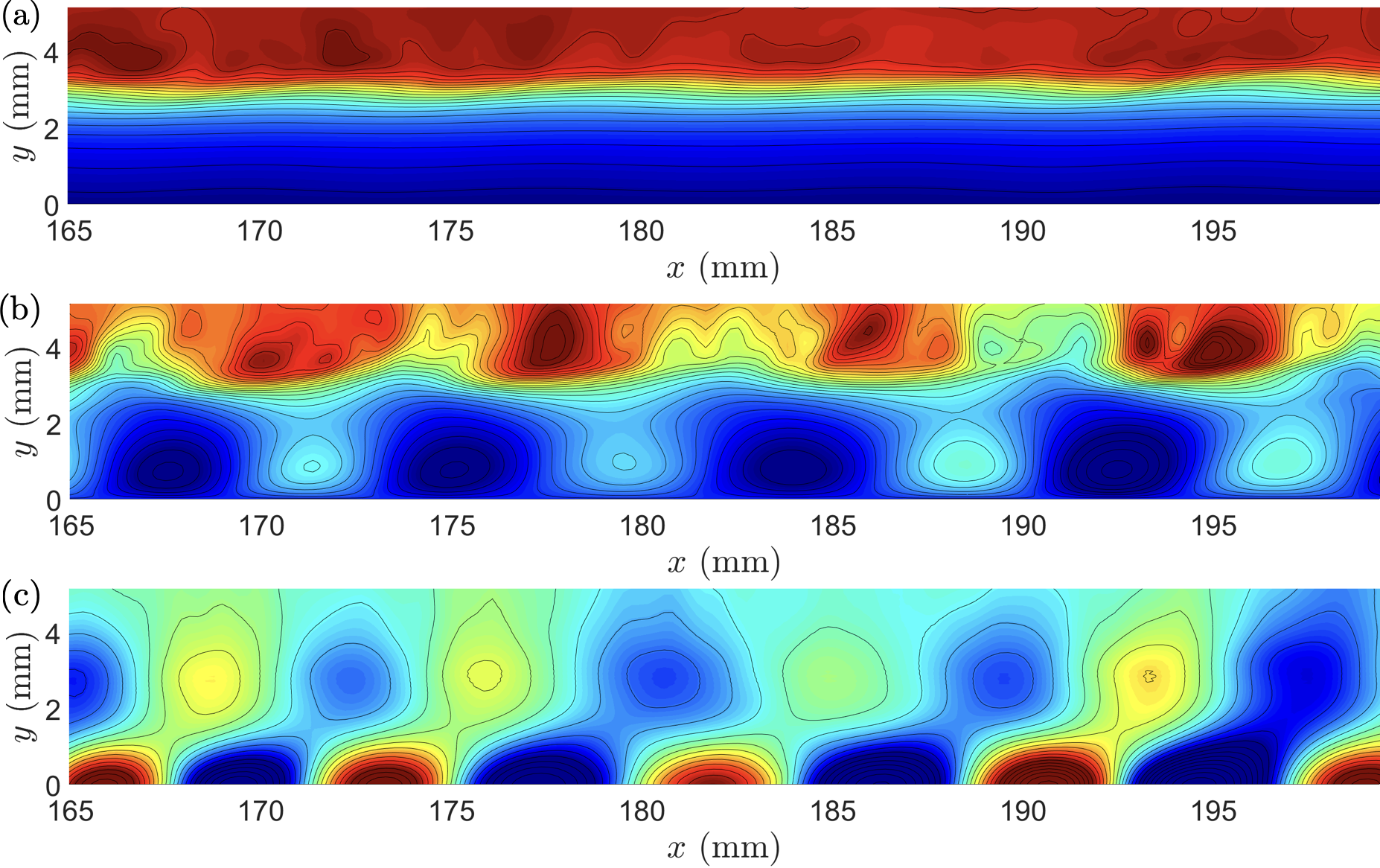}
\caption{Close-up view of the instantaneous solution in the laminar boundary layer: (a) instantaneous streamwise momentum $\rho_h u_h/u_\infty$, (b) instantaneous wall-normal momentum $\rho_h v_h/u_\infty$, (c) instantaneous pressure $(p_h-p_\infty)/p_\infty$ for the shock wave boundary-layer interaction.}
\label{swbli3}
\end{figure}

\begin{figure*}[ht]
	\centering	\includegraphics[width=0.985\textwidth]{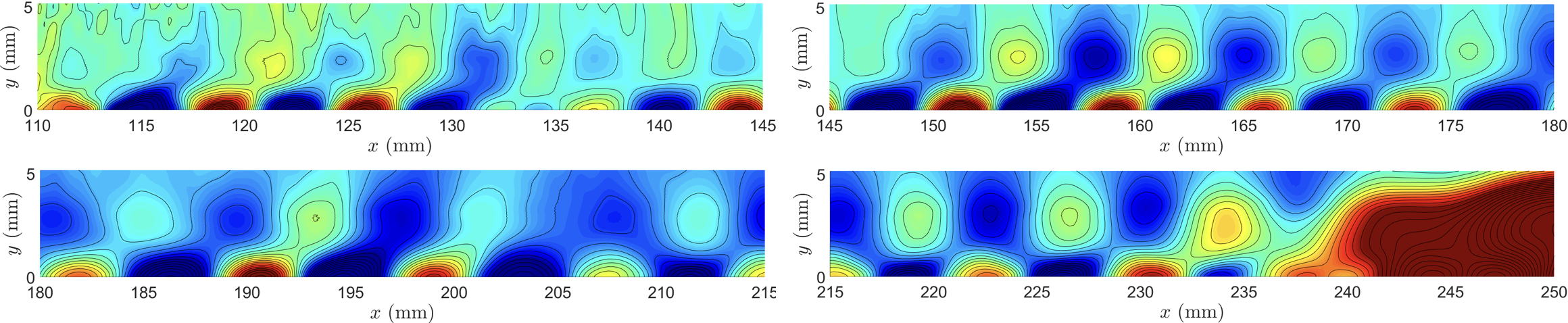}
\caption{Close-up view of the instantaneous pressure $(p_h-p_\infty)/p_\infty$ in the boundary layer at four different intervals of the flat plate for the shock wave boundary-layer interaction. The second mode waves are two-dimensional with approximately $3$ mm length and 1 mm height. The hypersonic laminar boundary layer receives free-stream disturbances which then interact with the second mode waves, resulting in the growth of sinusoidal acoustical waves above the second mode waves.}
	\label{swbli7}
\end{figure*}

The spatial structure of the two-dimensional second mode boundary layer instability wave can be readily seen in Figure \ref{swbli3}, which shows a close-up view of the instantaneous streamwise and wall-normal momentum as well as the pressure. The x-domain of this figure is well upstream of the shock impingement location and thus well before the boundary layer transitions to turbulence. At this point along the flat plate, the boundary layer is approximately $3 \mbox{mm}$ thick. Figure \ref{swbli7} shows a close-up view of just the instantaneous pressure at four different intervals along the flat plate surface. The second mode waves are two-dimensional with approximately $3$ mm length and 1 mm height. The hypersonic laminar boundary layer receives free-stream disturbances which then interact with the second mode waves, resulting in the growth of sinusoidal acoustical waves above the second mode waves. The large pressure rise that occurs at around $x = 240 \mbox{mm}$ is caused by the shock wave that occurs due to the boundary layer separating as a result of the SBLI. The interaction of this second mode wave with free-stream disturbances combined with the destabilization caused by the SBLI are what ultimately lead the boundary layer to transition.

\begin{figure}[htbp]
\includegraphics[width=0.485\textwidth]{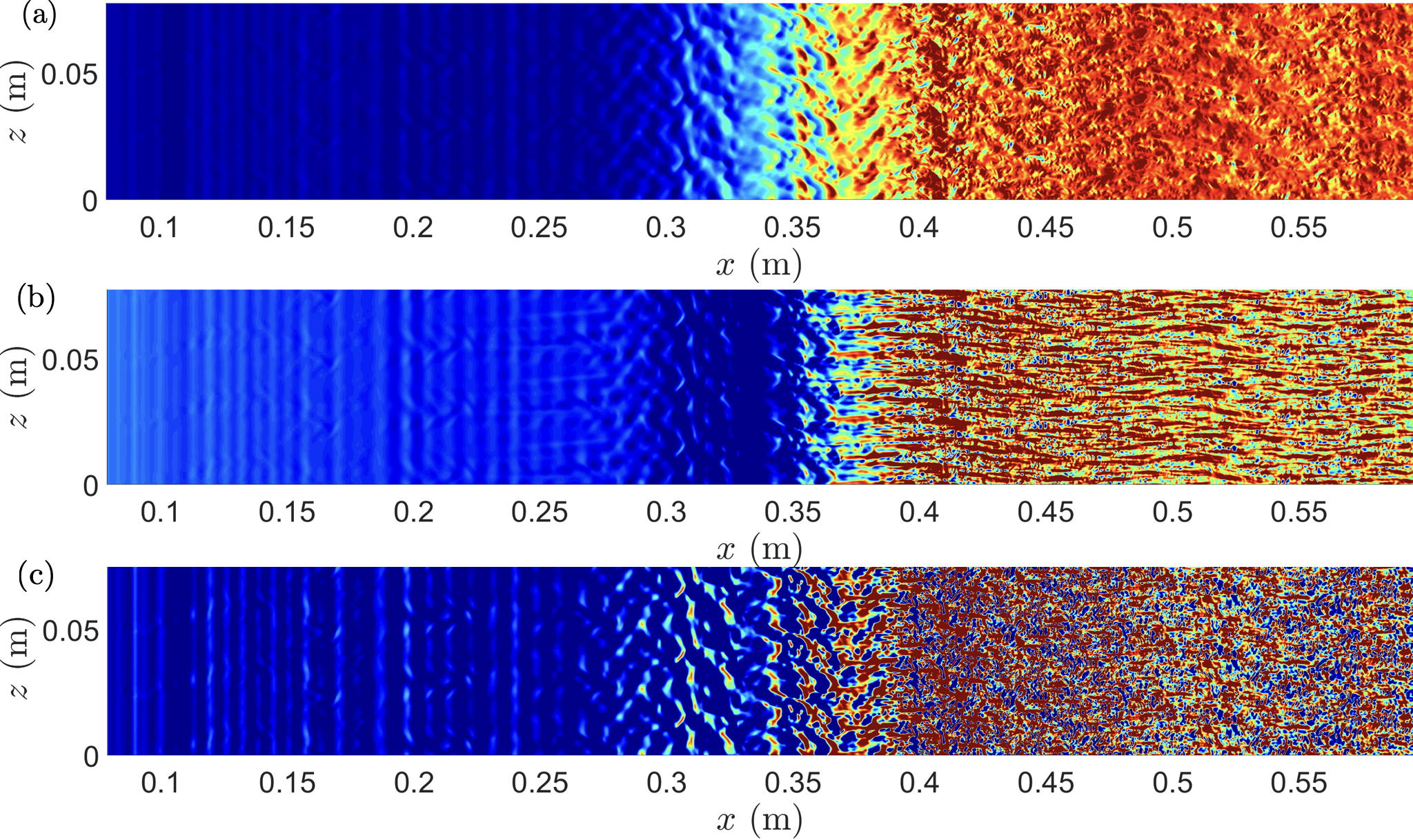}
\caption{Computed surface quantities: (a) instantaneous wall pressure $p_w$, (b) instantaneous skin friction coefficient $c_f$, (c) instantaneous Stanton number $St$ for the shock wave boundary-layer interaction.}
\label{swbli4}
\end{figure}

The macroscopic progression of the second mode wave can be seen in Figure \ref{swbli4}, which shows the instantaneous wall pressure, skin friction coefficient, and Stanton number for the entire computational domain. Near the domain inflow, the second mode remains largely two-dimensional. As it propagates downstream, the spanwise velocity perturbations begin to interact with the second mode wave and cause the boundary layer instabilities to become more and more three-dimensional. At around $x = 0.25 \mbox{m}$, very close to where the large pressure rise due to the separation shock was seen in Figure \ref{swbli7}, the boundary layer disturbance amplitude begins to increase more rapidly. The boundary becomes fully turbulent at around $x = 0.4 \mbox{m}$, which is downstream of the shock impingment location at $x = 0.337 \mbox{m}$

\begin{figure}[htbp]
\includegraphics[width=0.485\textwidth]{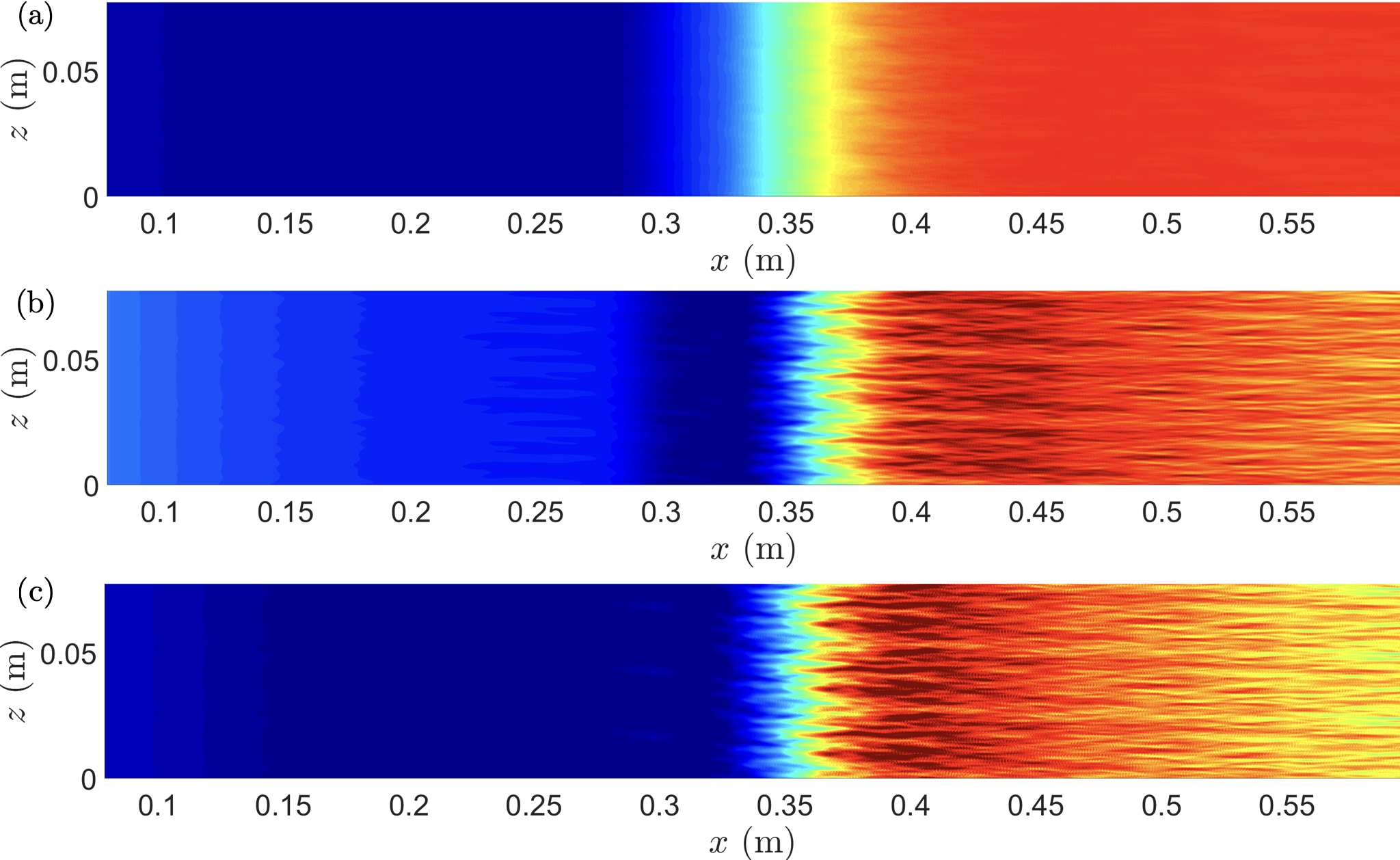}
\caption{Computed surface quantities: (a) time-average wall pressure $p_w$, (b) time-average skin friction coefficient $c_f$, (c) time-average Stanton number $St$ for the shock wave boundary-layer interaction.}
\label{swbli5}
\end{figure}

\begin{figure*}[ht]
	\centering	\includegraphics[width=0.985\textwidth]{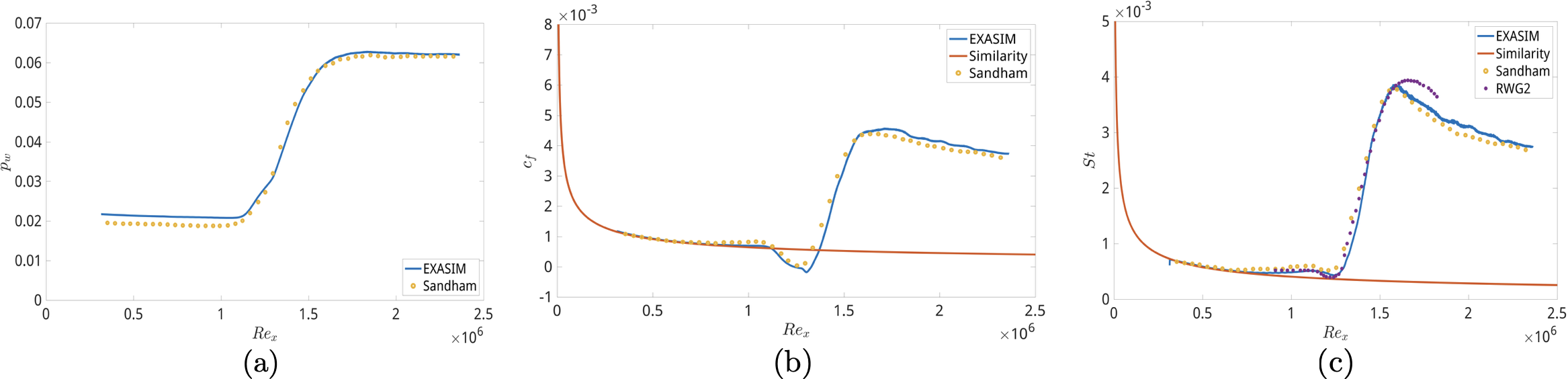}
	\caption{Comparison of time and spanwise average wall pressure (a), skin friction coefficient(b), and Stanton number as a function of the local Reynolds number $Re_x$ between Exasim DNS, Sandham DNS, and RWG2 experimental data. Here $Re_x$ is defined in Appendix A of \cite{Sandham2014}.
	\label{swbli6}}
\end{figure*}

Figure \ref{swbli5} shows the time average wall pressure, skin friction coefficient and Stanton number on the flat plate. These plots reveal a degree of spatial variation of the SBLI in the spanwise direction. To compare our results with those of \cite{Sandham2014} we take a spanwise average of our time averaged solution and plot it along with the Sandham DNS and the RWG2 experimental data in Figure \ref{swbli6}, again for the case with $I = 0.5\%$ and $C = 27.5$. The Exasim DNS results match the Sandham DNS as well as the RWG2 experimental results quite well. With careful tuning of the synthetic turbulence parameters, we were able to predict the minimum Stanton number of the RWG2 experiments better than the Sandham DNS, although both the Sandham and Exasim DNS underpredicted the broadness of the peak Stanton number in the RWG2 experiments.

\begin{figure*}[ht]
	\centering	
 \includegraphics[width=0.985\textwidth]{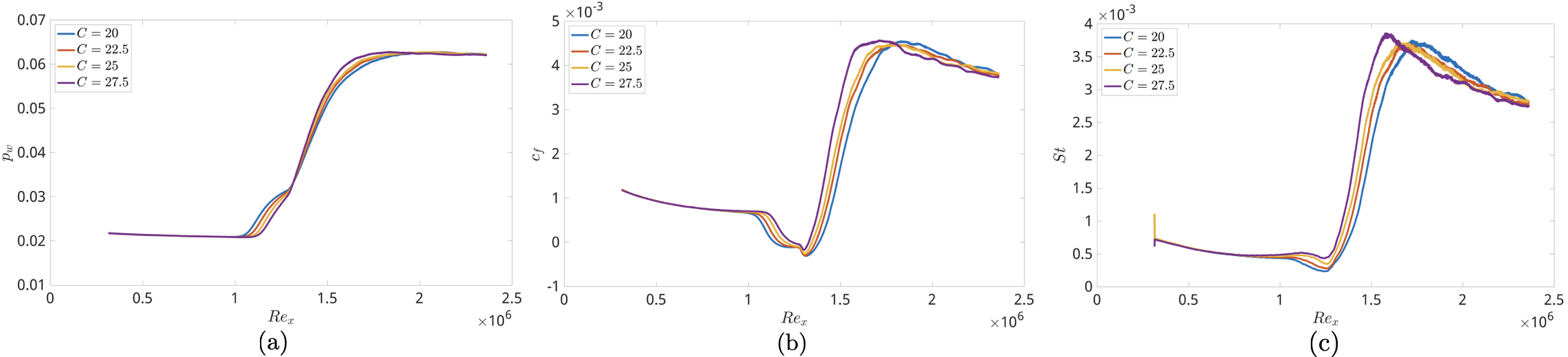}
 	\caption{Comparison of Exasim results for time and spanwise average surface quantities as a function of the local Reynolds number $Re_x$ for different perturbations of the spanwise  velocity amplitude.}
	\label{swbli9}
\end{figure*}

In order to match the RWG2 experimental transition location, the spanwise component of the inflow velocity perturbations had to be multiplied by a rather large factor of $C = 27.5$, while the streamwise and wall-normal perturbation velocity component coefficients are held at unity. The need for this large spanwise perturbation velocity component coefficient is likely due to the presence of the tunnel side walls, which are known to generate rather large acoustic fluctuations in wind tunnel experiments that are the dominant cause of boundary layer transition. Figure \ref{swbli9} shows a comparison of Exasim results of the time and spanwise averaged surface quantities for different values of $C$. As $C$ is increased, the boundary layer separation bubble reduces in size, and the point of rapid increase in pressure, skin friction, and heat transfer moves upstream.

\begin{figure*}[ht]
	\centering	
 \includegraphics[width=0.98\textwidth]{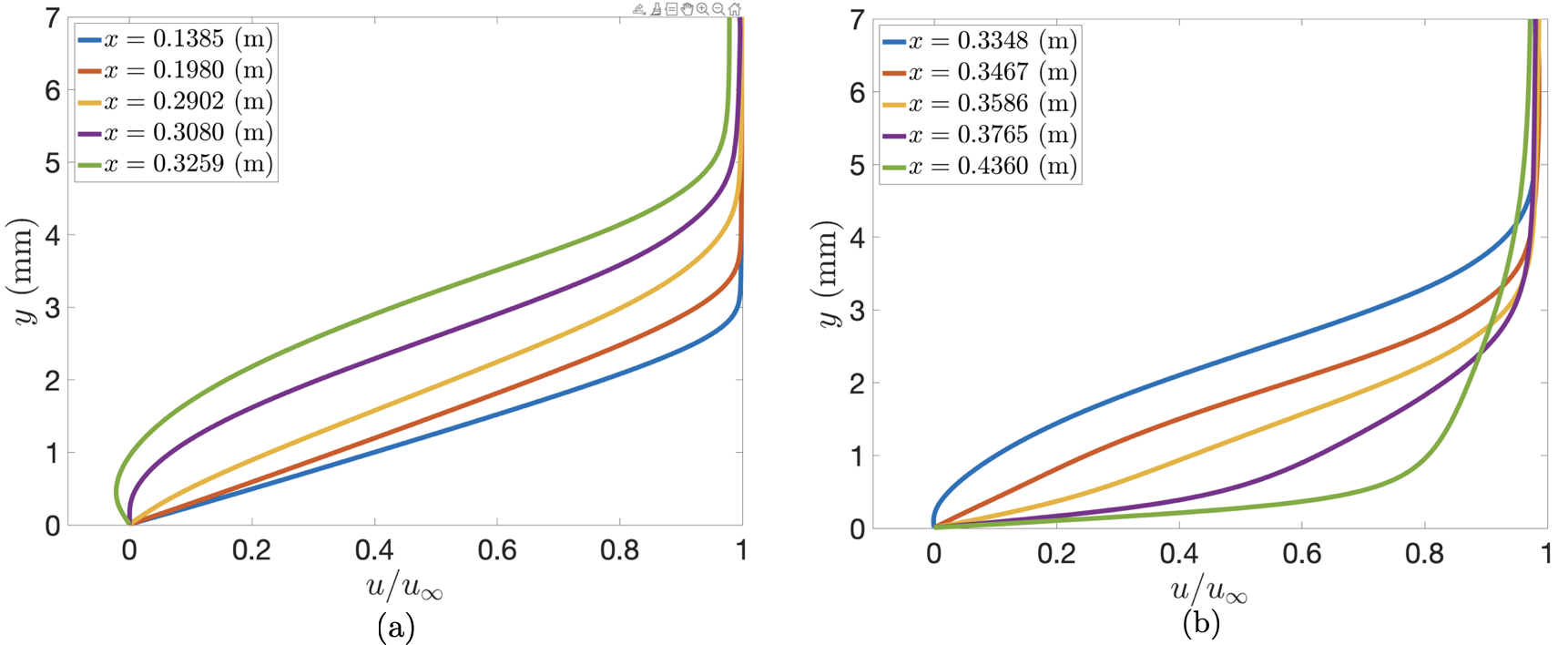}
	\caption{Boundary layer velocity profiles at upstream positions (a) and downstream positions (b) of the shock impingement location $x_{\rm imp} = 0.337$ m.}
	\label{swbli8}
\end{figure*}

Figure \ref{swbli8} shows boundary layer profiles both upstream and downstream of the shock impingement location. The figure shows that the boundary layer is fully laminar at $x = 0.1385 \mbox{m}$ and grows larger in size through $x = 0.1980 \mbox{m}$ and $x = 0.2902 \mbox{m}$. As the separation bubble begins to take effect, the boundary is on the verge of separating at $x = 0.3080 \mbox{m}$, and revered flow is observed at $x = 0.3259 \mbox{m}$. The boundary layer then reattaches at $x = 0.3348 \mbox{m}$, where the flow is no longer reversed. Hence, the interval of the separation bubble is between $x = 0.3080 \mbox{m}$ and  $x = 0.3348 \mbox{m}$, and boundary layer transition occurs within this interval and thus near the shock impingement location $x_{\rm imp} = 0.337$ m. As a result, SBLI has a significant impact on the onset of transition. Downstream of the reattachment, the boundary layer is turbulent through $x = 0.3467 \mbox{m}$, $x = 0.3586 \mbox{m}$, and $x = 0.3765\mbox{m}$. Finally, a fully turbulent boundary layer profile is seen at $x = 0.4360 \mbox{m}$.

\section{Perspectives}

The advancement of high-order DG methods, combined with the development of shock-capturing schemes, mesh adaptivity algorithms, and iterative solvers, has resulted in the successful application of such methods to very complex hypersonic flows. Although this article has covered a broad spectrum of topics and applications, a number of research areas in hypersonic flows have not been discussed in this context. Therefore, we would like to conclude the paper by offering our perspectives on critical aspects such as error estimation, turbulence modeling, and real gas effects.

\subsection{Output-based error estimation}
\label{sec:err}
The mesh adaptivity strategies described above can be highly effective but are critically dependent on the availability of a reliable error estimate. For certain problem classes, rigorous error estimation and even bounds for various measures of the error are achievable  \cite{becker_rannacher_2001, MADAY1999823, AINSWORTH19971}. However, these approaches tend to fail when considering more complex equations, such as those involved in hypersonic flows. In the realm of computational fluid dynamics, and especially in the context of high-speed flows, the application of output-based error estimation has demonstrated significant effectiveness \cite{VENDITTI2000204, Yano2012}. Adjoint-based techniques are particularly useful in this domain, as they can estimate errors in selected solution outputs and offer local indicators for mesh adaptivity. While these methods are commonly paired with h-adaptivity for high-speed flows, there is no fundamental reason why they cannot be effectively integrated with r-adaptivity approaches, as described earlier.

For steady-state problems, the method can be explained by considering the discretized system of equations resulting from our DG discretization: find  $\bm{u}_h \in \bm{\mathcal{V}}_h^k $ such that 
\begin{equation} {\cal R}(\bm{u}_h, \bm{w}_h) = 0, \qquad \forall \bm{w}_h \in \bm{\mathcal{V}}_h^k. \label{primaleq} \end{equation} In this section, we outline the approach for dealing with a straightforward inviscid scenario, but this approach can be similarly applied to more complex situations involving viscous and hybridized problems \cite{FidHDG}. From (\ref{primaleq}), it follows that perturbations to the vector of unknowns, $\delta \bm{u}_h$ and perturbations to the residual $\delta {\cal R} : \bm{\mathcal{V}}_h^k \to \mathbb{R}$ are related as
\begin{equation}  {\cal R}'[\bm{u}_h](\delta \bm{u}_h, \bm{w}_h) + \delta {\cal R}_h(\bm{w}_h) = 0, \qquad \forall \bm{w}_h \in \bm{\mathcal{V}}_h^k, \label{dprimaleq} \end{equation} where the prime denotes the Fr\'echet  derivative of ${\cal R}_h$ with respect to the argument in the square brackets.

If we now consider a scalar functional output of interest $\mathcal{J}: \bm{\mathcal{V}}_h^k \to \mathbb{R}$, the sensitivity of the output $\mathcal{J}_h(\bm{u}_h)$ to perturbations in $\bm{u}_h$, can be written as 
\begin{equation}
    \delta {\cal J} = {\cal J}'[\bm{u}_h](\delta \bm{u}_h) = \delta {\cal R}(\bm{\varphi}_h) =  -{\cal R}'[\bm{u}_h](\delta \bm{u}_h, \bm{\varphi}_h) \label{adjoint}
\end{equation}
where the adjoint $\bm{\varphi}_h \in \bm{\mathcal{V}}_h^k$, satisfies
\begin{equation}
{\cal R}'[\bm{u}_h]( \bm{w}_h, \bm{\varphi}_h) + {\cal J}'[\bm{u}_h]( \bm{w}_h) = 0, \qquad \forall \bm{w}_h \in \bm{\mathcal{V}}_h^k .
\end{equation}
The equation presented above enables us to understand the adjoint as a means to determine how residual perturbations should be weighted in order to calculate perturbations in the desired output. We note that some level of arbitrariness exists in defining the residual and output functional, ${\cal R}_h$ and	${\cal J}_h$, often stemming from the manner in which boundary conditions are imposed. While the impact of different formulations is typically minor for the calculation of the primal solution  $\bm{u}_h$, their influence on the adjoint solutions and the computed sensitivities can be considerable. It is essential that the discrete adjoint equation (\ref{adjoint}) aligns consistently with the continuous adjoint problem. A detailed discussion on issues related to adjoint consistency can be found in \cite{Hartmann07}.

It can be shown (see \cite{FidDar} for instance) that if we consider two discretizations $\bm{\mathcal{V}}_h^k$ and $\bm{\mathcal{V}}_{h_r}^{k_r}$ such that $\bm{\mathcal{V}}_h^k \subset \bm{\mathcal{V}}_{h_r}^{k_r}$ the difference in the output will be given by 
\begin{equation}
{\cal J}( \bm{u}_{h_r}) - {\cal J}( \bm{u}_{h}) = {\cal R}( \bm{u}_h, \bm{\varphi}^{mv}_{h_r} - \bm{\varphi}_h). \label{error_in_functional}
\end{equation}
where $\bm{\varphi}^{mv}_{h_r} \in \bm{\mathcal{V}}_{h_r}^{k_r}$ is an `improved' adjoint satisfying an equation analogous to (\ref{adjoint}) but using a mean-value linearization for the residual and output [refs]. Rather than solving for solving for $\bm{\varphi}^{mv}_{h_r}$, one can estimate the right hand side of (\ref{error_in_functional}) in an element-by-element fashion using only local computations to estimate the error in the ouptut. An added advantage of this approach is that it is very easy to bound the total error by the sum of the absolute values of the element contributions thus obtaining a local error measure that can be used to either refine the mesh \cite{FidDar}, or to provide source terms to drive the $r$-adaptive process in equation (\ref{r-density}).

For time-dependent problems, and in particular for chaotic flows, the adjoint problem is unstable and not useful to estimate errors in time-averaged quantities of interest. Getting sensitivities in such cases in a computationally efficient manner is an active area of research \cite{Qiqi17}. A more practical approach is  based on the use of differentiable dynamic closures \cite{fid2108, FIDKOWSKI2022107843} for the time-averaged solution and which can be effective to drive output-based adaptivity for the high fidelity time-dependent simulations.

\subsection{Turbulence modeling}

For practical hypersonic flow applications one must consider a range of Reynolds numbers which typically result in transitional and turbulent flows. RANS models with second-order finite volume codes have been ubiquitous for hypersonic flow applications \cite{Roy2006}. These RANS models have also been implemented in discontinous Galerkin methods and in some cases used for the soluiton of hypersonic flows \cite{bassi2005discontinuous, nguyen2007rans, OliverDarmofalRANS, IhmeRANS}. A significant drawback of RANS models is that they lack the capability to independently forecast the shift from laminar to turbulent flow, necessitating the integration of an additional transition model. Unfortunately, the effectiveness of transition models is often compromised due to the presence of several, sometimes simultaneous, transition mechanisms, as highlighted by Fedorov (2011). This complexity presents significant difficulties in precisely calibrating these models

As already highlighted in this article, implicit LES simulations can be used effectively to capture the process of transition, including the onset of primary and second-mode instabilities followed by transition to turbulence breakdown \cite{Nguyen2023a}.  One of the key aspects of the LES simulations presented is that while the flow remains laminar, all the relevant scales and the laminar-to-turbulence transition mechanisms are resolved. On the other hand, the implicit LES method utilized, relies on the numerical dissipation of the DG method to model the under-resolved scales. While we have not conducted detailed studies to date, we expect that for more complex flows involving higher Reynolds numbers more sophisticated sub-grid scale models will be required \cite{SMG-Germano, WALE, Vreman}.  

Moving forward, we foresee the development of LES models in conjunction with unified approaches combining subgrid-scale and wall data-driven models informed by machine learning methods trained on simpler canonical that have been successfully employed for low-order finite volume discretizations \cite{PARISH2016758, Adrian2023}..

\subsection{Nonequilibrium effects} 
The compressible Navier-Stokes equations are suitable for modeling a wide range of hypersonic flow configurations. 
For flows at moderate Mach number and altitude, ideal gas assumptions and Sutherland laws for viscosity can accurately model the relevant physical phenomena. 

The strong shocks induced by hypersonic flow around a blunt body convert large amounts of kinetic energy into thermal energy. 
At low enough temperatures, it can suffice to consider a calorically imperfect gas with temperature-dependent specific heat. 
For larger temperatures, the air mixture will undergo dissociation or ionization, its internal energy modes undergo excitation, and gas-surface interactions can occur. 

Flows undergoing significant dissociation or ionization are said to be in chemical nonequilibrium.
In this case, it is necessary to describe air as a mixture of different species and ions using a continuity equation for each species, with significant modifications needed for thermodynamic quantities and transport coefficients. 
Thermal nonequilibrium occurs when the timescales associated with the equilibration of translational, vibrational, rotational, and electronic energy modes are on the same order as the flow velocity timescale.  
This is commonly modeled using the two temperature (2T) models, most notably that of Park \cite{park1989assessment_N, park1988assessment_O}. 
It assumes that rotational and translation energy modes can be described by one temperature, while vibrational and electronic energy modes can be described by another. The highest fidelity description of thermal nonequilibrium can be achieved with state-to-state (StS) models where all internal energy levels are explicitly modeled with a continuity equation. 
These come with a massive computational overhead, so they have only recently been coupled with CFD codes. 
While these models have exhibited a shock standoff location more consistent with experiments for a hypersonic cylinder case \cite{colonna2019impact}, their impact on the wall quantities in double cone flows appears marginal considering the computational cost \cite{wang2023high}.

The particular modeling terms for closing the flow equations will depend on the regimes being simulated. 
Common choices for hypersonic modeling terms can be found in \cite{gnoffo1989conservation, josyula2015hypersonic, scoggins2017development, shang2012nonequilibrium}. 
Each species is typically treated as an ideal gas with the total mixture pressure following Dalton's law. 
Other thermodynamic quantities like specific heats are often modeled using the NASA polynomials or rigid-rotor harmonic-oscillator (RRHO) models. 
Species diffusion coefficients can be modeled using the curve fits of \cite{blottner:70cp} and Wilke mixing rules \cite{wilke1950viscosity} or with more sophisticated models using Chapmann-Enskog theory \cite{magin2004transport}. 
Chemical reactions are captured using laws of mass action using Arrhenius rate laws. 
For thermal nonequilibrium, the second temperature can be incorporated in reaction rates using the models of \cite{park1988assessment_O, park1989assessment_N} or Marrone-Treanor type models \cite{marrone1963chemical, chaudhry2020implementation}.

Interactions between the flow mixture and surface material can also become relevant.
These interactions are particularly important for the simulation of thermal protection systems (TPS) undergoing reentry. 
The incorporation of wall catalycity, ablation, or surface radiation can be necessary to accurately model the flow field and degradation of the material. 
Each of these gas-surface interactions can be captured using boundary conditions of varying complexity, as described in \cite{maclean2011finite, marschall2015surface}.
Wall catalysis is often captured with phenomenological specified reaction efficiency (SRE) models, though finite rate chemistry (FRC) models for wall catalysis have also been implemented for some common reactions and demonstrated in FVM codes \cite{josyula2015hypersonic, capriati2021development}. 
Ablation can be simulated using so-called $B'$ tables if the wall and flow are assumed to be in equilibrium, though more detailed models are available \cite{zhluktov1999viscous}.
For significant ablative effects, surface recession should also be accounted for. 
These effects can be modeled with boundary conditions to the flow, though they are more accurately modeled by coupling flow and surface response solvers or by modeling the flow through a porous medium, depending on the material \cite{schroeder2021coupled, stern2019nonequilibrium}
           
These models have long been incorporated into FV codes and have seen recent adoption in the stabilized finite element community \cite{sabo2022investigation, codoni2022streamline, seguin2019finite, gao2019finite_II, pointer2022influence}, but they are much less frequently seen in DG methods. 
A calorically imperfect perfect gas model was considered in \cite{ching2022computation} for cylinder computations. 
The Park 2T model was used in \cite{papoutsakis2014discontinuous} 
for reacting double cone flows. 
In \cite{may2021hybridized} chemical nonequilibrium models are incorporated with an HDG solver shown for 1D and 2D problems, while entropy stable fluxes are derived in \cite{peyvan2022high} for an entropy stable DGSEM scheme applied to 1D problems.
Chemical reactions in the shock layer tend to lead to extremely sharp jumps in temperature, features that may be well-suited for implicit shock tracking; indeed, \cite{luo2023moving} recently applied the MDG-ICE method to high-speed inviscid cylinder flows in thermochemical nonequilibrium using a 2T model.
In terms of DG solvers that use gas-surface interactions, the literature is very sparse; a notable exception is \cite{schrooyen2016fully}, which models the interaction of compressible reacting flow and porous media with a volume averaging method and an implicit BR2 scheme.

\begin{figure}[htbp]
	\centering
	\includegraphics[width=\columnwidth]{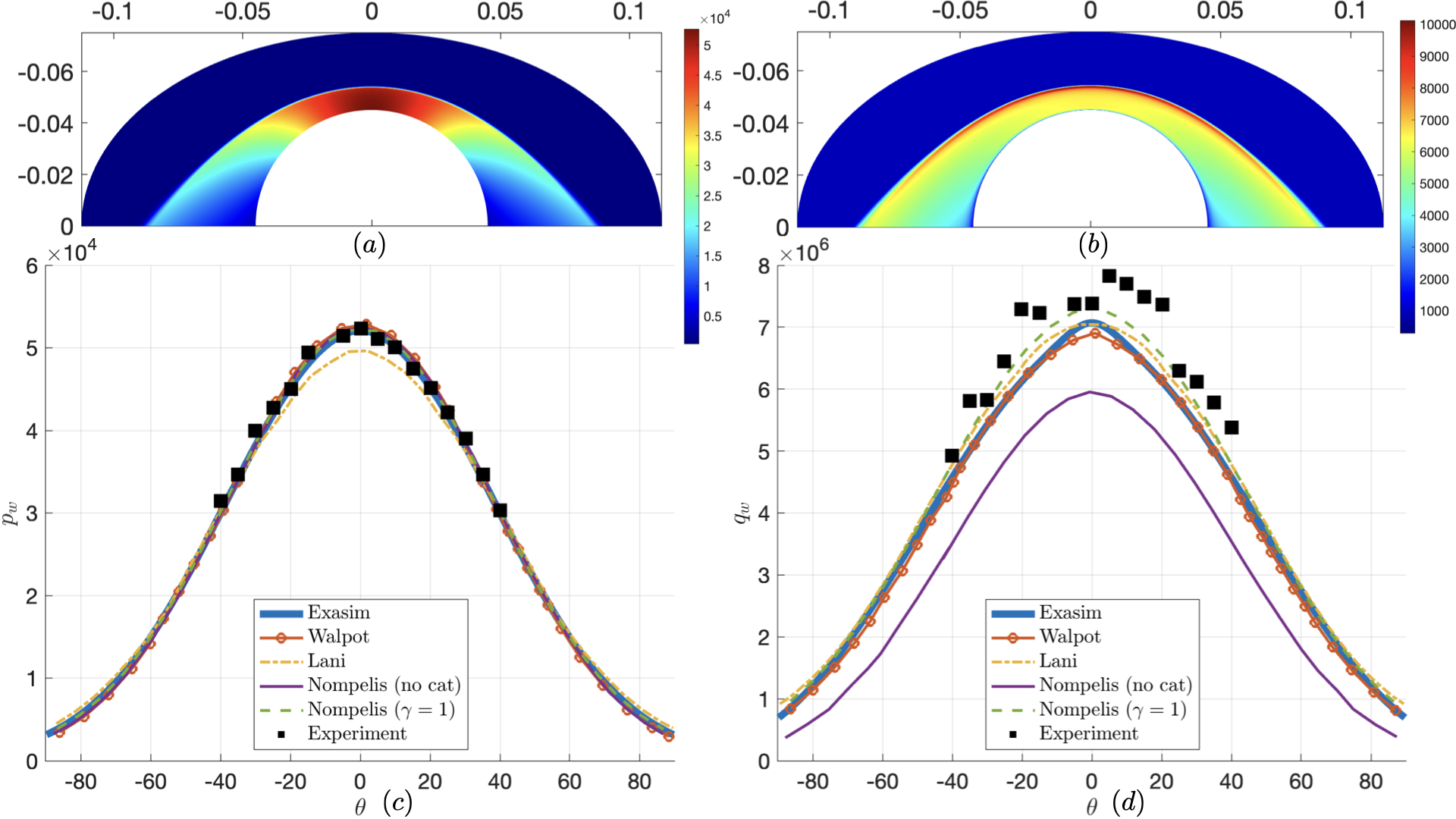}
	\caption{Exasim solutions for HEG-I hypersonic cylinder using Laplacian artificial viscosity and Mutation++ for modeling of chemical nonequilibrium effects. Plotted quantities are pressure (a), temperature (b), wall pressure (c), and wall heat flux (d) compared to simulation and experimental results from \cite{Knight2012}.}
	\label{fig:cyl_react}
\end{figure}

The latter four works all use the open-source software Mutation++ to compute the thermochemical closures \cite{scoggins2020mutation++}.
The transport models offered by Mutation++ based on perturbative Chapmann-Enskog expansions of the Boltzmann equation can be particularly challenging to implement while providing qualitatively different results. 
Exasim has been coupled with Mutation++ to simulate the high enthalpy shock tunnel (HEG) flow over a cylinder, Case I with an isothermal wall as described in \cite{Knight2012}. Laplacian artificial viscosity was used on a structured grid of 250 elements in the axial and radial directions with polynomial order $k=2$. 
A 5 species air mixture is used, with thermodynamic quantities given by the NASA-9 polynomials and transport coefficients calculated using Chapmann-Enskog formulations. 
The results are shown in Figure \ref{fig:cyl_react} and compared to the experimental and numerical results collected in \cite{Knight2012}.  
The numerical results and experimental values show good agreement for wall pressure. 
For simulations that use the curve fits of \cite{blottner:70cp} and mixing rules for transport terms a fully-catalytic wall is needed to match the experimental results. 
This is observed in the results from Nompelis, which use an SRE model with $\gamma=1.0$ to model catalysis. 
Solvers that use the more sophisticated transport models but without wall catalysis, including Exasim and the results by Lani, have generally larger heat flux values. This has also been observed in recent studies \cite{bacskaya2023assessment, maier2021su2}. Still, it is pointed out in \cite{bacskaya2023assessment} that a catalytic wall gives the closest match to the experimental values when using Chapman-Enskog methods for transport. 


The difference between transport models is not always substantial however, and numerous codes for steady and unsteady hypersonic flows report accurate results with simpler curve fits and mixture rules \cite{howard2017towards, di2020htr, passiatore2021finite}

The availability of open-source tools such as Mutation++ can hopefully spur more investigations of the coupling between DG methods and realistic thermochemical modeling for hypersonics. 
Best practices for AV methods and nonlinear and linear solvers may need to reevaluted for flows in thermochemical nonequilibrium.

At higher speeds and altitudes, the gas becomes rarefied; as the local Knudsen number increases and collisions become increasingly infrequent, the continuum assumptions of the Navier-Stokes equations can break down. 
Moderately rarefied flow can be modeled with particular slip boundary conditions detailed in \cite{candler2019rate}, 
though flows with an expected large variation in the mean-free path can be more consistently modeled with the Boltzmann equation. 
For large-scale hypersonic problems, probabilistic direct simulation Monte Carlo (DSMC) approaches are typically used, though recent works have shown promise in using DG methods for deterministic solutions of high-speed flows using the Boltzmann equation \cite{su2020implicit, dzanic2023towards}. 




\section*{Acknowledgements} \label{}

Dr. Nguyen and Prof. Peraire gratefully acknowledge the United States Department of Energy under contract DE-NA0003965 for supporting this work. Dr. Nguyen, Dr. Vila-P\'erez and Prof. Peraire also acknowledge the National Science Foundation, under grant number NSF-PHY-2028125, and the MIT Portugal program under the seed grant number 6950138. Dr. Nguyen also acknowledges the Air Force Office of Scientific Research under Grant No. FA9550-22-1-0356 for supporting this work. This research used resources of the Oak Ridge Leadership Computing Facility at the Oak Ridge National Laboratory, which is supported by the Office of Science of the U.S. Department of Energy under Contract No. DE-AC05-00OR22725. The authors would like to thank the Oak Ridge Leadership Computing Facility for providing access to their Summit GPU supercomputer. Dominique Hoskin acknowledges the support provided by a MathWorks Fellowship.  

\printcredits

\bibliographystyle{cas-model2-names}

\bibliography{library, library_loek}





\end{document}